\documentclass[prd,aps,twocolumn,a4paper,nofootinbib]{revtex4-1}

\newif\ifusesec
\usesectrue 

\usepackage{graphicx,psfrag}
\usepackage{mathrsfs}
\usepackage{amsmath,amsfonts,amssymb}
\usepackage{multirow}
\usepackage{comment}

\DeclareSymbolFontAlphabet{\mathrsfs}{rsfs}
\DeclareMathAlphabet\mathbfcal{OMS}{cmsy}{b}{n}

\newcommand{\be}{\begin{equation}}
\newcommand{\ee}{\end{equation}}
\newcommand{\bea}{\begin{eqnarray}}
\newcommand{\eea}{\end{eqnarray}}
\newcommand{\bel}{\begin{align}}
\newcommand{\eel}{\end{align}}

\newcommand{\scri}{{\mathrsfs{I}}}

\def\i{{\rm i}}

\def\GMc2{G M_{\odot} c^{-2}}

\def\O{\mathcal{O}}

\def\F{{\cal F}}
\def\lm{{\ell m}}
\def\teobLR{{t^{\rm EOB}_{\Omega \, \rm peak}}}
\def\tdomgmax{{t^{\rm NR}_{\dot{\omega}_{22}\, \rm peak}}}
\def\tAmax{{t_{A_{22} \, \rm peak}^{\rm NR}}}
\def\tnrLR{{t_{\rm extr}^{\rm NR}}}
\def\lm{{\ell m}}

\def\v{v_\varphi}

\def\de{\partial}
\def\lm{{\ell m}}

\def\l{{\ell }}

\def\F{{\cal F}}

\def\O{{\cal O}}

\def\k{{\hat{\hat{k}}}}

\DeclareSymbolFontAlphabet{\mathrsfs}{rsfs}
\DeclareMathAlphabet{\mathcal}{OMS}{cmsy}{m}{n}

\usepackage{color}

\definecolor{cyan}{rgb}{0,0.9,0.9}
\definecolor{orange}{rgb}{0.9,0.5,0}
\definecolor{magenta}{rgb}{1,0,1}
\definecolor{purple}{rgb}{0.8,0.4,0.8}
\definecolor{gray}{rgb}{0.8242,0.8242,0.8242}

\begin{document}

\title{Improved effective-one-body description of coalescing nonspinning  black-hole binaries and its numerical-relativity completion}

\author{Thibault \surname{Damour}$^1$}
\author{Alessandro \surname{Nagar}$^1$}
\author{Sebastiano \surname{Bernuzzi}$^2$}

\affiliation{$^1$Institut des Hautes Etudes Scientifiques, 91440 Bures-sur-Yvette, France}
\affiliation{$^2$Theoretical Physics Institute, University of Jena, 07743 Jena, Germany}

\date{\today}

\begin{abstract}
We improve the effective-one-body (EOB) description of nonspinning coalescing black hole binaries
by incorporating several recent analytical advances, notably: (i) logarithmic contributions 
to the conservative dynamics; (ii) resummed horizon-absorption contribution to the orbital 
angular momentum loss; and (iii) a specific radial component of the radiation reaction force implied
by consistency with the azimuthal one. We then complete this analytically improved EOB model by
comparing it to accurate numerical relativity (NR) simulations performed by the Caltech-Cornell-CITA
group for mass ratios $q=(1,2,3,4,6)$. In particular, the comparison to NR data allows us to determine
with high-accuracy ($\sim 10^{-4}$) the value of the main EOB radial potential: $A(u;\,\nu)$, where 
$u=GM/(R c^2)$ is the inter-body gravitational potential and $\nu=q/(q+1)^2$ is the symmetric mass ratio.
We introduce a new technique for extracting from NR data an intrinsic
measure of the phase evolution, 
($Q_\omega(\omega)$ diagnostics). Aligning the NR-completed EOB quadrupolar waveform and the NR one at
low frequencies, we find that they keep agreeing (in phase and amplitude) within the NR uncertainties
throughout the evolution for all mass ratios considered. We also find good agreement for several 
subdominant multipoles without having to introduce and tune any extra parameters.  
\end{abstract}


\maketitle

\section{Introduction}
\label{sec:intro}

The Effective One Body (EOB) formalism 
\cite{Buonanno:1998gg,Buonanno:2000ef,Damour:2000we,Damour:2001tu,Damour:2008gu}
has been proposed as a new analytical method for describing the motion and radiation of
coalescing black hole binaries. 
One of its main aims is to provide analytical~\footnote{Here we use the
adjective ``analytical'' (instead of ``semi-analytical'') for methods that
are based on solving analytically given ordinary differential equations, even if one needs to use 
numerical tools to solve them.} 
gravitational wave (GW) templates covering the full coalescence process,
from early inspiral to ringdown, passing through late inspiral, plunge
and merger.
The definition of the EOB formalism mainly relies on two sources of information:

(i) high-order results of post-Newtonian (PN) theory; 

(ii) high-accuracy results from Numerical Relativity (NR) 
simulations of coalescing black hole binaries
(both in the comparable-mass case, $\nu =\O(1)$, and in the extreme-mass-ratio limit, $\nu \ll1$). 
[Here, $\nu \equiv m_1 m_2/(m_1+m_2)^2$ denotes the symmetric mass ratio.]

In addition, EOB theory has recently tapped useful information out of Gravitational Self Force (GSF)
computations at order $\O(\nu)$.  All this information is not used in its original form, but rather
as a way to determine, or at least constrain, the structure of the few basic functions that enter 
the definition of the EOB formalism. For recent general reviews of the
EOB formalism and its historical roots,
see~\cite{Damour:2009ic,DamourPrague:2012}. 

The EOB formalism has been developed in a sequence of papers, both for
nonspinning black hole binaries~\cite{Buonanno:1998gg,Buonanno:2000ef,Damour:2000we,Damour:2007xr,Damour:2007yf,Damour:2008gu}
and for spinning ones~\cite{Damour:2001tu,Buonanno:2005xu,Damour:2008qf,Barausse:2009xi,Nagar:2011fx, Barausse:2011ys}. 
In addition, it has been extended
to the case of tidally interacting neutron star binaries~\cite{Damour:2009wj,Bini:2012gu}. 
For all those types of systems,  many comparisons between the predictions
of  EOB theory and the results of NR simulations have been performed~\cite{Damour:2002qh,Buonanno:2006ui,Buonanno:2007pf,
Damour:2007yf,Damour:2007vq,Damour:2008te,Damour:2009kr,Buonanno:2009qa,Baiotti:2010xh,Baiotti:2011am,Damour:2011fu,Bernuzzi:2012ci,Pan:2011gk,Taracchini:2012ig}
 and have demonstrated  that it is possible to devise accurate EOB
 waveforms by combining improved resummation 
methods~\cite{Damour:2007xr,Damour:2007yf,Damour:2008gu},  high-order
PN results (see~\cite{Blanchet:2006zz} for a review), 
and some nonperturbative information coming from high-accuracy NR
results. These EOB waveforms can be used both in GW detection and in GW
parameter-estimation protocols. The EOB formalism can thereby crucially help detecting the
GWs emitted by coalescing black hole binaries, since many thousands of waveform templates need
to be computed to extract the signal from the broad-band noise, an impossible task for NR alone.
The EOB formalism might also be crucial in allowing one to extract information on the equation of
state of nuclear matter from observations of coalescing neutron star binaries~\cite{Damour:2012yf}.
An early version of the EOB waveform~\cite{Pan:2011gk} has already been 
incorporated\footnote{See https://www.lsc-group.phys.uwm.edu/daswg/projects/lalsuite.html.},
and used~\cite{Brown:2012nn}   in the LIGO and Virgo search pipeline.

In addition, some recent comparisons between NR studies of the {\it dynamics} of black hole binaries
and its EOB description,   have directly confirmed the ability of EOB theory to accurately describe 
several (gauge-invariant) aspects of the conservative dynamics of binary systems, such as 
periastron precession~\cite{LeTiec:2011bk}  and  the relation between energy and 
angular momentum~\cite{Damour:2011fu}.

The aim of the present paper  is to improve the definition of some of the basic elements
of the EOB formalism both by including for the first time recently obtained analytical
information, and by  extracting, in a new way, nonperturbative information 
from accurate NR simulations performed by the Caltech-Cornell-CITA group~\cite{Buchman:2012dw}.
Though our study will be limited to nonspinning binaries, the  EOB structures we shall
improve (such as the basic EOB radial potential $A(R)$)  are central, and should then be
included both in the spinning and tidal extensions of the EOB formalism.

The recent analytical progresses that we shall incorporate here in EOB theory are:
\begin{itemize}
\item[(i)] 4PN and 5PN logarithmic contributions to the conservative 
   dynamics~\cite{Damour:2009sm,Blanchet:2010zd,Damourlogs,Barausse:2011dq};

\item[(ii)]  the  $\O(\nu)$ 4PN nonlogarithmic contribution to 
     the conservative dynamics ~\cite{Blanchet:2010cx,Blanchet:2010zd,LeTiec:2011dp,Barausse:2011dq};

\item[(iii)] resummed horizon-absorption contributions to angular momentum loss ~\cite{Nagar:2011aa,Bernuzzi:2012ku};

\item[(iv)] the radial component of the radiation reaction force implied by consistency with the azimuthal one~\cite{Bini:2012ji};

\item[(v)] an additional 3.5PN contribution to the phase of the (factorized~\cite{Damour:2007xr,Damour:2007yf,Damour:2008gu}) 
           quadrupolar waveform~\cite{Faye:2012we}.
\end{itemize}

In addition, we shall bring up some novelties in the definition of the EOB formalism, and in the way to extract
information from (comparable-mass) NR data. Namely:
\begin{itemize}
\item[(a)]  we introduce a Pad\'e resummation of the additional tail phases $\delta_\lm$ of the factorized EOB
waveform;

\item[(b)] we show how to accurately extract from NR data the $Q_\omega(\omega)$  function measuring,
in an {\it intrinsic} way, the phase evolution of the (curvature) quadrupolar waveform;

\item[(c)] we introduce a new way to improve the EOB waveform during plunge and merger by matching it
to the NR one at a specifically chosen ($\nu$-dependent) NR time $\tnrLR(\nu)$ around merger.
More precisely, we impose [by using six next-to-quasi-circular (NQC) parameters]
a $C^2$ contact condition between the amplitudes and the frequencies of the
NR and EOB waveforms at an NR instant $\tnrLR(\nu)$, corresponding to the maximum of the
EOB orbital frequency $\teobLR$.
\end{itemize}
The paper is organized as follows. In Sec.~\ref{sec:EOBtheory}  we present, in a self contained manner, 
the detailed definition of our improved EOB formalism. Section~III explains how to extract the $Q_\omega(\omega)$  
function from NR data while Sec.~IV revisits the extreme-mass-ratio case. In Sec.~V we then complete 
our new EOB formalism by comparison with several comparable-mass simulations performed by 
the Caltech-Cornell-CITA group. Section~VI studies the structure of the main EOB radial potential 
($A(u)$ function) obtained from the latter NR comparison and Sec.~VII discusses how to compute 
EOB waveforms for arbitrary values of $\nu$. We summarize our main conclusions in Sec.~VIII, 
while some supplemental material is presented in several Appendixes. In particular, 
Appendix~\ref{sec:rholm} gives the explicit expressions of  the $\rho_\lm$ 
and $\delta_\lm$ bricks of the EOB factorized waveform we use.

\section{Effective-One-Body analytical framework}
\label{sec:EOBtheory}

In this section we shall present in detail the definition of the 
new (nonspinning) EOB formalism,
incorporating several recent analytical improvements that we shall use in this paper. 
Our presentation will be self-contained so as to allow readers to generate for
themselves all our EOB results. We also intend to make available soon a public
version of our EOB codes.  

The EOB formalism is made of  three basic building blocks: (i) a EOB Hamiltonian that resums 
the conservative two-body dynamics; (ii) a resummed EOB radiation reaction force that completes
the conservative dynamics by causing the system to inspiral down to merger, and
(iii) a resummed EOB inspiral-plus-plunge waveform, together with a 
prescription for extending the waveform through merger and ringdown.  Each one of these building
blocks has been developed in previous papers. In particular, the construction of the EOB Hamiltonian
was initiated in Refs.~\cite{Buonanno:1998gg,Damour:2000we}, while the definition of the resummed,
factorized inspiral waveform was initiated in Refs.~\cite{Damour:2007xr,Damour:2007yf,Damour:2008gu}.
Here we bring new (recently derived) theoretical improvements to each element of the formalism, namely:  
(i) we include logarithmic contributions~\cite{Damour:2009sm,Blanchet:2010zd,Damourlogs,Barausse:2011dq}
to the EOB Hamiltonian;  (ii) we include the effect of a resummed version of horizon absorption~\cite{Nagar:2011aa,Bernuzzi:2012ku}
in the radiation reaction; (iii) we add a recently derived~\cite{Bini:2012ji} radial component of radiation reaction; 
(iv)  we include the 3.5PN contribution~\cite{Faye:2012we} to the phase
 $\delta_{22}$ of  the factorized quadrupolar waveform; (v)  we resum  $\delta_{22}$, 
as well as some higher-multipoles $\delta_{\l m}$'s, by Pad\'e methods.
All these improvements either add some new physics  that was not included in
the previous EOB models~\cite{Damour:2009kr,Pan:2011gk}, or improve [in the case of (v)]  
the robustness of the EOB resummations. We shall discuss them in detail in the sections below.

\subsection{Improved Hamiltonian: logarithmic contributions to the A function}

The conservative (nonspinning) two-body dynamics is described, within the EOB formalism, by a Hamiltonian
$ H_{\rm EOB} (Q^i, P_i)$,  describing the {\it relative} motion $Q^i = Q_1^i - Q_2^i$ of the binary, and
depending on two radial functions, $A(R)$ and $B(R)$, where $R\equiv |Q^i|$ is the binary separation (in EOB coordinates).  
We are using phase space variables $(R , P_R , \varphi , P_{\varphi})$ associated to polar coordinates in the equatorial plane 
$\theta = \pi/2$. Actually it is useful  to replace the radial momentum $P_R$ by the momentum 
$P_{R_*} = (A/B)^{1/2} \,P_R$ conjugate to the ``tortoise'' 
radial coordinate $R_* = \int dR (B/A)^{1/2}$. Furthermore, it is convenient to
use suitably rescaled dimensionless  variables: 
\be \label{rescale}
r = \frac R{GM}, \,  p_{r_*} = \frac{P_{R_*} }{ \mu}, \, p_{\varphi} = \frac{P_{\varphi} }{ \mu \, 
GM}, \; t=\frac{T}{GM} \; .
\ee
Here,  and in the following, we use the notation
\be
M \equiv m_1 + m_2, \; \mu \equiv \frac{m_1 m_2}{m_1+m_2}, \; \nu \equiv \frac{\mu}{M}, \; q \equiv \frac{m_1}{m_2}.
\ee
Note that the dimensionless {\it symmetric mass ratio} $\nu = m_1 m_2/(m_1+m_2)^2 = q/(q+1)^2$  varies between $0$ (extreme mass-ratio case) 
and $\frac14$ (equal-mass case), and that we shall conventionally consider that $m_2 \leq m_1$, so that $ q \geq 1$.
In  addition we generally set $c=1$, and shall also often set $G=1$ in the following.

With the above notation, the $\mu$-rescaled (real) EOB Hamiltonian reads
\begin{equation}
\label{eq_H_EOB}
\hat H_{\rm EOB} (r,p_{r_*} , p_\varphi) \equiv \frac{H_{\rm EOB}}{\mu} 
= \frac{1}{\nu}\sqrt{1 + 2 \nu \, (\hat H_{\rm eff} - 1)} \, ,
\end{equation}
where $\hat H_{\rm eff}$ denotes the ($\mu$-rescaled) {\it effective} EOB Hamiltonian, given by
\begin{equation}
\label{eqn11}
\hat H_{\rm eff} = \sqrt{p_{r_*}^2 + A(r) \left( 1 + \frac{p_{\varphi}^2}{r^2} + z_3 \, \frac{p_{r_*}^4}{r^2} \right)} \, ,
\end{equation}
with $z_3 = 2\nu \, (4-3\nu)$. 

The (rescaled) EOB Hamiltonian \eqref{eq_H_EOB} leads to equations 
of motion for $(r,\varphi,p_{r_*},p_{\varphi})$  with respect to the {\it rescaled time} $t = T/GM$,  Eq. \eqref{rescale} , 
of the form\footnote{For clarity, we shall  sometimes restore the $M$'s in the text below, as well as in the figures. }
\begin{subequations}  \label{EOM1}
\begin{eqnarray}
\label{eqn12}
\frac{d \varphi}{dt}     & \equiv&   \Omega = \frac{\partial \, \hat H_{\rm EOB}}{\partial \, p_{\varphi}}  \, ,\\
\label{eqn12a}
\frac{dr}{dt}           & =  & \left( \frac{A}{B} \right)^{1/2} \,
\frac{\partial \, \hat H_{\rm EOB}}{\partial \, p_{r_*}} \, ,\\
\label{eqn12b}
\frac{d p_{\varphi}}{dt} & = &  \hat{\F}_{\varphi} \,,\\ 
\label{eqn12c}
\frac{d p_{r_*}}{dt}    & = &- \left( \frac{A}{B} \right)^{1/2} \, \frac{\partial \, \hat H_{\rm EOB}}{\partial \, r} + \hat{\F}_{r_*} \, ,
\end{eqnarray}
\end{subequations}
which explicitly read
\begin{subequations}
\begin{align}
\label{eob:1}
&\frac{d\varphi}{dt}  \equiv \Omega  = \frac{A p_\varphi}{\nu r^2\hat{H}_{\rm EOB}\hat{H}_{\rm eff}} \ , \\
\label{eob:2}
&\frac{dr}{dt}         = \left(\frac{A}{B}\right)^{1/2}\frac{1}{\nu\hat{H}_{\rm EOB}\hat{H}_{\rm eff}}\left(p_{r_*}+z_3\frac{2A}{r^2}p_{r_*}^3\right) \ , \\
\label{eob:3}
&\frac{dp_{\varphi}}{dt} = \hat{\cal F}_{\varphi} \ , \\
&\frac{dp_{r_*}}{dt}    = -\left(\frac{A}{B}\right)^{1/2}\frac{1}{2\nu\hat{H}_{\rm EOB}\hat{H}_{\rm eff}} \nonumber \\
\label{eob:4}
&\left\{A'+\frac{p_\varphi^2}{r^2}\left(A'-\frac{2A}{r}\right)+z_3\left(\frac{A'}{r^2}-\frac{2A}{r^3}\right)p_{r_*}^4 \right\}+ \hat{\F}_{r_*} \ ,
\end{align}
\end{subequations}
where $A'=dA/dr$. In these equations,   $\hat{\F} \equiv \F/\mu$ denotes the $\mu$-rescaled radiation-reaction force.
Its explicit form will be given in Sec.~\ref{sec:radiation_reaction} below.

Let us now define the explicit forms of the two basic EOB radial functions $A(r)$ and $B(r)$ entering the
Hamiltonian \eqref{eq_H_EOB}.
One of the main theoretical novelties of  the EOB model used in the present work is the inclusion in $A(r)$ (which
plays the role of the main {\it radial potential} in the EOB Hamiltonian) of the
recently computed {\it logarithmic} contributions appearing at the 4PN and 
5PN levels  \cite{Damour:2009sm,Blanchet:2010zd,Damourlogs,Barausse:2011dq}. 
If we first focus on the {\it Taylor-expanded} version of the $A$ potential, it has, when considered at the
5PN level, the form
\begin{align} \label{ATaylor}
&A^{\rm Taylor}(u) = 1-2u + 2\nu u^3 + \left(\dfrac{94}{3}-\dfrac{41}{32}\pi^2\right)\nu u^4 \nonumber\\
     & + \nu\left[a_5^c(\nu) + a_5^{\ln}(\nu)\ln u\right]u^5 + \nu\left[a_6^c(\nu) + a_6^{\ln}(\nu)\ln u\right]u^6,
\end{align}
where $u \equiv  GM/R \equiv 1/r$ denotes the (EOB) dimensionless gravitational potential,  and where
\begin{align}
\label{eq:a5ln}
a_5^{\ln}(\nu) & = \frac{64}{5},\\
\label{eq:a6ln}
a_6^{\ln}(\nu) & = -\dfrac{7004}{105}- \dfrac{144}{5}\nu,
\end{align}
denote the analytically known logarithmic contributions, 
while  $a_5^c(\nu)$ and  $a_6^c(\nu)$ represent 
currently unknown, {\it nonlogarithmic} $\nu$-dependent 
4PN and 5PN contributions to $A(u)$.  Following the
EOB methodology initiated in Ref. ~\cite{Damour:2000we}, 
we do not use the Taylor-expanded radial
potential  $A^{\rm Taylor}(u)$ to define the EOB Hamiltonian, 
but use instead its  (1,5) Pad\'e approximant, namely
\begin{align} 
\label{Alogs}
&A(u; \,  a_5^c(\nu) ,a_6^c(\nu) ;\,\nu)\equiv P^1_5[A^{\rm Taylor}(u)]\nonumber\\
&= \frac{1+ n_1 u}{1 + d_1 u + d_2 u^2 + d_3 u^3 + d_4 u^4 + d_5 u^5},
\end{align}
where the coefficients $n_1$ and $d_i$ appearing in the numerator and 
the denominator of the Pad\'e approximant depend rationally on $a_5^c,a_6^c,\nu$ 
and  $\ln u$. 

As is well known, Pad\'e approximants can sometimes
exhibit ``spurious poles'' in $u$. The appearance of such poles 
was emphasized by Pan et al.~\cite{Pan:2009wj} within the
context of an EOB model for spinning black holes where the $A(u)$ 
radial potential is defined by Pad\'eing a Taylor-expanded $A$ function
augmented by Kerr-like spin-dependent terms (as suggested in Ref.~\cite{Damour:2001tu}).
In the case we shall investigate here (with $a_5^c$ fixed to the value in 
Eq.~\eqref{a5c} below) we found that such a spurious pole is present even
in the absence of spin, but that it is always located behind a horizon
(i.e. a zero of $A(u)$). However, when $\nu=0.25$  and $a_6^c\lesssim -130$, 
the presence of this pole (even ``hidden'' behind the horizon) starts visibly 
affecting the position of the adiabatic light-ring (i.e. the location 
of the maximum of $u^2 A(u)$), and thereby the late-plunge 
dynamics\footnote{By contrast, for $\nu \lesssim 0.2$  the spurious 
pole still exists but has nearly no effect neither on the location of 
the adiabatic light-ring nor on the late-plunge dynamics.}. 
This hidden pole will not affect our analysis below because we 
shall work in the range $a_6^c\geq -110$. 
We note in this respect that the presence of spurious poles in
the context of a spinning EOB model has motivated Barausse 
and Buonanno~\cite{Barausse:2009xi} to propose a different 
resummation of the $A$ potential which does not rely on Pad\'e
approximants, but imposes by hand the presence of a horizon.

The logarithmic-dependent 5PN-Pad\'e-resummed radial potential 
$A(u; a_5^c,a_6^c;\nu)$  will play in our work the role played by the nonlogarithmic 5PN Pad\'eed potentials
$A^{\rm no-log}(u; a_5,a_6;\nu)$ (obtained by replacing  $a_5^c(\nu) + a_5^{\ln}(\nu)\ln u \to a_5(\nu)$
and $a_6^c(\nu) + a_6^{\ln}(\nu)\ln u \to a_6(\nu)$ in the formulas above) used in the previous 
EOB works~\cite{Damour:2009kr,Damour:2009ic,Pan:2011gk}. 
As in those references, we shall use NR data to constrain, for each value
of the symmetric mass ratio $\nu$, the values of  $ a_5^c(\nu)$, and  
$a_6^c(\nu)$.  To simplify this task,  we shall take into account from the 
beginning a finding of Refs.~\cite{Damour:2009kr,Damour:2009ic,Pan:2011gk}.
The latter references found that there is, for each value
of $\nu$, a good EOB/NR agreement within a long and thin banana-like region 
in the $(a_5,a_6)$ plane.
In view of this degeneracy between $a_5$ and $a_6$, 
we shall then fix the value of $a_5^c$ and fit 
only for the ($\nu$-dependent) value of $a_6^c(\nu)$. 

Recent works connecting PN and/or  EOB theory to gauge-invariant observables 
computable from GSF theory have succeeded in determining the $\nu$-linear 
contributions to the two EOB potentials $A(u;\nu)$ and 
$B(u;\nu)$~\cite{Damour:2009sm,Barack:2010ny,LeTiec:2011ab,LeTiec:2011dp,Barausse:2011dq,Akcay:2012ea}.  
In particular, the limiting values as $\nu \to 0$ of the Taylor 
value of $a_5^c(\nu)$ and $a_6^c(\nu)$ (defined from Eq.~\eqref{ATaylor}) 
were found by Barausse~et~al.~\cite{Barausse:2011dq} to be 
$a_5^{c \, \rm Taylor}(0)= 23.50190(5)$  and $a_6^{c \, \rm Taylor}(0)= -131.72(1)$.
It is important to note here that these values  correspond to the ``true'' 
Taylor coefficients of the PN-expansion of the $A(u)$ function
when $u \to 0$, i.e., the coefficients of $u^5$ and $u^6$ in an expansion 
in powers of $u$ around $u=0$~\footnote{By contrast, note that 
Ref.~\cite{Akcay:2012ea} obtained  slightly different values of $a_5^c(0)$ 
and  $a_6^c(0)$, namely  $a_5^{c \, \rm eff}(0)= 23.47267$ and  $a_6^{c \, \rm eff}(0)= -127.154$,
because they were derived from $u$-global fits instead of an expansion
around $u= 0$.}.

However, within our present EOB model the meaning of 
the parameters $(a_5^c(\nu),a_6^c(\nu))$ is different.
First, when $\nu\to 0$, as the expansion~\eqref{ATaylor} 
does not include powers of $u$ beyond $u^6$, any attempt
at determining values of $(a_5^c(0),a_6^c(0))$ by comparing
Eq.~\eqref{ATaylor} to GSF data will strongly depend on
the $u$-interval where this comparison is done. For instance,
we might want to require that the function ${\mathsf a}(u)$~\cite{Damour:2009sm}
takes at $u=1/6$, i.e. at the unperturbed $\nu=0$ LSO, the numerical value 
corresponding to periastron precession, as determined by 
GSF calculations~\cite{Barack:2009ey,Akcay:2012ea}.
This would lead (similarly to what is done 
in Ref.~\cite{Damour:2009sm} which did not take into account logarithmic
contributions) to determining values of  $(a_5^c(0),a_6^c(0))$ such
that Eq.~(92) of~\cite{Akcay:2012ea}, namely
\be
{\mathsf a}(1/6,a_5^c,a_6^c)=0.795883004(15)
\ee
is satisfied. Taking for instance $a_5^c(0)=23.50190(5)$~\cite{Barausse:2011dq},
we would then get the following ``effective'' value of $a_6^c(0)$
\be
\label{eq:a6c_from_GSFLSO}
a_6^c(0) =+39.1223\quad\text{[from GSF LSO precession]}.
\ee
Note that this value is completely different, even in sign, from the value
$a_6^{c \, \rm Taylor}(0)=- 131.72(1)$ which refers to the Taylor expansion around
$u=0$.

A second reason why the meaning of $(a_5^c(\nu),a_6^c(\nu))$ is different in
our framework than in the GSF one is that the function 
$A(u; \,  a_5^c(\nu) ,a_6^c(\nu) ;\,\nu)$ defined by Eq.~\eqref{Alogs} is the 
Pad\'e-resummed version of the Taylor polynomial given in Eq.~\eqref{ATaylor},
which does not contain any term beyond $u^6$. This implies that, when $\nu\neq 0$,
the Taylor-expansion of $A(u; \,  a_5^c(\nu) ,a_6^c(\nu) ;\,\nu)$ does contain
higher-order terms in $u$ which are all expressed in terms of $(a_5^c,a_6^c)$ 
and $\nu$. Therefore, the values of $(a_5^c(\nu),a_6^c(\nu))$ extracted 
by comparison with NR data (for $\nu\neq 0$) represent a kind of 
mix between the true Taylor values and a plethora of higher-order 
PN corrections.
In other words $(a_5^c(\nu),a_6^c(\nu))$ represent an effective 
parametrization of the global shape of the $A$ potential.

Summarizing, in view of the effective character of the parameters  
$(a_5^c(\nu),a_6^c(\nu))$ there is no necessity to impose that
their $\nu\to 0$ limits coincide with those of Ref.~\cite{Barausse:2011dq}.
However, due to the strong degeneracy between $(a_5^c(\nu),a_6^c(\nu))$, 
it is convenient to fix $a_5^c(\nu)$ to some fiducial value.
We then decided to use the following simple, 
$\nu$-independent, fiducial value
\be \label{a5c}
a_5^{c \, \rm fiducial}(\nu) = 23.5\, .
\ee
We will see later that we could have replaced this value (which is
compatible with the rounded-off Taylor value of $a_5^c(0)$) 
with a significantly different one.

Finally,  as  the other EOB potential $B(u)$, or equivalently the 
associated potential
\be
 D(u;\nu) \equiv A(u;\nu) B(u,\nu),
\ee
plays only a secondary role in the dynamics of coalescing binaries,  and is therefore difficult to
probe by using NR data, we used its 3PN-resummed value as obtained in Ref.~\cite{Damour:2000we}, namely
\be \label{defD}
 D(u;\nu) = \frac1{1+ 6 \nu u^2 + 2(23-3\nu) \nu u^3},
 \ee
 without trying to improve it by including the known logarithmic contributions 
appearing at 4PN and 5PN~\cite{Damourlogs,Barausse:2011dq} 
(which mix with unknown nonlogarithmic contributions).
\newline
\paragraph*{Summarizing:} Our EOB Hamiltonian  
$H_{\rm EOB} (r,p_{r_*} , p_\varphi)$ contains only one free ($\nu$-dependent) 
parameter, namely $a_6^c(\nu)$. The Hamiltonian $H_{\rm EOB} (r,p_{r_*} , p_\varphi; a_6^c(\nu))$  
is defined by Eqs.~\eqref{eq_H_EOB},~\eqref{eqn11},~\eqref{Alogs},~\eqref{defD}, 
with Eqs.~\eqref{ATaylor},~\eqref{eq:a5ln},~\eqref{eq:a6ln}, and~\eqref{a5c},
together with $p_{r_*} = (A/B)^{1/2} \,p_r$ , $B(u)\equiv D(u)/A(u)$ and $u\equiv 1/r$.

\subsection{Improved EOB waveform during inspiral and plunge}
\label{sec:EOB_waveform}

Following Refs.~\cite{Damour:2007xr,Damour:2007yf,Damour:2008gu}, we describe the inspiral-plus-plunge
multipolar waveform by the factorized structure
\be
\label{eq:hlm}
h_{\ell m}^{\rm insplunge} = h_\lm^{(N,\epsilon)}(\v)  S_{\rm eff}^{(\epsilon)}\hat{h}_\lm^{\rm tail}(y) \left[\rho_\lm(\v^2)\right]^\ell \hat{h}_\lm ^{\rm NQC},
\ee
where  we indicated the (main) arguments used in several factors of the waveform.
Here $\epsilon=0,1$ is the parity of the considered
multipole (i.e. the parity of $\ell+m$) ,   $h_\lm^{(N,\epsilon)}$  the Newtonian waveform, 
$\hat{S}_{\rm eff}^{(\epsilon)}$  a source factor, with
$\hat{S}_{\rm eff}^{(0)}=\hat{H}_{\rm eff}$ or $\hat{S}_{\rm eff}^{(1)}=p_\varphi/(r_\omega v_\varphi)$ 
according to the parity of the multipole (see below for definitions), 
\be \label{htail}
\hat{h}^{\rm  tail}_\lm(y) \equiv T_\lm(y) e^{i\delta_\lm (y)},
\ee 
 the tail factor  \cite{Damour:2007xr,Damour:2007yf,Damour:2008gu}, $\rho_\lm$ 
 the resummed modulus correction and $\hat{h}^{\rm NQC}_\lm$  a 
next-to-quasi-circular correction.  The precise definitions of the factors 
 entering Eq.~\eqref{eq:hlm}, and of their arguments is given next.
 
The Newtonian contribution reads
\be
\label{eq:hlm_Newt}
h_\lm^{(N, \epsilon)}(\v)=\dfrac{M\nu}{R}n_\lm^{(\epsilon)} c_{\ell +\epsilon}(\nu) \v^{\ell+\epsilon} Y^{\ell -\epsilon,-m}\left(\dfrac{\pi}{2},\varphi\right),
\ee
where $\varphi$ is the orbital phase, $\v=r_\omega \Omega$ a suitably defined azimuthal velocity,
and $r_\omega\equiv r\psi^{1/3}$  a modified EOB radius with $\psi$  defined as
\begin{align}
\psi(r,p_\varphi) &= \dfrac{2}{r^2}\left(\dfrac{dA}{dr}\right)^{-1}\nonumber\\
                &\times\left[1+2\nu\left(\sqrt{A\left(1+\dfrac{p_\varphi^2}{r^2}\right)}-1\right)\right].
\end{align} 
The definitions of $\v$ and  $r_\omega$ are such that they satisfy  Kepler's law, 
 $ 1= \Omega^2 r_\omega^3 = \v^2 r_\omega$, during the adiabatic inspiral~\cite{Damour:2006tr}.  
In Eq.~\eqref{eq:hlm_Newt}, $n_\lm^{(\epsilon)}$ and $c_{\ell+\epsilon}(\nu)$ are numerical 
coefficients  given by~\cite{Damour:2008gu}
\begin{align}
n_\lm^{(0)} & =(im)^\ell \dfrac{8\pi}{(2\ell+1)!!}\sqrt{\dfrac{(\ell+1)(\ell+2)}{\ell(\ell-1)}}\ ,\\
n_\lm^{(1)} & = -(im)^\ell \dfrac{16\pi
  i}{(2\ell+1)!!}\sqrt{\dfrac{(2\ell+1)(\ell+2)(\ell^2-m^2)}{(2\ell-1)(\ell+1)\ell(\ell-1)}}\ ,\\
\label{eq:clm_LO}
c_{\ell + \epsilon} & =X_2^{\ell+\epsilon-1}+(-1)^m X_1^{\ell +\epsilon-1} \ ,
\end{align}
where $X_{1,2}\equiv m_{1,2}/M$. [Note that, in our EOB/NR comparisons below, we shall often
work with a ``Zerilli-normalized'' waveform, denoted $\Psi_\lm$, whose normalization differs from that of $h_\lm$
by a factor $R/(M \sqrt{(l+2)(l+1)(l)(l-1)})$.]
For what concerns the tail factor $\hat{h}_\lm^{\rm tail}$, 
Eq.~\eqref{htail}, its main contribution,
$T_\lm$, is written as
\be
\label{eq:Tlm}
T_\lm(y) = \dfrac{\Gamma(\ell+1-2\i\k)}{\Gamma(\ell+1)}e^{\pi \k} e^{2\i\k\ln(2kr_0)},
\ee
with $\hat{\hat{k}}\equiv m G H_{\rm EOB} \Omega$, $k\equiv m\Omega$ and $r_0=2 G M /\sqrt{e}$.
Note that, apart from the logarithm term  $\ln(2kr_0)$, the main tail contribution $T_\lm$ depends on the
dimensionless argument  $y \equiv (G H_{\rm EOB}\Omega)^{2/3}$, which differs from the usual dimensionless
frequency parameter $x\equiv (G  M \Omega)^{2/3}$ by the replacement $M\to H_{\rm EOB}$.

 \subsubsection{Further resummation of the residual tail phase  $\delta_{\lm}(y)$.}
\label{sec:resum_deltalm} 

The main factorized tail term $T_\lm(y)= |T_\lm| e^{i \tau_{\lm}}$ is a complex quantity whose modulus $|T_\lm| $
describes the tail amplification of the waveform modulus, and whose phase $\tau_{\lm}$ describes the main
part of the dephasing caused by tails. There are, however, additional dephasings caused by tails, which
are described by the supplementary phase factor $e^{ i \delta_{\lm}}$ in Eq.~\eqref{htail}.
The residual phase corrections  $\delta_{\lm}(y)$ entering the tail factor~\eqref{htail} were obtained 
in Ref.~\cite{Damour:2008gu} as a PN series in the variable $y=(G H_{\rm EOB}\Omega)^{2/3}$.  
Here we shall use for $\delta_{\lm}(y)$ an expression that differs both from the one originally 
given in Ref.~\cite{Damour:2008gu}, and from its test-mass-higher-PN completion
given in Ref.~\cite{Fujita:2010xj}. More precisely: 
(i) we do not include the highest-order $O(y^{9/2})$ test-mass ($\nu=0$) PN corrections because of  their  PN-gap with respect 
to the last known comparable-mass terms; (ii) we include the 3.5PN, $\nu$-dependent, contribution 
to $\delta_{2 2}(y)$ that can be  deduced
from a recent analytical computation of the PN-expanded waveform at 3.5PN accuracy~\cite{Faye:2012we}; 
and 
(iii) we Pad\'e-resum the Taylor series in powers of $y^{1/2}$ giving $\delta_\lm(y)$. Indeed, we found that
the PN-expanded version of  $\delta_\lm(y)$ presents some unpleasant features (discussed below in the $\ell=m=2$ case)
that are avoided if one resums  $\delta_\lm(y^{1/2})$ by  factorizing the leading-order term and  
and replacing the rest with a suitable Pad\'e approximant $N(y^{1/2})/ D(y^{1/2}) $.

\begin{figure}[t]
 \includegraphics[width=0.48\textwidth]{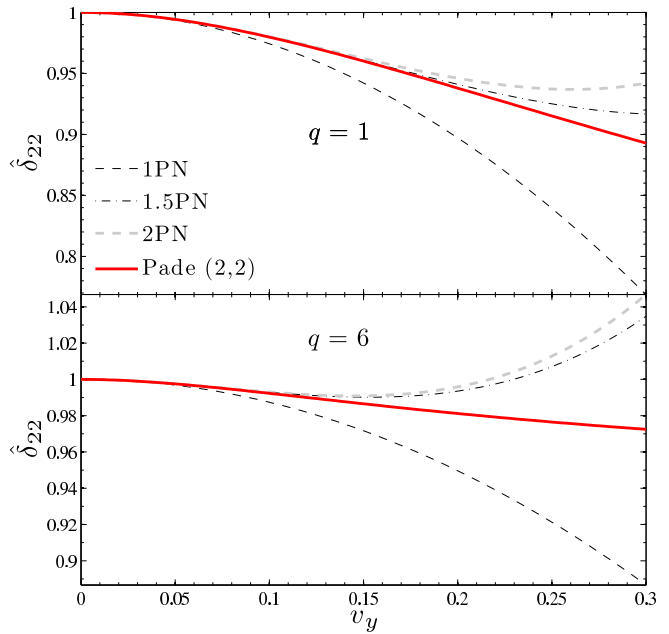}
    \caption{ \label{fig:resum_delta22} (color online) Comparing the Taylor-expanded $\hat{\delta}_{22}$ 
    with its (2,2) Pad\'e approximant for two mass ratios.} 
\end{figure}

Let us explain our new procedure on the (most important) example of the  $\ell=m=2$ phase 
(the others are listed in Appendix~\ref{sec:rholm}). Let us start from 
its Taylor-expanded form 
\begin{align}
\label{eq:delta22_35PN}
\delta_{22}^{\rm Taylor}(y) & = \dfrac{7}{3} y^{3/2}-24\nu y^{5/2} + \dfrac{428}{105}\pi y^3 \nonumber\\
                        & + \left(\dfrac{30995}{1134}\nu + \dfrac{962}{135}\nu^2\right)y^{7/2}.
\end{align}
Here we did not include the highest-order test-mass term
$\left(-2203/81 + 1712/315\pi^2\right)y^{9/2}$ that was obtained in Ref.~\cite{Fujita:2010xj}. 
On the other hand, the  3.5~PN $\nu$-dependent term proportional to $y^{7/2}$ is a new contribution that is obtained by 
applying the factorization of~\cite{Damour:2008gu} to the results of~\cite{Faye:2012we}. 
Note that this is the only genuinely new information given by this calculation; indeed, 
the real 3.5PN contributions to $h_{2 2}$ are already contained in the modulus of the 
EOB-resummed tail factor  $\hat{h}_\lm^{\rm tail}$.  For the comparable-mass cases that are of
primary concern for upcoming GW observations (say for $ \nu \gtrsim 0.1$) the $O(y^{7/2})$
contribution is numerically quite significant compared to the lower-order terms.
To better appreciate the relative importance of the successive PN corrections
we  factorize Eq.~\eqref{eq:delta22_35PN} in a leading order (LO) part, 
$\delta_{22}^{\rm LO}(y) \equiv (7/3) y^{3/2}$  and a fractional PN-correction 
term, $\hat{\delta}_{22} \equiv \delta_{22}^{\rm Taylor}/\delta_{22}^{\rm LO}$. 
In terms of $v_y \equiv \sqrt{y}$, the latter fractional PN-correction has the structure 
\be
\hat{\delta}_{22}= 1 + c_2 v_y^2 + c_3 v_y^3 + c_4v_y^4.
\ee
We plot, in Fig.~\ref{fig:resum_delta22} the successive truncated PN approximants, at 
1PN, 1.5PN and 2PN accuracy (i.e. up to $v_y^2$, $v_y^3$ and $v_y^4$) for $q=1$ ($\nu=1/4$) 
and $q=6$ ($\nu=6/49\approx 0.1224$).
This figure illustrates two facts: (i) the successive PN approximants to $\hat{\delta}_{22}= 1 + c_2 v_y^2 + c_3 v_y^3 + c_4v_y^4 + \cdots $ 
are suspiciously different from each other; and (ii) they introduce rather large fractional modifications
of the LO phase  $\delta_{22}^{\rm LO}(y) \equiv (7/3) y^{3/2}$ when  $v_y \gtrsim 0.3$ (which is reached during the late
plunge).
 This suggests a nonrobust behavior of the Taylor approximants in the high-velocity regime.
In addition we have found that using $\delta_{22}^{\rm Taylor}(y)$ in the generation
of the EOB waveform generates pathological features in the waveform phase in the very late plunge phase, 
compromising the accuracy of the phasing in a crucial region. To overcome this difficulty, we replace 
$\hat{\delta}_{22}^{\rm Taylor}(v_y)$  with its $(2,2) $ Pad\'e approximant, i.e. 
we take $P^2_2[\hat{\delta}_{22}(v_\Omega)]$. Finally, we use in defining the factorized EOB waveform 
the following resummed version of the  $\delta_{22}(y)$ phase:
\begin{align}
\label{eq:delta22_pade}
\delta_{22}(y)& \equiv \delta_{22}^{\rm LO}(y)P^2_2\left[\hat{\delta}_{22}\left(\sqrt{y}\right)\right] =   \dfrac{7}{3} v_y^{3} \dfrac{p_0 + p_1 v_y + p_2 v_y^2}{p_0 + p_1 v_y + p'_2 v_y^2},
\end{align}
where $v_y \equiv y^{1/2}$. The explicit expressions of the $\nu$-dependent Pad\'e coefficients 
$p_0(\nu), p_1(\nu), p_2(\nu), p'_2(\nu)$ will be found in
Appendix~D. Note that this Pad\'e  representation degenerates
as $\nu \to 0$, and yields $P^2_2\left[\hat{\delta}_{22}\left( v_y \right)\right]  \to 1$; this occurs because
the definition of this Pad\'e approximant crucially depends on having a non-vanishing 3.5PN contribution.
Figure~\ref{fig:resum_delta22}  compares the Pad\'e-resummed $\hat{\delta}_{22}(v_y)$ to its successive Taylor
approximants. This figure suggests that the Pad\'e approximant represents a reasonable ``average'' of the
successive Taylor approximants.

We found that the (known) successive PN approximants to 
$\hat{\delta}_{21}^{\rm Taylor}$, $\hat{\delta}_{33}^{\rm Taylor}$ and 
$\hat{\delta}_{31}^{\rm Taylor}$, 
exhibited a rather nonrobust  behavior similar to that of $\hat{\delta}_{22}^{\rm Taylor}$.  
We therefore decided to Pad\'e resum them,
using now $(1,2)$ Pad\'e approximants, in view of the available PN knowledge.
For the other residual phase corrections, $\delta_{32}$, $\delta_{4m}$ with $m=1,\dots,4$
and $\delta_{55}$, there is too little PN information to try a resummation, 
so that we keep them in their  unresummed Taylor-expanded form. 
See Appendix~\ref{sec:rholm} for details.

 \subsubsection{Further factorized corrections to the waveform:  $\rho_\lm(\v^2)$  and  $\hat{h}_\lm^{\rm NQC}$.}

Let us first emphasize that, as in our previous work~\cite{Damour:2009kr},  we shall use as argument 
in the modulus correction $\rho_\lm$ (to replace the generic variable $x$ used in \cite{Damour:2008gu}) 
the quantity  $x_{\varphi} = \v^2 = (r_\omega \Omega)^2$ defined
above. By contrast, Ref.~\cite{Pan:2011gk} uses
$x=(M \Omega)^{2/3}$ as argument in the  $\rho_\lm$'s. 
The $\rho_\lm$'s that enter Eq.~\eqref{eq:hlm} are taken at the complete $3^{+2}$~PN approximation
(as done in previous work~\cite{Damour:2010zb,Damour:2011fu,Damour:2012yf,Bernuzzi:2012ci,Bernuzzi:2012ku}), 
i.e., by completing the 3PN-accurate, $\nu$-dependent results of Ref.~\cite{Damour:2008gu} 
by the $\nu=0$, 5PN-accurate, terms obtained\footnote{In successive steps, this computation has been 
recently pushed to the remarkable 22~PN order by Fujita~\cite{Fujita:2011zk,Fujita:2012cm}.} 
by Fujita and Iyer~\cite{Fujita:2010xj}. 
Note that in doing so we are taking into account
more test-mass terms in the $\rho_\lm$'s than was done in Ref.~\cite{Pan:2011gk},
which was stopping one PN order earlier for $\rho_{33}$, $\rho_{31}$, $\rho_{4m}$, and two PN orders
earlier for $\rho_{5m}$, $\rho_{6m}$ and $\rho_{7m}$. 
For completeness we list in Appendix~\ref{sec:rholm} the explicit 
expressions of the $\rho_\lm$'s that we use. 
As we said, one must replace the generic variable $x$
used in these expressions by $x_{\varphi} = \v^2 = (r_\omega
\Omega)^2$. 

Let us now discuss the structure of  the final, NQC factor  $\hat{h}_\lm^{\rm NQC}$ in the
factorized waveform, Eq. \eqref{eq:hlm}, as well as the procedure we shall use to determine
(from NR data) the values of the coefficients $a_i^{\lm}$ and $b_i^{\lm}$ entering this NQC
correction factor $\hat{h}_\lm^{\rm NQC}$.
We shall adopt here a more elaborate  NQC factor $\hat{h}_\lm^{\rm NQC}$ 
than what was considered in previous EOB literature.
In particular, for each multipole 
$(\ell,m)$ this NQC factor depends on 6 real parameters, 3 for the amplitude, 
$a_i^{\ell m}$, $i=1,\dots,3$, and 3 for the phase $b_i^{\lm}$, $i=1,\dots,3$ and reads
\be
\label{eq:hNQC}
\hat{h}_\lm^{\rm NQC} = \left(1 + \sum_{j=1}^{3} a_j^{\lm} n_j \right)\exp\left(i\sum_{j=1}^{3}b_j^{\lm} n_{j+3}\right),
\ee
where the $n_i$'s factors are chosen here to be
\begin{subequations}
\label{eq:allnqc}
\begin{align}
\label{eq:n1_nqc}
n_1 &= \left(\dfrac{p_{r_*}}{r\Omega}\right)^2\\
\label{eq:ddotr}
n_2 &= \dfrac{(\ddot{r})^{(0)}}{r\Omega^2},\\
n_3 &= n_1 p_{r_*}^2,\\
n_4 &= \dfrac{p_{r_*}}{r\Omega},\\
n_5 &= n_4 (r\Omega)^2,\\ 
n_6 &= n_5 p_{r_*}^2 .
\end{align}
\end{subequations}
Here, the superscript $(0)$ on the right-hand side of 
the definition of $n_2$ means that the second time derivative of 
$r$ is evaluated along the conservative dynamics 
(i.e. neglecting the contributions proportional to $\F$, see 
Appendix~\ref{sec:ddotr} for a discussion).

One should keep in mind that the EOB (dynamical) time $t^{\rm EOB}$ 
differs from the NR (retarded) time $t^{\rm NR}$ by an apriori 
unknown constant shift: $t^{\rm EOB} = t^{\rm NR}+\tau$. 
Determining $\tau$ is equivalent to the problem of aligning 
the NR and EOB waveforms. Physically, determining $\tau$ is equivalent 
to identifying one specific feature in the EOB waveform to a 
corresponding specific feature in the NR one.
This choice has been different in various EOB-related works.
From the beginning, i.e.~\cite{Buonanno:2000ef}, it was emphasized
that a good marker on the EOB time axis of the ``moment of merger'' 
was the time $\teobLR$ where the EOB orbital frequency reaches its
maximum. The issue is then to select the corresponding 
moment on the NR time axis. In all early EOB studies, it was assumed
that the NR correspondent of $\teobLR$ is $\tAmax$, i.e. the NR instant
when the $\ell=m=2$ amplitude reaches its maximum. However, several
recent EOB-related works~\cite{Bernuzzi:2010xj,Barausse:2011kb,Taracchini:2012ig}
gave evidence that, in the test-mass limit, the two instants 
$\teobLR$ and $\tAmax$ do not exactly correspond to each other.

In this work, we shall define the correspondence between 
$t^{\rm EOB}$ and $t^{\rm NR}$ by requiring that the correspondent
on the NR time axis of the EOB instant $\teobLR$ is a specific time  $\tnrLR$
which will be defined in Eq.~\eqref{deftnrLR} below.
In addition, we shall use this time  $\teobLR  \leftrightarrow \tnrLR$
both as NQC determination point and as QNM attachment 
one~\footnote{Note that this choice differs from the one used in 
Refs.~\cite{Barausse:2011kb,Taracchini:2012ig}. In these references the
NQC and QNM EOB instant is chosen to be  {\it earlier}  than  $\teobLR$ and to 
correspond to the NR instant $\tAmax$.}.
More precisely, for each multipole, the 6 parameters $a_i^{\lm}$ and $b_i^{\lm}$ 
entering Eq.~\eqref{eq:hlm} are determined from NR data by imposing that the 
EOB waveform $h^{\rm EOB}_\lm(t^{\rm EOB})$ (which is a function of the EOB dynamical time $t^{\rm EOB}$)
``osculates''  the NR waveform  $h^{\rm NR}_\lm(t^{\rm NR})$  
(which is a function of the NR retarded  time $t^{\rm NR}$)
around  the NQC-determination point $\teobLR  \leftrightarrow \tnrLR$.

Note again that in this work we shall use as NQC-determination point 
on the EOB time axis the EOB dynamical time $\teobLR$ when 
the EOB orbital frequency $\Omega(t^{\rm EOB})$ reaches 
its (first) maximum\footnote{This EOB time was often referred to, in previous works, as 
the ``effective EOB light-ring crossing time'', because, in the 
test-mass limit, it does correspond to the dynamical time when 
$R(t^{\rm EOB}) = 3M$, and, in the comparable-mass case, it is very 
close to the time when  $R(t^{\rm EOB}) $ crosses the formal EOB analog of
the light ring. Here, to avoid confusion, we shall call it the 
$\Omega$-peak time, and denote it as $\teobLR$.}.
The degree of osculation between the EOB and NR waveforms is defined by separately 
imposing  a  $C^2$ contact between the amplitudes, $A_{\lm}$, and  the frequencies, $\omega_{\lm}$, of the two waveforms
at the NQC-determination point  $\teobLR  \leftrightarrow \tnrLR$.
We do not constrain the relative phase of the EOB and NR waveforms.
Explicitly,  we impose the following six conditions
\begin{subequations}
\label{eq:C2_cond}
\begin{align}
\label{comp_1}
A_{\lm}^{\rm EOB}(\teobLR)             & = A_{\lm}^{\rm NR}(\tnrLR),\\ 
\label{comp_2}
\dot{A}_{\lm}^{\rm EOB}(\teobLR)       &= \dot{A}_{\lm}^{\rm NR}(\tnrLR),\\ 
\label{comp_3}
\ddot{A}_{\lm}^{\rm EOB}(\teobLR)      &= \ddot{A}_{\lm}^{\rm NR}(\tnrLR),\\ 
\label{comp_4}
\omega_{\lm}^{\rm EOB}(\teobLR)        &= \omega_{\lm}^{\rm NR}(\tnrLR),\\ 
\label{comp_5}
\dot{\omega}_{\lm}^{\rm EOB}(\teobLR)  &= \dot{\omega}_{\lm}^{\rm NR}(\tnrLR),\\ 
\label{comp_6}
\ddot{\omega}_{\lm}^{\rm EOB}(\teobLR) &= \ddot{\omega}_{\lm}^{\rm NR}(\tnrLR),
\end{align}
\end{subequations}
which yield two separate $3\times 3$ linear systems to be solved to
obtain the $a_i^\lm$'s, and, separately, the $b_i^\lm$'s.

Note that the values of the  $a_j^\lm$'s affect the modulus of the inspiral-plus-plunge
waveform, which then affects the computation of the radiation reaction force (through
the  angular momentum flux, see below). In turn, this modifies the EOB dynamics itself,
and, consequently, the determination of the $(a_j^\lm,b_j^\lm)$'s.
This means that one must bootstrap, by iteration, the determination
of the $(a_j^\lm,b_j^\lm)$'s until convergence (say at the third decimal digit)
is reached. This typically requires three iterations. In previous
work only the dominant $(2,2)$ NQC  correction was included in the radiation
reaction (though they were all taken into account when finally comparing EOB and
NR waveforms). Here we shall follow the same simplifying prescription, though we have
explored the effect of including also the subdominant $(2,1)$ and 
$(3,3)$ NQC corrections to the flux. We found that their effect amounts
only to a small change in the NR determination of the ``good values'' of $a_6^c$
(see Appendix~\ref{sec:Flm_high}).
\newline
\paragraph*{Summarizing:} Our EOB waveform is given by Eq.~\eqref{eq:hlm} 
and employs the resummation of residual phases $\delta_\lm$ as 
in Eq.~\eqref{eq:delta22_pade}. The NQC correction is defined by 
Eqs.~\eqref{eq:hNQC}-\eqref{eq:allnqc} with constants determined
from NR data by Eqs.~\eqref{eq:C2_cond}.

\subsection{EOB waveform during merger and ringdown}

One of the specificities of the EOB formalism is to construct a {\it complete} waveform, covering
the full process from early inspiral to ringdown, passing through late inspiral, plunge, and merger. 
This is done by attaching a sum of
quasi-normal modes (QNM) to the end of the plunge waveform. The procedure for doing so has
improved over the years~\cite{Buonanno:2000ef,Damour:2007xr,Pan:2009wj}. Here, we use a new way of
extending the inspiral-plus-plunge waveform to describe the merger-plus-ringdown subsequent signal,
which fits with the NQC-determination procedure we have explained above.  Our new procedure for,
simultaneously, determining NQC corrections, and attaching QNM's,   
is motivated by the findings of  Bernuzzi, Nagar and Zenginoglu~\cite{Bernuzzi:2010xj}
in the extreme mass ratio limit ($\nu \ll 1$). 
We shall discuss the rationale for this procedure in the next Section.

The  merger-plus-ringdown signal is described, for each multipole $\lm$, by a sum of $N$ 
QNM signals of a final Kerr black hole (of mass $M_f$ and spin parameter $a_f$), say 
\begin{equation}
\label{eqn17}
\left( \frac{R c^2}{GM} \right) h_{\ell m}^{\rm ringdown} (t) = \sum_{n=0}^{N-1} C_n^\lm \, e^{-\sigma_n^{+ , \, \lm}(t-\teobLR)} \, ,
\end{equation}
where $ \sigma_n^{+ , \, \lm}= \alpha_n^\lm + i \, \omega_n^\lm$ is the complex frequency 
of the $n$th QNM of  multipolarity $\lm$ and $C_n^\lm$ are complex constants.

In this work, we use $N=5$  positive frequency ($\omega_n^\lm >0$) QNM's. 
These complex frequencies are functions of  the mass $M_f$ and spin parameter 
$a_f$ of the final hole~\cite{Berti:2005ys}. 
For $M_f$ and $a_f$ we adopt the fit to the numerical
results given in Eqs.~(29) of~\cite{Pan:2011gk},
\begin{align} \label{Mfaf}
\dfrac{M_f}{M} = 1 + \left(\sqrt{\dfrac{8}{9}}-1\right)\nu - 0.4333\nu^2 - 0.4392\nu^3,\\
\dfrac{a_f}{M} = \sqrt{12}\, \nu - 3.871\nu^2 + 4.028\nu^3 .
\end{align}

The procedure we shall use here for matching the ringdown signal~\eqref{eqn17} to the inspiral-plus-plunge 
signal~\eqref{eq:hlm}  is similar to the ones used in previous EOB work~\cite{Bernuzzi:2010xj} 
though it differs in a significant way from the one used in~\cite{Pan:2011gk}. Namely, contrary to 
the latter reference, the attachment (along the EOB dynamical time axis $t^{\rm EOB}$) of  the QNM 
signal~\eqref{eqn17} to the NQC-corrected inspiral-plus-plunge signal~\eqref{eq:hlm}  is done, 
for each multipole $\lm$, at the time
\be
 t^{\rm EOB}_{\lm \rm QNM attachment }= t^{\rm EOB}_{\lm \rm matching} =\teobLR,
\ee
where we recall that $\teobLR$ denotes the EOB dynamical time where the EOB orbital frequency reaches
its (first) maximum. 
Note in particular that  $\teobLR$  does not depend on the considered multipolarity $\lm$, so that we are 
attaching the QNM's  corresponding to all the different multipolarities at the  {\it same} EOB dynamical time.

To complete the description of our QNM attachment procedure it remains to say that we determine, for
each multipolarity $\lm$, the
values of the $N$ complex coefficients   $C_n^\lm $ by requiring that the (NQC-corrected) EOB inspiral-plus-plunge
waveform $h_{\ell m}^{\rm insplunge}(t^{\rm EOB})$, Eq.~\eqref{eq:hlm}, coincides with the QNM sum~\eqref{eqn17}
at $N$ points, 
say $t_1, t_2, \cdots, t_N$, forming a  regularly spaced ``comb'' on the $t^{\rm EOB}$ axis, 
centered on $\teobLR$. Such a ``matching comb''
is specified by choosing its total length, say
\be
\Delta^{\rm match} = t_N - t_1 \, .
\ee

\subsection{Improved radiation reaction: Including horizon absorption and  a radial component $\F_{r_*}$ }
\label{sec:radiation_reaction}

Let us now turn to our improved description of the radiation reaction force  $\cal F$ entering the EOB dynamics.
Note that we have included in the equations of motion~\eqref{EOM1} not only an azimuthal radiation 
reaction $ \hat{\cal F}_{\varphi} $ (as in all previous EOB works), but also an explicit {\it radial} 
contribution $\hat{\F}_{r_*}$. We have improved the analytical description of both components 
of $\cal F$. Let us discuss them in turn.

The azimuthal component, $\F_\varphi$, of the radiation reaction force describes the loss of the
orbital angular momentum  $p_\varphi$ of the system during evolution.  Indeed,  Hamilton's equation 
for $p_\varphi$ reads
\be
\dfrac{dp_\varphi}{dt} = \hat{\F}_\varphi,
\ee
where $\hat{\F}_\varphi=\F_\varphi/\nu$. 

Following a standard EOB practice (since Ref.~\cite{Buonanno:2000ef}),  we require that the loss of {\it orbital}
angular momentum  be balanced by the instantaneous flux of angular momentum leaving the orbital system.
In previous EOB work, one took into account only the flux of angular momentum in the form of GWs at infinity.
However, there is also a flux of angular momentum which is drained out of the two-point mass orbital system by penetrating
within the two horizons of the moving black holes. [The latter flux is transformed from the orbital form measured by   
$p_\varphi$ to some intrinsic spin-angular momentum
of the holes; from the point of view of the orbital  $p_\varphi$ this represents a loss that must be accounted for
by an additional contribution to  $\F_\varphi$.]  We shall include here such an additional horizon-absorption
flux by using the recent work of Nagar and Akcay~\cite{Nagar:2011aa}. The corresponding 
effect is rather small and, in a PN sense, starts only at the 4PN level~\cite{Poisson:1994yf,Alvi:2001mx}. 
Reference~\cite{Alvi:2001mx}, using a leading-order (Newtonian) approximation both to the phase evolution
and to the horizon flux had estimated that, in the nonspinning case (that we consider here), 
the inclusion of the horizon flux entails an additional dephasing at $R\approx 6M$ 
smaller than 0.01~rad for mass ratios $1\leq q \leq 4$.
On the other hand, recently Bernuzzi, Nagar and Zenginoglu~\cite{Bernuzzi:2012ku}, using an EOB
description of the phase evolution together with an improved estimate of the horizon flux (resumming
higher effects), have found significantly larger dephasings (accumulated over the last 20-30 orbits) 
than those estimated in~\cite{Alvi:2001mx}. Within the EOB model that we use here we confirmed the
findings of Ref.~\cite{Bernuzzi:2012ku}. For instance taking the most relevant 
case $q=6$ with initial separation $r_0=15$ (corresponding to $\sim 27$ orbits 
up to merger, see Table~\ref{tab:EOB_ID} below) the effect of horizon  absorption 
entails a dephasing $\Delta^H\phi\equiv \phi^{H+\scri}-\phi^\scri\sim 0.12$~rad 
at  $\teobLR$, that increases up  to $0.18$~rad during ringdown\footnote{Note that
one has $\Delta^H\phi\sim 1.6\times 10^{-4}$ at the initial separation $r_0=15$, 
which is negligible compared to the dephasing accumulated during the subsequent evolution.}.
Such dephasings are quite significant for the EOB/NR comparison that we shall perform below.
This is why we decided to include the horizon contribution to the angular momentum flux.

It is convenient to decompose $\F_\varphi$ as the product of the usual quadrupolar GW flux
(expressed in terms of $r_\omega$ and of the orbital frequency $\Omega=d\varphi/dt$) and 
of a supplementary dimensionless correction factor (of the $1 + O(x)$-type) :
\be
\label{eq:Fphi}
\hat{\F}_\varphi = -\dfrac{32}{5} \nu r_\omega^4 \Omega^5 \hat{f}(\v^2;\,\nu).
\ee 
Here the function $\hat{f}(x;\,\nu)= 1 + O(x)$ (taken with the argument $x=v_\varphi^2$) 
is the {\it reduced flux function}. It
can be defined, for a circularized binary, as the ratio between the
total energy flux (including the horizon flux) and the $\ell=m=2$ asymptotic energy flux. In our
case this function is given by the sum of an asymptotic (labeled by $\scri$)
and a horizon (labelled by $H$) contribution, and can be further written as 
\be
\label{eq:fhat}
\hat{f}(x;\,\nu) = \hat{f}^\scri(x;\,\nu) + (1-4\nu+2\nu^2)x^4 \hat{f}^H(x;\,\nu),
\ee
where each function $ \hat{f}^{(\scri,H)}(x;\,\nu) $ is of the  $1 + O(x)$ type 
and is defined by dividing by the corresponding $\ell=m=2$ LO contribution, namely
\begin{align}
\hat{f}^{(\scri,H)}(x;\,\nu) &= F^{(\scri,H)}_{\ell_{\rm
    max}}/F^{(\scri,H), {\rm LO}}_{22}. 
\end{align}
Here, $F^{(\scri,H)}_{\ell_{\rm max}}$ is either the total asymptotic
($\scri$) or horizon ($H$) energy flux for circular orbits summed
up to multipole $\ell=\ell_{\rm max}$,  
while $F_{22}^{\scri, {\rm LO}} = (32/5) \nu^2
x^5$ is the LO (or ``Newtonian'') quadrupolar (asymptotic) energy flux, 
and $F_{22}^{H, {\rm LO}} = (32/5) \nu^2(1-4\nu+2\nu^2)x^9= x^4(1-4\nu+2\nu^2)F_{22}^{\scri, N} $ 
the LO quadrupolar horizon flux~\cite{Poisson:1994yf,Alvi:2001mx}.  
In the EOB model one uses suitably factorized expressions for the
multipolar fluxes $F^{(\scri,H)}_{\lm}$ to resum and improve them
with respect to standard PN-expanded expressions in the strong-field,
fast-velocity regime. In the case of the multipolar asymptotic flux $F^{\scri}_{\lm}$, 
this factorized flux is simply defined (as first proposed in~\cite{Damour:2009kr})
by squaring the corresponding factorized multipolar waveform of~\cite{Damour:2008gu},
recalled above.
An analogous procedure for the multipolar horizon fluxes, $F^{H}_{\lm}$
 was introduced in Ref.~\cite{Nagar:2011aa} and compared with
Regge-Wheeler-Zerilli numerically computed horizon fluxes in 
Ref.~\cite{Bernuzzi:2012ku}. [Here, we are considering nonspinning binaries.]

The horizon and asymptotic energy fluxes along circular orbits 
are then written as multipolar sums, say
\be
\label{eq:FH}
F^{[(\scri,H),\ell_{\rm max}]}(x;\,\nu) = \sum_{\ell =2}^{\ell_{\rm max}}\sum_{m=1}^{\ell} 
F_\lm^{(\scri,H,\epsilon)}(x;\,\nu),
\ee
where $F_\lm^{(\scri,H,\epsilon)}=F_{\ell|m|}^{(\scri,H,\epsilon)}$ sums
the two equal contributions corresponding to $+m$ and
$-m$ ($m\neq 0$ as the $m=0$ contributions vanish for circular orbits).

Inserting in the (circular) asymptotic multipolar flux contribution, 
\be \label{Flm1}
F^{(\scri,\epsilon)}_\lm = \dfrac{1}{8\pi} (m\Omega)^2| R h_\lm^{(\epsilon)}|^2\ ,
\ee
the factorized waveform~\eqref{eq:hlm} yields
\be
F_{\lm}^{\scri, \epsilon} = F_\lm^{(N,\epsilon)}   \left(\hat{S}_{\rm eff}^{(\epsilon)}\right)^2|T_\lm(y)|^2\left(\rho_\lm(x;\nu)\right)^{2\ell} \hat{F}_\lm^{(\scri,\epsilon){\rm NQC}},
\ee
where $F_\lm^{(N,\epsilon)}$ is defined by inserting the Newtonian-order waveform in~\eqref{Flm1},
and where each subsequent factor is the squared modulus of a corresponding PN-correction factor entering  \eqref{eq:hlm};
e.g., $\hat{F}_\lm^{(\scri,\epsilon){\rm NQC}}= \left|  \hat{h}_\lm ^{\rm NQC} \right|^2= \left(1 + \sum_{j=1}^{3} a_j^{\lm} n_j \right)^2$.
Let us mention that $|T_\lm(y)|^2$ can be explicitly written in
the simple form
\be
\left|T_\lm(y)\right|^2=
\dfrac{1}{\left(\ell!\right)^2}\dfrac{4\pi\k}{1-e^{-4\pi\k}}\prod_{s=1}^{\ell}\left(s^2
  + (2\k)^2\right) \ .
\ee

\begin{table}[t]
  \caption{\label{tab:clm} Coefficients of our hybrid
    1$^{+3}$PN-accurate $\rho_\lm^H(x;\,\nu)$ functions as given by
    Eq.~\eqref{eq:all_rho}.} 
  \begin{center}
    \begin{ruledtabular}
      \begin{tabular}{cccccc}
        $\ell$   &   $m$  &    $c_1^\lm$   &   $c_2^\lm$  &   $c_3^\lm$   &   $c_4^\lm$ \\
        \hline 
        2    &    2   &  $\frac{4-21\nu + 27\nu^2
          -8\nu^3}{4(1-4\nu+2\nu^2)}$   &  4.78752  & 26.760136  &
        43.861478  \\ 
        2    &    1   &   0.58121   &  1.01059  & 7.955729    & 1.650228 \\
   \end{tabular}
  \end{ruledtabular}
\end{center}
\end{table}

Similarly the horizon partial multipolar fluxes are written in factorized form~\cite{Nagar:2011aa} 
\be
\label{eq:FlmH}
F_\lm^{(H,\epsilon)}(x;\,\nu)=F_{\lm}^{(H_{\rm
    LO},\epsilon)}(x;\,\nu)\left[\hat{S}_{\rm eff}^{(\epsilon)}(x;\nu)
  \left(\rho_\lm^H(x;\,\nu)\right)^\ell\right]^2. 
\ee
where $\rho_\lm^H(x;\,\nu)= 1 + O(x)$ are the residual amplitude
corrections to the horizon waveform. 
Following Refs.~\cite{Nagar:2011aa,Bernuzzi:2012ku} we use a $1^{+3}$~PN
approximation for $\rho_\lm^H(x;\,\nu)$ and we include only the $\ell=2$
contribution in Eq.~\eqref{eq:FlmH} (i.e., we fix $\ell_{\rm max}=2$ 
in Eq.~\eqref{eq:FH}).

Finally, this means that the fractional horizon correction (before multiplication by
the additional factor $1-4\nu+2\nu^2$) in Eq.~\eqref{eq:fhat} is of the form
\be
 x^4 \hat{f}^H(x;\,\nu)= x^4 \left[\hat{S}_{\rm eff}^{(0)}
  \left(\rho_{22}^H(x;\,\nu)\right)^2\right]^2 +  x^5 \left[\hat{S}_{\rm eff}^{(1)}
  \left(\rho_{21}^H(x;\,\nu)\right)^2\right]^2
\ee
where $\hat{S}_{\rm eff}^{(0)}= \hat H_{\rm eff}$,   $\hat{S}_{\rm eff}^{(1)}=p_\varphi/(r_\omega v_\varphi)$,
and where we use  4PN accurate expressions for  $\rho_\lm^H(x;\,\nu)$,
\be
\label{eq:all_rho}
\rho_\lm^H(x;\,\nu) = 1 + c^\lm_1 x +  c^\lm_2 x^2 +  c^\lm_3 x^3 + c^\lm_4 x^4 ,
\ee
with  values for the needed $\ell=2$ coefficients $c^\lm_i$, $i=1,\dots,4$  
listed in Table~\ref{tab:clm}.

Let us finally come to discussing the radial component $\F_{r_*}$ of the radiation reaction force.
 Such a contribution was generally neglected in previous EOB papers, or replaced 
(e.g. in~\cite{Buonanno:2005xu,Pan:2011gk}) by an expression which was not consistently derived.
Recently, Bini and Damour~\cite{Bini:2012ji} (building on previous work by Iyer and 
collaborators~\cite{Iyer:1995rn,Gopakumar:1997ng,Gopakumar:1997bs}) have shown that 
consistency with the usual EOB definition of  $ \hat{\cal F}_{\varphi} $ 
(as being equal to minus the instantaneous flux of angular momentum) required a 
{\it specific} form for $\hat{\F}_{r_*}$ which differed from previously used expressions.

The result of Ref.~\cite{Bini:2012ji}  that we use here has the form
\be
\label{eq:Fr}
\hat{\F}_{r_*}=-\dfrac{5}{3}\dfrac{p_{r_*}}{p_\varphi}\hat{\F}_\varphi (1 + c_1(\nu) u + c_2(\nu)u^2 )
\ee
where the  coefficients entering the 2PN correction read~\cite{Bini:2012ji}  
\begin{align}
c_1(\nu)&= -\dfrac{227}{140}\nu + \dfrac{1957}{1680}, \\
c_2(\nu)& = \dfrac{753}{560}\nu^2 + \dfrac{165703}{70560}\nu - \dfrac{25672541}{5080320}.
\end{align}

\subsection{Post-post-circular initial data}
\label{sec:post_post_circular}
The construction of initial data for the EOB dynamics has been 
refined in a series of works~\cite{Buonanno:2000ef,Buonanno:2005xu,Damour:2007yf,Damour:2007vq}.
Here we shall use the {\it post-post-circular} prescription, 
introduced in 2007 (see Sec.~III~B of~\cite{Damour:2007yf}), and 
then used in all subsequent EOB-related works by our 
group~\cite{Damour:2007vq,Damour:2008te,Damour:2009kr,Damour:2009wj,Damour:2010zb,
Baiotti:2010xh,Damour:2011fu,Baiotti:2011am,Damour:2012yf,Bernuzzi:2012ku}.
This choice allows one to start the EOB dynamics (with negligible initial eccentricity)
at a frequency that is compatible with the initial frequency of the NR waveforms 
we shall use here ($M\omega_{22}\approx 0.0345$ approximately corresponding to 
initial separation $R_0\approx 15M$, see Table~\ref{tab:EOB_ID} below).
Note that, by contrast, Pan et al.~\cite{Pan:2011gk}, who use the less accurate 
post-circular initial data of Ref.~\cite{Buonanno:2005xu}, start their EOB runs
at an initial radius $R_0\gtrsim 50M$ (corresponding to an initial GW frequency $M\omega_{22}\leq 0.005$) 
in order to get a good circularization of the dynamics at the 
frequency where numerical simulations start. 
 
For completeness, let us review here the construction of post-post-circular
initial data for a given relative initial separation $r_0$.
 We introduce a formal bookkeeping parameter $\varepsilon$ 
(to be set to 1 at the end) 
in front of the radiation reaction $\F_\varphi$ in the EOB equations of motion. 
The quasi-circular inspiralling solution of the EOB equations of motion can then
be formally expanded in powers of $\varepsilon$ as
\begin{align}
p_{\varphi}^2& = j_0^2(r)\left(1 + \varepsilon^2 k_2(r) + \O(\varepsilon^4)\right), \\
p_{r_*}    & =\varepsilon\pi_1(r) + \O(\varepsilon^3).
\end{align}
Here, $j_0^2(r)$ is the usual {\it circular} approximation to
the inspiralling squared angular momentum as explicitly given by
\be
\label{eq:j02}
j_0^2(r) = -\dfrac{A'(u)}{\left[u^2 A(u)\right]'}
\ee
where the prime means $d/du$ (recall $u\equiv 1/r$). 
The order $\varepsilon$-approximation to $p_{r_*}$, i.e. $\pi_1(r)$ 
(``post-circular'') is then obtained by approximating 
the left-hand side (l.h.s) of Eq.~\eqref{eob:3} by 
$dp_\varphi/dt\approx dj_0(r)/dt=(dj_0(r)/dr)(dr/dt)$.
This determines $dr/dt$ and thereby a corresponding 
value of $p_{r_*}$ using Eq.~\eqref{eob:2} (where we neglect 
the $p_{r_*}^3$ contribution).
This leads to the following explicit expression for $\pi_1(r)$:
\begin{align}
\label{pi1}
\varepsilon \pi_1(r)
= \left[\nu\hat{H}_{\rm EOB}\hat{H}_{\rm eff}\left(\dfrac{B}{A}\right)^{1/2}\left(\dfrac{dj_0}{dr}\right)^{-1}\hat{\cal
    F}_\varphi\right]_0,
\end{align} 
where the subscript $0$ indicates that the r.h.s. is evaluated
at the leading circular approximation $\varepsilon\to 0$.
The {\it post-post-circular} approximation to $p_\varphi^2$
(term $\varepsilon^2k_2$ above) is then obtained by approximating 
the l.h.s. of Eq.~\eqref{eob:4} by
\be
\dfrac{dp_{r_*}}{dt}\approx \varepsilon \dfrac{d\pi_1(r)}{dr}\dfrac{dr}{dt},
\ee
where the radial derivative $d\pi_1(r)/dr$ is numerically computed.
This transforms Eq.~\eqref{eob:4} in a linear equation for $p_\varphi^2$,
which leads to an explicit expression for the $r$-dependent correction 
$\varepsilon^2 k_2(r)$ introduced above. In solving for  $p_\varphi^2$ we 
keep, for additional accuracy, the contribution proportional 
to $p_{r_*}^4\simeq \varepsilon^4\pi_1^4(r)$. 

Table~\ref{tab:EOB_ID} lists the post-post-circular data (as a function of $r_0$) 
obtained by this procedure, as we have used them in the present study. 
Note that these values mainly depend on the parameters entering the 
$A$ function, $(a_5^c,a_6^c)$, and depend almost negligibly 
on the values of the NQC parameters $a^{\lm}_i$ entering $\F_\varphi$ 
that appears on the r.h.s. of Eq.~\eqref{pi1}.
Actually, the values listed in Table~\ref{tab:EOB_ID} were computed 
by keeping only the $(a_1^{22},a_3^{22})$ NQC contributions.

\begin{table}[t]
  \caption{\label{tab:EOB_ID}Post-post-circular initial data for EOB dynamics that we shall consider in this paper
  to assure negligible initial eccentricity. They are obtained with the choice $a_5^c=23.5$ and 
  $a_6^c(\nu)=[-110.5+129(1-4\nu)]\left[1-1.5\times 10^{-5}/(\nu-0.26)^2\right]^{1/2}$.} 
  \begin{center}
    \begin{ruledtabular}
      \begin{tabular}{cccccc}
        $q$   &   $\nu$  &    $r_0$  & $p_\varphi$ & $p_r$ & $p_{r_*}$  \\
        \hline 
        1        & 0.25           & 16  & 4.42467206 & -0.00101207 & -0.00088970 \\
        2        & $0.\bar{2}$    & 15  & 4.31684166 & -0.00113064 & -0.00098466 \\
        3        & 0.1875         & 15  & 4.31889270 & -0.00096445 & -0.00083930 \\
        4        & 0.1600         & 15  & 4.32052406 & -0.00083018 & -0.00072202 \\
        6        & 0.1224         & 15  & 4.32276101 & -0.00064296 & -0.00055874 \\
   \end{tabular}
  \end{ruledtabular}
\end{center}
\end{table}

\subsection{Analytically unknown parameters,  choices to be made, NR-completion of the EOB model}
\label{sec:summary_unknown}

Let us summarize the parameters entering the construction of our EOB model, emphasizing which parameters 
contain important dynamical information, which ones are already known with sufficient accuracy,
which ones depend on reasonable choices we can make, and how NR data
can be used to complete the EOB model by determining the various parameters.

At face value, the EOB model defined above depends on quite a few analytically unknown parameters,
namely:  $a_5^c(\nu)$, $a_6^c(\nu)$,  the 6 NQC
parameters $(a_i^\lm,b_i^\lm)$ for each waveform multipole, the values of the mass $M_f$ and spin $a_f$
of the final black hole, the number $N$ of QNM modes used in
the ringdown signal, and the width $\Delta^{\rm match}$
of the QNM matching comb.

Our attitude towards the use of NR data to complete the EOB model by determining these parameters
is the following:
\begin{itemize}
\item[(i)]  As already said, we think (in view of previous EOB results~\cite{Damour:2009kr,Pan:2011gk})
that it is a reasonable choice to impose some a priori relation between $a_5^c(\nu)$ and $a_6^c(\nu)$,
so as to look only for one free dynamical parameter. Here we shall fix  $a_5^c(\nu)$ 
to the simple value $a_5^c(\nu)= 23.5 $, Eq.~\eqref{a5c}. This leaves only  $a_6^c(\nu)$ 
as free parameter. We shall discuss below (see Section~\ref{sec:a6c_determination})
how the nonperturbative information contained in NR phasing data can be used to determine the value of
$a_6^c(\nu)$, in a way which is nearly decorrelated from the uncertainties in the determination of
other parameters.  Let us already indicate here a possible analytical fit to represent 
the, essentially linear in $\nu$, final result we shall get for this EOB parameter
\begin{align}
\label{a6_final}
a_6^c(\nu)  &= \left(-110.5 + 129(1-4\nu)\right)\nonumber\\
            &\times \left(1-\dfrac{0.000015}{(\nu-0.26)^2}\right)^{1/2} \\  
&\text{from Caltech-Cornell-CITA data.} \nonumber
\end{align}
We think that the NR determination of  $a_6^c(\nu)$ leads to important information
about the conservative dynamics of binary black holes (as we shall illustrate below);

\item[(ii)]  Concerning the NQC parameters $(a_i^\lm,b_i^\lm)$, the procedure  explained above
reduces their determination from nonperturbative NR data to a {\it single} choice, namely that
of the time  $\tnrLR$ on the NR (retarded) time axis corresponding to the EOB time  $\teobLR$ (which can
be thought of as defining the ``EOB merger time''). The choice of $\tnrLR$ on the NR time axis is not
a matter of convention, but has (a priori) important physical consequences. It must be done by combining
information coming both from comparable-mass NR simulations, and from extreme-mass-ratio ones.
For reasons that shall be discussed below, we shall choose, for each mass ratio $\nu$, a specific value of   $\tnrLR(\nu)$
given by
\begin{align}
\label{deftnrLR}
\tnrLR(\nu) & =  \tAmax(\nu) \nonumber \\
            & +  f(\nu)\left(\tdomgmax(\nu)-\tAmax(\nu)\right)
\end{align}
where 
\be
f(\nu) = \frac{1}{6} (1 + 3(1-4\nu)),
\ee
and where $ \tAmax(\nu)$  is the NR time  when the NR quadrupolar amplitude reaches its peak, 
and $\tdomgmax(\nu)$  the NR time when the quadrupolar frequency has an inflection point.
Here $f(\nu)$ varies between $f(0) =2/3$ and $f(1/4)=1/6$ as $\nu$ varies between $0$ and $1/4$.
 $\tnrLR(\nu)$ always lies on the right of (i.e. later than) the NR time  $ \tAmax(\nu)$. 
We shall extract nonperturbative information from NR data by computing from the various multipolar 
NR waveforms a certain number of
derivatives of their amplitudes and frequencies at the extraction point   $\tnrLR(\nu)$.
 
\item[(iii)] Building on previous work, we shall use the simple (NR-based) analytical fits~\eqref{Mfaf} for the mass
 and spin of the final black hole. Note, however, that, in principle, the EOB model (when NR-completed 
 by NQC corrections up to merger) does yield, by itself, predictions
 for $M_f$ and $a_f$~\cite{Buonanno:2000ef,Damour:2007cb}.
 This might be useful  in cases (e.g. with large, precessing spins) where one does not have in 
 hand accurate analytical fits for the 
characteristics of the final black hole.
 
\item[(iv)] We shall use here $N= 5$ QNM's, and 
as explained below, we shall fix $\Delta^{\rm match} = 0.7 M $ for all multipoles.
 Note, that by contrast, Ref.~\cite{Pan:2011gk} uses $N=8$ QNM's, 
 introduces ``pseudo QNM's", and employs much larger matching
 intervals, which also vary with $\lm$. [E.g., the latter reference
 uses  $\Delta^{22}= 5 M$ and  $\Delta^{33}= 12 M$.]

\item[(v)]  Let us finally note that (contrary to~\cite{Pan:2011gk}) we shall not introduce adjustable 
      parameters in the waveforms, nor shall we introduce special modifications to improve the 
      behavior of some subdominant multipoles.
\end{itemize}

\section{Numerical relativity information, and ${Q}_\omega$ diagnostic}
\label{secQomega}

\subsection{Overview of numerical waveforms data}

The NR data we use here to complete the EOB waveform were obtained 
with the  Spectral Einstein Code 
(SpEC) developed by the Caltech-Cornell-CITA collaboration~\cite{Scheel:2006gg,Boyle:2006ne,Boyle:2007ft,Boyle:2008ge,Scheel:2008rj}. 
Specifically, we used the waveforms  recently published in 
Ref.~\cite{Buchman:2012dw}, coming from 
simulations of nonspinning black hole binaries with mass ratios
$q=m_1/m_2=(1,2,3,4,6)$. Before
their publication, these data were already used in some EOB/NR and PN/NR 
comparisons~\cite{LeTiec:2011bk,Pan:2011gk,MacDonald:2012mp}.
We address the reader to Ref.~\cite{Buchman:2012dw} for all technical details about the
numerical setup and estimates of the accuracy. Here we only recall that these are the 
longest published  waveforms to date (together with the 33 orbits, equal-mass waveform 
of Ref.~\cite{MacDonald:2012mp}), with a number of gravitational wave cycles up to 
merger (here conventionally defined as the maximum of the modulus of the quadrupolar
 metric waveform $|h_{22}^{\rm NR}|$) respectively $N_{\rm GW}=\{33,31,31,31,43\}$.
We made use of two different types of waveform data:  curvature, $\psi^4_\lm$, 
and metric,  $h_\lm$, extrapolated to infinite extraction radius.
Indeed, the metric waveform $h_\lm$ was also directly extracted from 
the numerical spacetime using a Regge-Wheeler-Zerilli-based (RWZ) 
approach~\footnote{This type of RWZ approach was initiated by
Abrahams and Price~\cite{Abrahams:1995gn} and first implemented in 
the form of Ref.~\cite{Buonanno:2009qa} in Refs.~\cite{Pazos:2006kz,Baiotti:2008nf}.},
see Appendix of Ref.~\cite{Buonanno:2009qa} for a discussion.

\subsection{Estimating the NR $Q_\omega(\omega)$ function for the curvature waveform}
\label{sec:Qomg_NR}
\begin{figure}[t]
 \includegraphics[width=0.48\textwidth]{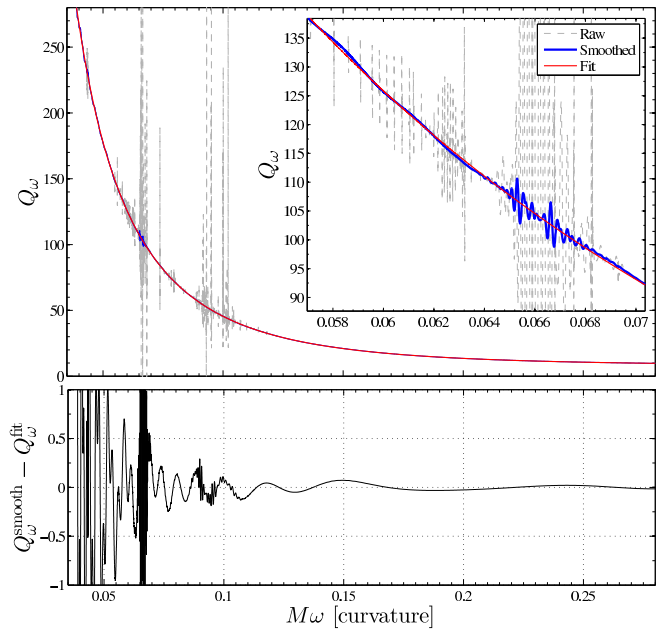}
    \caption{ \label{fig:q1_Qomg_raw_smooth} (color online) Top panel: Raw NR, curvature waveform, data; 
      smoothed data and fit. Bottom panel: difference between smoothed data and the fit.} 
\end{figure}
\begin{figure}[t]
 \includegraphics[width=0.48\textwidth]{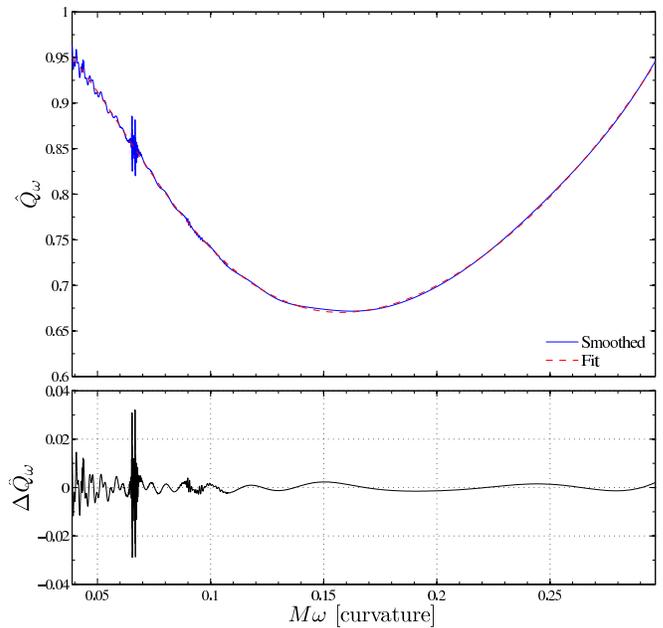}
    \caption{\label{fig:q1_hat_Qomg} (color online) Fitting the Newton-rescaled $\hat{Q}_\omega$, curvature waveform, function. 
      The top panel contrasts the smoothed data with the outcome of the fit. 
      The bottom panel shows their difference.} 
\end{figure}
\begin{table*}[t]
  \caption{ \label{tab:Qomg_fit}Coefficients entering the fitting function for the Newton-rescaled, curvature-waveform, 
    $\hat{Q}_\omega$, Eq.~\eqref{Qhat_fit}. In the second column we also report the frequency interval $M(\omega_1,\omega_2)$
  on which the fit was performed.}
  \centering  
  \begin{ruledtabular}  
  \begin{tabular}{ccccccccccc}        
    \hline
    $q$ & $M\omega_1$ & $M\omega_2$  & $n_1$ & $n_2$ & $n_3$ & $n_4$ & $n_5$ & $d_1$ & $d_2$ & $d_3$ \\
   \hline
    1 &  0.03877 & 0.29654  & -27.88757 & 256.94609  & -1053.85269 &  1926.40123  & -1274.57280 & -6.60927    & 47.87468  & -104.35366 \\
    2 &  0.04133 & 0.29709  & 15.51565  & -372.20973 &  1725.17714 &  -3145.40474 & 2105.30901  & -15.77371   &  90.80420 & -103.95952\\
    3 & 0.04476  & 0.29642  &  6.50413  & -243.11043 & 1108.15054  & -1913.96522  & 1193.58571  &-14.56312    & 79.22950  & -111.26577\\
    4 &  0.04819 & 0.29671  &  0.52391  &-172.68858  & 806.19352   & -1350.57200  & 797.20936   & -14.39733   & 78.72314  & -124.83300 \\
    6 & 0.04280  & 0.29720  &  7.18353  & -247.53679 & 1096.28420  & -1833.39721  & 1090.77236  & -14.59256   & 75.97063  & -113.64331 \\
    \hline
  \end{tabular}
  \end{ruledtabular}  
\end{table*}

In this subsection we shall explain how we extracted from NR data a useful, intrinsic measure 
of the NR phase evolution, namely the  $Q_\omega(\omega)$ function. This function is a 
convenient version of the ``intrinsic phase acceleration'' function $\alpha(\omega)$ introduced 
in Ref.~\cite{Damour:2007yf}, which was defined such that $d \omega/ dt = \alpha(\omega)$.  
This function is an {\it intrinsic} measure of the time-domain phase evolution
in the sense that it is independent of the two shift ambiguities that affect any time-domain phase, $\phi(t)$:
an arbitrary phase shift  $\phi \to \phi + c$, and an arbitrary time shift $t \to t+ \tau$.  The  $Q_\omega(\omega)$ function is defined as
\be
\label{eq:Qomg}
Q_\omega(\omega) \equiv   \dfrac{\omega^2}{\alpha({\omega})}= \dfrac{\omega^2}{\dot{\omega}}
\ee
Note that this definition is equivalent to saying that the time-domain phase accumulated in the frequency interval $(\omega_1,\omega_2)$ is given by the integral
\be
\phi_{(\omega_1,\omega_2)} = \int_{\omega_1}^{\omega_2} Q_\omega d\ln\omega.
\ee

The function  $Q_\omega(\omega)$ has proven to be a very useful diagnostic of phase evolution in recent EOB/NR comparisons of binary 
neutron stars~\cite{Baiotti:2010xh,Baiotti:2011am,Bernuzzi:2012ci}. 
Note that, in the definition,  $\omega$ can be  the frequency either of the curvature waveform, or of the
metric one (thereby defining two different, though numerically close, functions).  In general, one only
considers the frequency of the dominant quadrupolar waveform, though one can
also study the $Q_\omega(\omega)$ function of any $(\ell,m)$ multipole.
Note also that we are here considering the phase acceleration of a time-domain phase.
One can also usefully consider the frequency-domain counterpart of  $Q_\omega(\omega)$, defined
as ${Q^{\rm FD}}_\omega(\omega) \equiv \omega^2 d^2 \psi(\omega)/ d\omega^2$, where $\psi(\omega)$ denotes
the phase of the Fourier-transformed waveform. In the stationary phase approximation,
$Q^{\rm FD}_\omega(\omega)$ is simply equal to the time-domain $Q_\omega(\omega)$ 
(see, e.g.,  Eq.~(17) in \cite{Damour:2012yf}).

Let us now discuss  how to  accurately estimate $Q_\omega(\omega)$ from the numerical data, in
spite of the loss of accuracy associated to the fact that its definition~\eqref{eq:Qomg} involves 
the computation of two derivatives of the phase $\phi(t)$.
We consider the $\psi^4_{22}$ curvature waveform extrapolated to
infinite extraction radius, decompose it in amplitude and phase with the convention
\be
\psi^4_{22} = |\psi^4_{22}| e^{-\i\phi_{22}} \ ,
\ee
and consider as frequency in the definition~\eqref{eq:Qomg} 
 the curvature quadrupolar frequency:   $\omega \equiv \dot{\phi}_{22}$.

It is somewhat of a challenge  to get an accurate $Q_\omega$ out from numerical data. For example, 
in the case of binary neutron star waveforms, Refs.~\cite{Baiotti:2010xh,Baiotti:2011am,Bernuzzi:2012ci} 
argued that the successive straightforward differentiation (using finite-differencing, 4th-order stencils) 
of the numerical data is unable to get this information correctly, so that a suitable fitting of
the GW phase was necessary to obtain something qualitatively and quantitatively correct.
For general binary black hole simulations, due to the much higher resolution involved as well as 
due to the higher finite differencing operators used, direct differentiation could be more meaningful
than in the binary neutron star case. 
This should be even more true for SpEC data, since they are expected to be particularly accurate.

Therefore, as a first step we directly computed $Q_\omega$ from the raw data simply by 
finite-differencing $\phi$ twice to get $\omega$ and $\dot{\omega}$, i.e., applying twice
a 1st-derivative finite-differencing operator with 4th-order stencil.
The result of this first step is shown, for $q=1$ data, as a dashed, light-gray line 
in Fig.~\ref{fig:q1_Qomg_raw_smooth} (see also the close up). The figure shows 
the presence of high frequency noise  which prevents one from using this diagnostics  as is
for reliable quantitative estimates.

To improve on this, and  get a quantitatively useful estimate of the $Q_\omega$ curve, 
we applied three more steps.  First, in order to eliminate the high-frequency noise, 
we smoothed  $\omega(t)$ with a Sgolay filter. Second,  we computed the time derivative 
of the smoothed $\omega(t)$, and then smoothed again that derivative with a Sgolay filter. 
These two steps succeeded in strongly reducing the high-frequency noise 
in the curve (thick line in Fig.~\ref{fig:q1_Qomg_raw_smooth}, blue online). 
However, there remained  a low-frequency residual oscillation in the resulting $Q_\omega$curve
(evident in the inset of Fig.~\ref{fig:q1_Qomg_raw_smooth}).
We do not know the precise origin of this residual oscillation (it might either be related 
to some small residual eccentricity in the waveform or connected to the extrapolation procedure), 
but we think it is of spurious numerical origin and that it does not have any actual physical 
content (note that such an oscillation is not present in the EOB $Q_\omega$ curve).

This led us to our third step: a {\it fitting procedure} of the $Q_\omega(\omega)$ function.
 To implement such a fitting procedure, it is convenient to first
normalize the $Q_\omega$ curve with respect to its leading-order, Newtonian part,
\be
Q_\omega^N(\omega) = \dfrac{5}{3\nu}2^{-7/3} \omega^{-5/3},
\ee
thereby factoring out the blowing up of  $Q_\omega(\omega)$ at low frequency. 
The normalized function
\be
\label{eq:Qhat_def}
\hat{Q}_\omega(\omega) = Q_\omega/Q_\omega^N
\ee
stays of order unity on the full frequency range (and $\hat{Q}_\omega\to 1$ for $\omega\to 0$)
and  is a better starting point for any fitting 
procedure (see Fig.~\ref{fig:q1_hat_Qomg} for $q=1$). Then we use as fitting 
template for $\hat{Q}_\omega$ a general analytical structure consistent with the 
structure of $\hat{Q}_\omega$ predicted by PN theory in the adiabatic approximation. 
More precisely, the 3.5PN-accurate expansion of  $\hat{Q}_\omega$ is a Taylor 
expansion in half-integer powers of   $x=(M \Omega)^{2/3}$ (modulo some logarithmic corrections) 
that reads
\be
\label{eq:Qomg_PN}
\hat{Q}_\omega^{\rm PN}(x) = 1 + b_2 x + b_3 x^{3/2}+b_4 x^2 + b_5 x^{5/2}+b_6 x^3 + b_7 x^{7/2},
\ee
where
\begin{subequations}
\begin{align}
b_2 &= \dfrac{743}{336} + \dfrac{11}{4}\nu,\\
b_3 &=-4\pi , \\
b_4 &=  \dfrac{3058673}{1016064} + \dfrac{5429}{1008}\nu + \dfrac{617}{144}\nu^2,\\
b_5 &= \pi\left(-\dfrac{7729}{672} + \dfrac{13}{8}\nu\right),\\
b_6 &=  -\dfrac{10817850546611}{93884313600}+\dfrac{32}{3}\pi^2\nonumber\\
     & + \left(\dfrac{3147553127}{12192768} -\dfrac {451\pi^2}{48}\right)\nu 
       - \dfrac{15211}{6912}\nu^2       
       + \dfrac{25565}{5184}\nu^3 \nonumber      \\
      & + \dfrac{1712}{105}(\gamma_E+2\log x + 2\log 2),\\
b_7 &= -\pi\left(\dfrac{15419335}{1016064} + \dfrac{75703}{6048}\nu + \dfrac{14809}{3024}\nu^2\right).
\end{align}
\end{subequations}

This motivated us to  fit the smoothed version (coming out of the first two steps) 
of the numerically computed $\hat{Q}_\omega(\omega)$ with a Pad\'e-type function of 
the form
\be
\label{Qhat_fit}
\hat{Q}_\omega^{\rm fit} (x_\omega) = \dfrac{1 + n_1 x_\omega + n_2 x_\omega^{3/2} + n_3 x_\omega^2 + n_4 x_\omega^{5/2}+n_5 x_\omega^3}{1+d_1 x_\omega + d_2 x_\omega^2 +d_3 x_\omega^3},
\ee
where  $x_\omega \equiv(M \omega/2)^{2/3}$.

Let us now illustrate the result of performing this three-step evaluation of the numerical  
$Q_\omega(\omega)$ function. The top panel of Fig.~\ref{fig:q1_Qomg_raw_smooth} shows, for $q=1$, 
the three successive estimates of the numerical $\hat{Q}_\omega$: the raw one 
(dashed line, featuring many large spikes), the smoothed one (solid line),
and finally the fit obtained using the template~\eqref{Qhat_fit}. Note that all those 
curves are plotted versus $M \omega$. The bottom panel of the same figure shows the 
{\it difference} $\Delta \hat{Q}_\omega(\omega)=\hat{Q}_\omega^{\rm smoothed}(\omega) - \hat{Q}_\omega^{\rm fit}(\omega)$
between the smoothed data and the fit. Note that this difference is oscillating 
around 0, which indicates that the fit has been effective in averaging away the 
low-frequency oscillation remaining after having smoothed the high-frequency noise.
The procedure works in the same way for the other mass ratios, and for each one the 
difference $\Delta \hat{Q}_\omega(\omega)$ nicely oscillates around zero. 

We list in in Table~\ref{tab:Qomg_fit}, for all mass ratios, the 
fitting coefficients of  the smoothed numerical $\hat{Q}_\omega$ to the 
template Eq.~\eqref{Qhat_fit}. Note that this list of coefficients provides 
a convenient way of condensing the information contained in the NR phasing 
during most of the inspiral and plunge (indeed, our fit worked well up to 
frequency $ M \omega \simeq 0.3$, which is quite close to the merger). 
This packaging of the NR phasing information might be useful for
many purposes,  e.g., comparing various numerical simulations, computing 
the Fourier transform in the stationary-phase approximation, etc.

\section{Revisiting test-mass limit results}
\label{sec:test_mass}

\begin{figure}[t]
\begin{center}
 \includegraphics[width=0.45\textwidth]{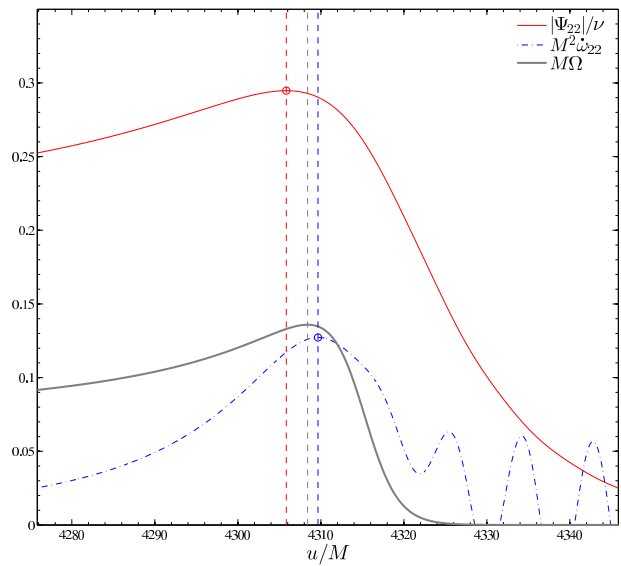}
    \caption{ \label{fig:merger_test_mass} 
      (color online) Hierarchy of important points of the test-mass (Zerilli-normalized) quadrupolar metric waveform
      (divided by $\nu$),
      $\Psi_{2 2}/\nu \equiv  (R/M) h_{2 2} /( \nu \sqrt{24}) $ around the merger point.
      The orbital frequency $\Omega$ peaks at approximately 2/3 of the time interval between the peak
      of the metric amplitude and the inflection point of the GW frequency, i.e. the first peak of 
      $\dot{\omega}_{22}$.}
\end{center}
\end{figure}

\begin{table}[t]
  \caption{\label{tab:DT_merger}Time intervals $\tdomgmax-\tAmax$ 
  for all numerical waveforms considered in this paper.} 
  \begin{center}
    \begin{ruledtabular}
      \begin{tabular}{cccccc}
        $q$   &   $\nu$  &    $\tdomgmax-\tAmax$  \\
        \hline 
        1        & 0.25              &   3.2493   \\
        2        & $0.\bar{2}$       &   3.4426   \\
        3        & 0.1875            &   3.3261   \\
        4        & 0.1600            &   3.5714   \\
        6        & 0.1224            &   3.5681   \\
        $\infty$ & 0                 &   3.8158   
   \end{tabular}
  \end{ruledtabular}
\end{center}
\end{table}
\begin{figure*}[t]
\begin{center}
 \includegraphics[width=0.45\textwidth]{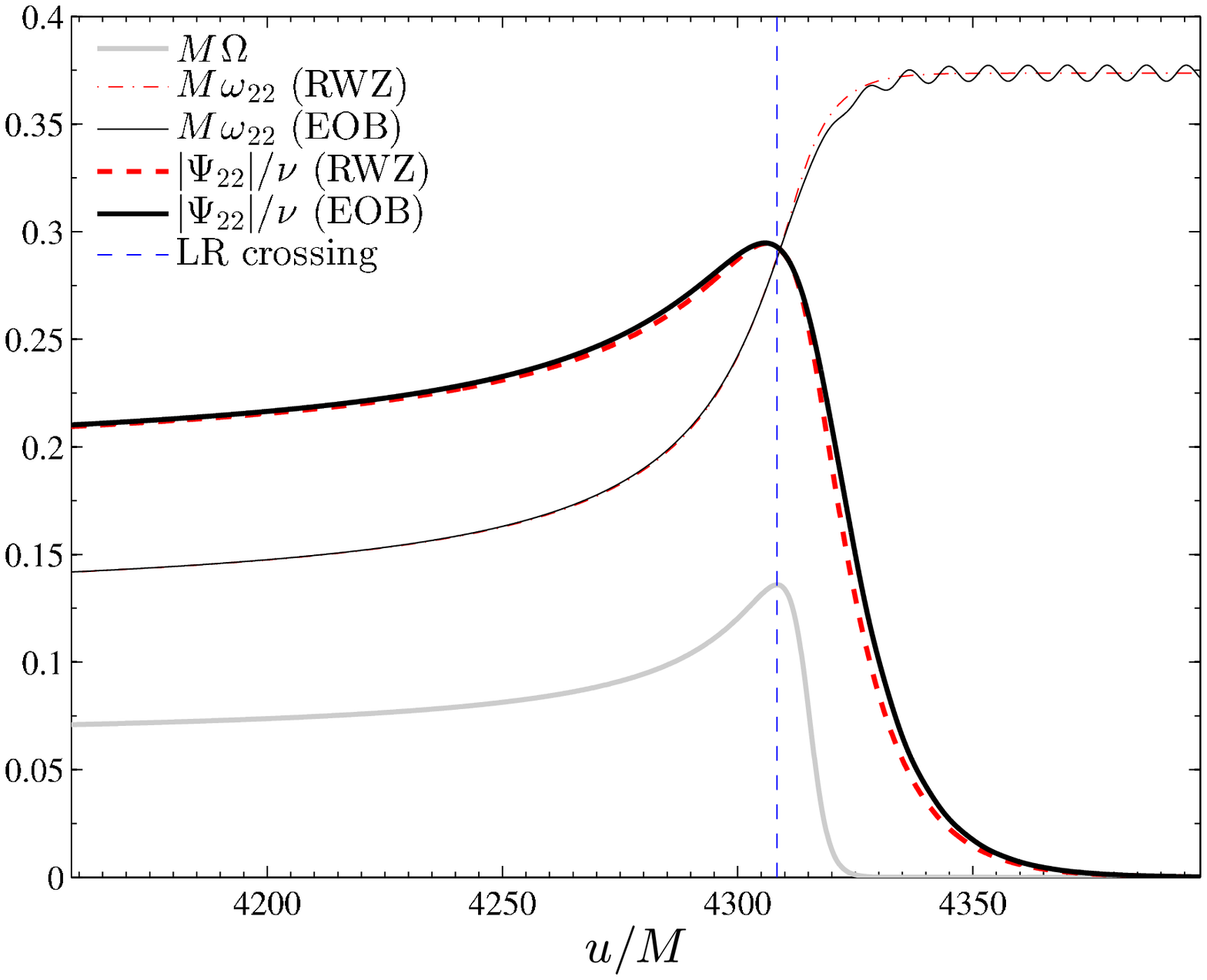}
\hspace{5mm}
 \includegraphics[width=0.45\textwidth]{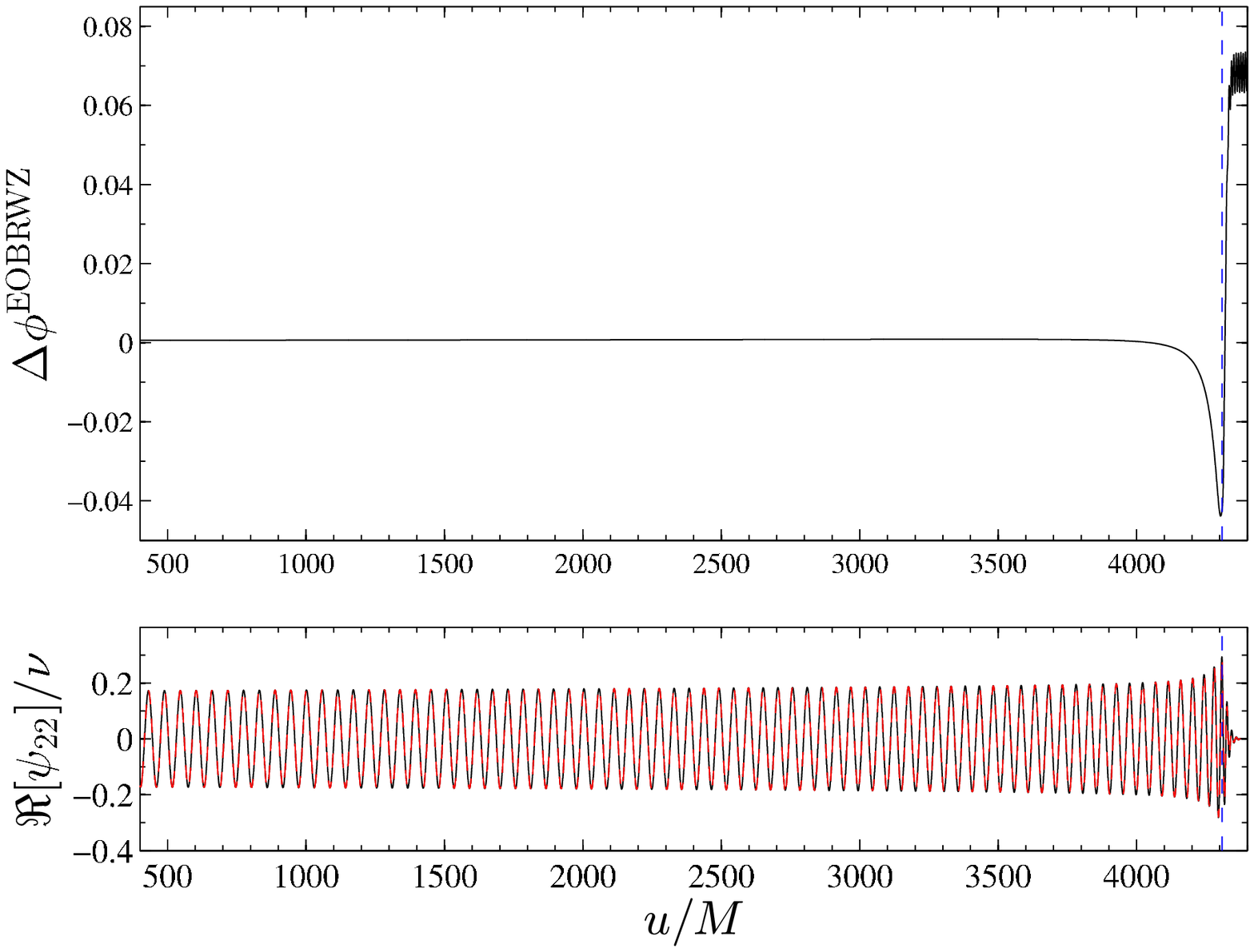} \\
\vspace{5mm}
 \includegraphics[width=0.45\textwidth]{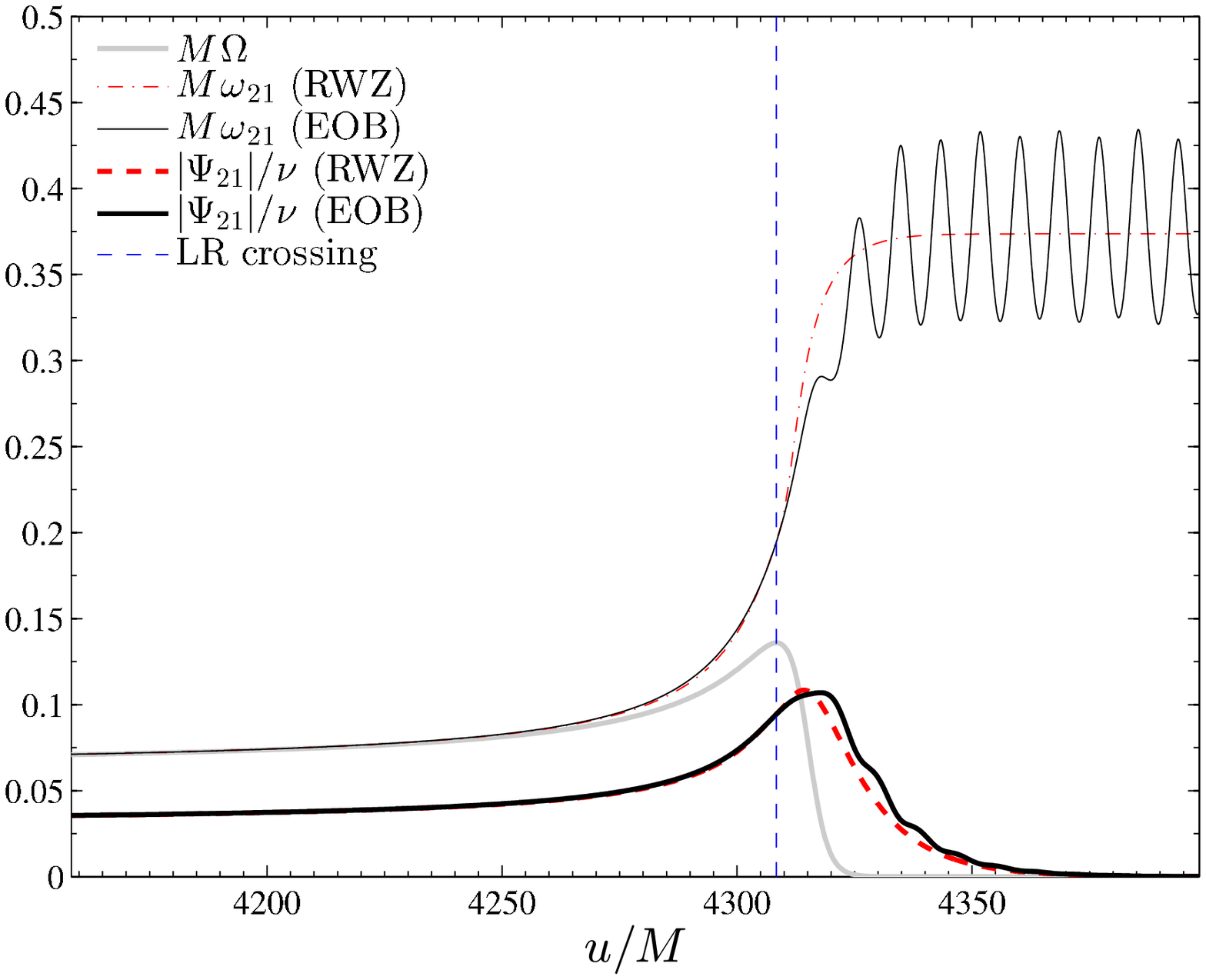}
\hspace{5mm}
 \includegraphics[width=0.45\textwidth]{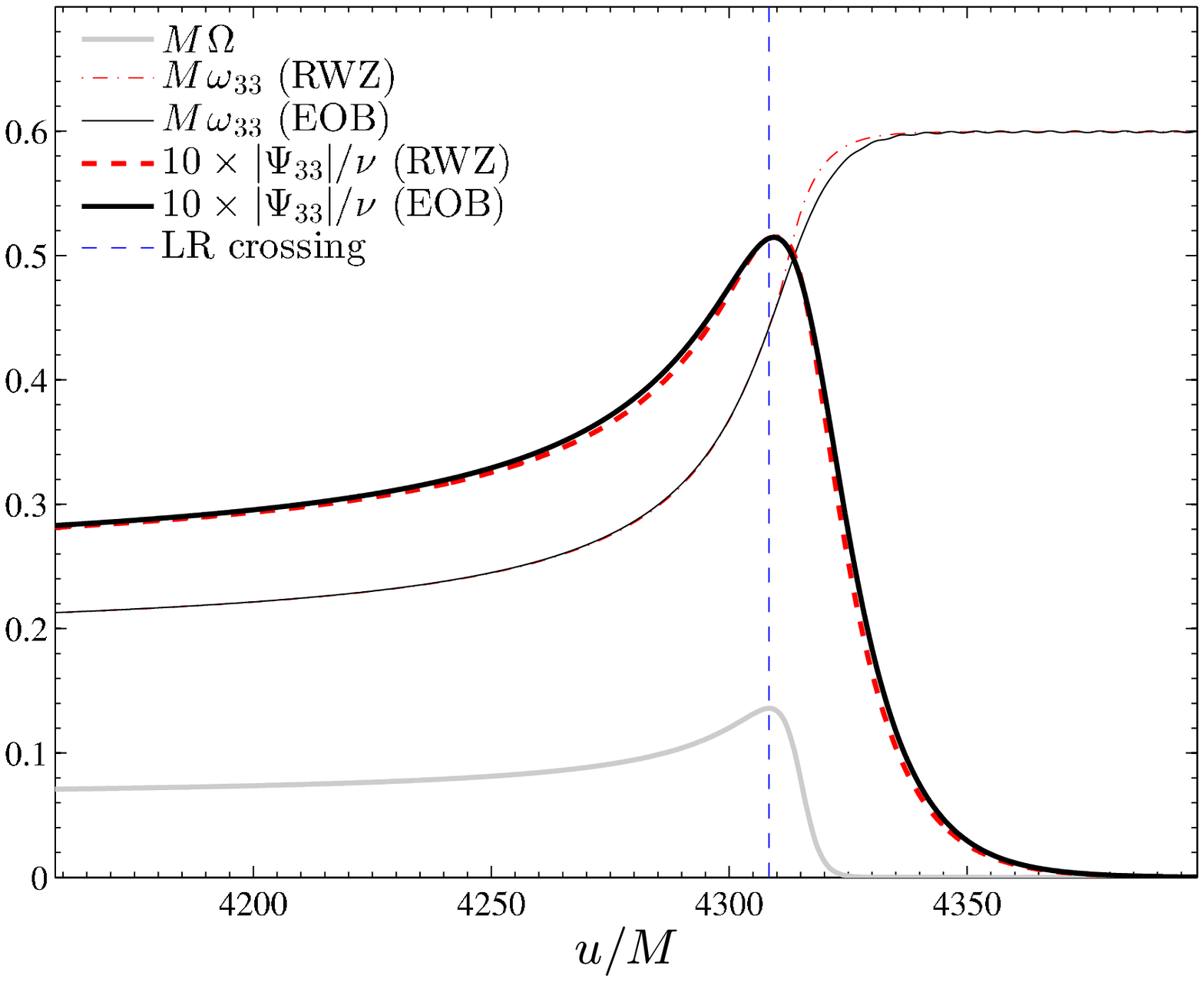}
    \caption{ \label{fig:q1000_l2}(color online) Test-mass waveform: comparison between RWZ waveform
    extracted at $\scri^+$ and EOB waveform completed by the 6-parameter NQC correction 
    factor to the waveform, Eq.~\eqref{eq:hNQC}. Top panels: $\ell=m=2$ multipole, modulus and frequency (left) and phasing (right).
    Bottom panels: $\ell=2$ and $m=1$ and $\ell=m=3$ frequency and modulus. The ringdown is modeled using 5 (positive-frequency only) QNMs.}
\end{center} 
\end{figure*}
\begin{figure}[t]
\begin{center}
 \includegraphics[width=0.45\textwidth]{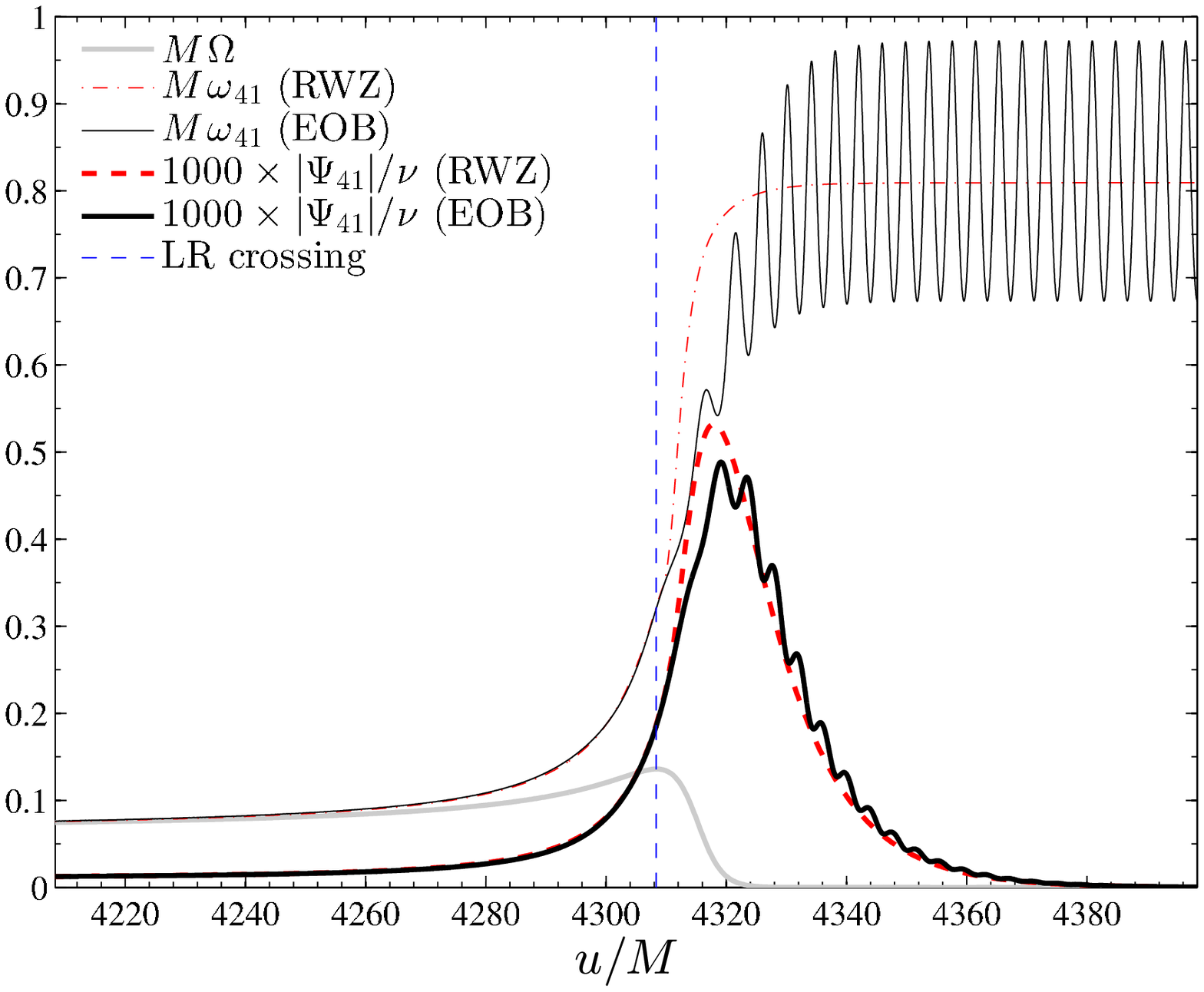}
    \caption{ \label{fig:tp_l4m1} 
      (color online)Test-mass limit: comparison between $\ell=4$, $m=1$ EOB and RWZ modulus and frequency} 
\end{center}
\end{figure}

\subsection{The new information acquired from test-particle computations}
\label{sec:test_mass_summary}

Before dealing with the Caltech-Cornell-CITA comparable-mass waveforms, we shall
revisit in this Section the test-mass limit case $\nu \ll1$ both to motivate our introduction
of an NR extraction point $\tnrLR$ differing from the peak of the waveforms, and to test
the performance of the basis of functions $n_i$'s that we shall use in our NQC correction factor,~\eqref{eq:hNQC}.

State-of-the-art computations of multipolar RWZ waveforms for the plunge
and merger of a test particle (of mass $\mu$), moving in a Schwarzschild background (of mass $M$),
 and submitted to a leading-order EOB resummed radiation-reaction force,
have been presented  in a recent series of 
works~\cite{Bernuzzi:2010ty,Bernuzzi:2010xj,Bernuzzi:2011aq,Bernuzzi:2012ku}.  
These works have used a recently developed method~\cite{Zenginoglu:2007jw,Zenginoglu:2009ey,Zenginoglu:2010cq}
allowing one to combine  an accurate treatment of the particle
motion in the strong field region, with the extraction of the waveforms directly at null
infinity ($\scri$). The findings of Ref.~\cite{Bernuzzi:2010xj} that will be of direct 
interest for our present study are:
\begin{itemize}
\item[(i)] The  extraction of the waveforms at $\scri$ allows one
to relate the retarded time $t^{\rm NR}$ used as argument of the waveforms
to the  EOB time $t^{\rm EOB}$ used in the dynamics of the particle (namely,
one has simply  $t^{\rm NR}= t^{\rm EOB}$).
This allows one to connect without ambiguity features in the waveform
(such as, say, a peak in the modulus of $h_{22}(t^{\rm NR})$)
with features in the dynamics (such as, say, the location
along the $t^{\rm EOB}$ axis of the maximum of the orbital frequency $\Omega(t^{\rm EOB})$).
Such a possibility is not available in comparable-mass NR simulations,
because they do not track the light cones emitted by the center of mass of the
binary system. In addition, even if they did, this would not allow one to relate
the dynamical EOB time $t^{\rm EOB}$ to the waveform time $t^{\rm NR}$, because we would not know
the exact relation between $t^{\rm EOB}$ and the NR coordinate time relevant for the NR dynamics.

\item[(ii)] Using the connection between the waveform time $t^{\rm NR}$ and the dynamical time
 $t^{\rm EOB}$  offered by (i), it was found that   the waveform amplitude $A_{22}$ 
 peaks approximately $\approx 2.56M$   {\it earlier} than the orbital frequency $\Omega$,
 i.e. $ \teobLR \approx  \tAmax + 2.56 M$ .
 This is new information which conflicts with the standard simplifying EOB assumption
 of a coincidence between the peaks of $A_{22}$ and of $\Omega$.
 The existence of a difference between $\teobLR$ and $\tAmax$ was later confirmed in 
 Ref.~\cite{Barausse:2011kb} and extended to the case of a spinning central black hole.

\item[(iii)] Using this new information, Ref.~\cite{Bernuzzi:2010xj}
suggested to incorporate it in a new prescription for the determination
of the EOB NQC correction factor based on extracting numerical data at the
NR point $\tnrLR$ corresponding to $\teobLR$, rather than\footnote{Note that, by contrast,
Ref.~\cite{Taracchini:2012ig} has chosen to keep, for $\ell=m=2$ the 
NR extraction point at $\tAmax$ and to map it to an EOB time earlier than $\teobLR$.} 
at  $\tAmax$. They implemented such a prescription by imposing a $C^1$ contact at 
$\tnrLR  \leftrightarrow \teobLR$ both (for the first time) 
between the modulus and the frequency of the waveform.
They then showed that such a procedure produced NQC-corrected EOB waveforms
which had an excellent agreement with the numerical RWZ waveforms up to merger.

\end{itemize}

The procedure we indicated in Eqs.~\eqref{comp_1}-\eqref{comp_6} above
is a generalization of this prescription to a  $C^2$ contact requirement.
We shall test below the increased accuracy brought by using such a $C^2$ contact requirement,
involving six NQC-parameters, instead of the $C^1$ contact requirement used in 
Ref.~\cite{Bernuzzi:2010xj}, which involved only four NQC parameters.
This test will also probe the new basis of NQC correction functions $n_i$'s
used in  Eq.~\eqref{eq:hNQC}.

\subsection{Zooming on the structure of the test-mass waveform near merger}

Before doing the latter test,  let us display the finding (ii) of 
Ref.~\cite{Bernuzzi:2010xj} by investigating in detail the structure 
of the $\ell=m=2$ RWZ waveform around the peak of the modulus, with 
the idea that a similar structure might hold in the comparable mass case.

Figure~\ref{fig:merger_test_mass}  shows together (as functions of  the waveform retarded time $u$,
which can be identified with the EOB dynamical time): the waveform modulus $A_{22}/\nu$; the orbital 
frequency $\Omega$; and  the {\it derivative} of the GW frequency $\dot{\omega}_{22}$.
Here,  $A_{22}$ is the modulus of the Zerilli-normalized quadrupolar metric test-mass 
waveform, $\Psi_{2 2} \equiv  (R/M) h_{2 2} / \sqrt{24} $. [For a general multipole the Zerilli
normalized metric waveform is   $\Psi_{\lm} \equiv  (R/M) h_{\lm} / \sqrt{ (\ell +2)(\ell+1)(\ell)(\ell-1)} $.]
The figure clearly illustrates how the orbital frequency peaks at a time $\teobLR$ that
is between the locations of the maxima of $A_{22}$ and $\dot{\omega}_{22}$, i.e. we
have the relation $\tAmax < \teobLR < \tdomgmax$.
Quantitatively, given that we have $\teobLR- \tAmax=2.565388 M$ and $\tdomgmax-\tAmax=3.815784 M$
we have that
\be
\frac{\teobLR - \tAmax}{\tdomgmax-\tAmax}=\frac{2.565388}{3.815784}=0.6723096\approx \dfrac{2}{3}.
\ee
The comparable-mass NR simulations show that the ordering  $\tAmax < \tdomgmax$ remains true for all
values of $\nu$ (for nonspinning binaries). By continuity, one then also expects  that the EOB orbital 
frequency will continue to peak between these two points for any value of $\nu$.  In other words, 
one expects that the correspondence between the EOB and NR time axes should be such that the EOB 
dynamical time $\teobLR(\nu)$ corresponds to an NR waveform time  $\tnrLR(\nu)$ such that
$\tAmax(\nu) < \tnrLR(\nu) <\tdomgmax(\nu)$ for any $\nu$.  It is convenient to rewrite these 
inequalities as
\be
\label{eq:NR_lr_position}
 \tnrLR(\nu) -  \tAmax(\nu)= f(\nu)\left(\tdomgmax(\nu)-\tAmax(\nu)\right)
\ee 
where $f(\nu)$ is an unknown function satisfying the condition that $f(0)=2/3$, and
expected to remain positive for any $\nu$. 
The intervals $\tdomgmax-\tAmax$ as measured on the numerical waveforms 
are listed in Table~\ref{tab:DT_merger}. 
We shall discuss our choice for the  function $f(\nu)$
in the following Section.

\subsection{Testing the improvements brought by requiring a $C^2$ contact when using the NQC factor Eq.~\eqref{eq:hNQC}}

Reference~\cite{Bernuzzi:2010xj} was able to build a rather
satisfactory EOB waveform modulus and frequency up to merger for the $\ell=m=2$ mode 
(and in general for all $\ell=m$ modes) by using four NQC parameters (two for the amplitude and two for the phase).
However, their results for the modulus were much less
satisfactory for the other ($\ell \neq m$) subdominant multipoles, such as the $\ell=2$, $m=1$ one. 
Let us show here how the use of the  new NQC factor, Eq.~\eqref{eq:hNQC} (which contains six NQC parameters, and uses
different choices for the NQC functions  $n_3$ and $n_4$) improves the closeness of the EOB waveform
to the numerical (RWZ) one.
To be consistent with Ref.~\cite{Bernuzzi:2010xj}, the EOB dynamics used for this comparison
is slightly different from the one we discussed above. Namely: (i) we set to zero $\F^H_\varphi$, i.e. the 
horizon-absorption part of  the radiation reaction; (ii) we also set $\F_{r_*}=0$; (iii)
in addition, the residual phase corrections $\delta_\lm$ for $\nu=0$ are considered 
in their Taylor-expanded form and all terms (up to 4.5PN accuracy) are included
(see Appendix~\ref{sec:rholm}).

The improved EOB waveform obtained by using the new six-parameter NQC factor is illustrated in
Fig.~\ref{fig:q1000_l2}.
 The top panels refer
to the $\ell=m=2$ mode: frequency and modulus (left) and phasing (right). The bottom panels 
compare EOB and RWZ frequency and modulus for $\ell=2$, $m=1$ (left) and $\ell=m=3$ (right).
For all waveforms the QNM matching comb has a total width  $\Delta =0.7 M$ and we use five, positive frequency, 
QNMs. The restriction to positive frequency QNMs is the reason why one 
 cannot reproduce the oscillations during ringdown in the $\ell=2$,
 $m=1$ mode.
The improvement with respect to Fig.~3 of Ref.~\cite{Bernuzzi:2010xj} is evident. Notably, the $\ell=2$ $m=1$
modulus comes out extremely well (modulo the absence of negative-frequency modes to model the ringdown).
The $\ell=m=2$ phasing remains good also during merger and ringdown $-0.05<\Delta\phi^{\rm EOBRWZ}<+0.05$
(while the QNM matching of  Ref.~\cite{Bernuzzi:2010xj}  led to significantly larger dephasings during ringdown).
Note on the top right panel of Fig.~\ref{fig:q1000_l2} the behavior of the phase difference:  it dips just before merger
 down to $-0.04$ rad, and then jumps up to $+ 0.06$ rad during ringdown. Such a behavior is a useful 
compromise for keeping, on average, a good phasing through inspiral, plunge, merger and ringdown.

Finally, to prove the robustness of the NQC determination procedure and the accuracy of the EOB waveform 
for  higher multipoles, we show in Fig.~\ref{fig:tp_l4m1} the $\ell=4$, $m=1$ frequency and modulus. 
The agreement between EOB and RWZ waveform is again very good, modulo the absence of negative modes in 
the ringdown modelization.

\section{Comparable mass case:  $a_6^c(\nu)$, $\tnrLR(\nu)$, and phasing performance}
\label{sec:a6c_determination}

\subsection{Iterative procedure for determining  
  $\tnrLR(\nu)$ and $a_6^c(\nu)$: overview}

After having tested  the performance of the NQC factor~\eqref{eq:hNQC} in 
the test-mass limit, we now move to the comparable-mass case. 
Let us explain how we distilled crucial nonperturbative information out of  the Caltech-Cornell-CITA waveform data.
Our aim was to determine good values of  the 5PN parameter $a_6^c(\nu)$, 
and of the NR time $\tnrLR(\nu)$ corresponding to
the EOB time $\teobLR$.  We recall that  $\tnrLR(\nu)$ is parametrized by a function $f(\nu)$, according to
Eq.~\eqref{eq:NR_lr_position}.
Actually, the determinations of   $a_6^c(\nu)$, and of $\tnrLR(\nu)$ are correlated,
and must be done essentially simultaneously. From a practical point of view, we used 
an iterative, trial and error method.

First, for a given mass ratio $\nu$, and a given choice of
NR extraction time $\tnrLR$ (chosen around merger), we extract, 
from the behavior of the waveform in the immediate vicinity of the 
retarded time, $\tnrLR$, a collection of NR waveform quantities
$(A_\lm^{\rm NR},\dot{A}_\lm^{\rm NR},\ddot{A}_\lm^{\rm NR},\omega_\lm^{\rm NR},\dot{\omega}_\lm^{\rm NR},\ddot{\omega}_\lm^{\rm NR})$. 
[As mentioned above, these quantities are then used, for any given value of  $a_6^c(\nu)$,
to determine the parameters $(a_i^\lm,b_i^\lm)$ entering the EOB NQC factor; i.e. the last factor in 
the pre-merger EOB waveform~\eqref{eq:hlm}.]
Second, we study how the phase difference $\Delta \phi^{\rm EOB NR}$ between the so determined NQC-corrected
EOB waveform and the NR waveform evolves (either as a function of frequency, or of time) from
the beginning of the simulation up to $\tnrLR$. The evolution of the phase difference $\Delta \phi^{\rm EOB NR}$ depends
(after having chosen $\tnrLR$, and having implemented  the previous step) only  on  the 5PN ($\nu$-dependent) 
parameter $a_6^c(\nu)$. We then search (for each $\nu$) whether there exist  values of $a_6^c(\nu)$ 
which entail that $\Delta \phi^{\rm EOB NR}(a_6^c(\nu))$ remains within the numerical uncertainty during the
full simulation (up to   $\tnrLR$).  If  such a tuning of   $a_6^c(\nu)$  does not seem to lead to a 
a satisfactorily small phase discrepancy during the whole evolution, we try another value of the
NR extraction time and repeat the two steps above, until we end up with a better pair $(\tnrLR, a_6^c(\nu))$.

\begin{table*}[t]
  \caption{\label{tab:nqc_ab}Next-to-quasi-circular $(a_i^\lm,b_i^\lm)$ coefficients needed to complete
    the EOB for the five mass ratios considered. They are obtained by imposing $C^2$ conditions to the
    waveform amplitude and frequency around the merger.}  
  \centering  
  \begin{ruledtabular}  
  \begin{tabular}{cccccccc}        
    \hline
    $q$ & $\ell$ $m$ & $a_1^\lm$ & $a_2^\lm$ & $a_3^\lm$ & $b_1^\lm$ & $b_2^\lm$ & $b_3^\lm$ \\
    \hline    
    \hline
    1 & 2  2 &      -0.0577 &       1.8127 &      -0.1205 &       0.0794 &      -0.9164 &      -2.5890 \\
      & 3  2 &       0.0987 &       2.4076 &      -0.4987 &       0.0490 &       1.0532 &      -2.9188 \\
    \hline
    \hline
    2 & 2  1 &      -0.0656 &       0.4871 &       0.2959 &       0.2544 &       0.9033 &       1.1975 \\
      & 2  2 &      -0.0602 &       1.7571 &      -0.0646 &       0.0963 &      -0.8789 &      -2.0165 \\     
      & 3  2 &      -0.0658 &       2.7289 &      -0.2130 &       0.0864 &       1.2601 &      -2.7701 \\
      & 3  3 &      -0.0068 &       2.1915 &      -0.1837 &       0.2300 &      -1.2604 &      -2.2847 \\
    \hline
    \hline
    3 & 2  1 &      -0.0566 &       0.2988 &       0.3668 &       0.2636 &       0.8883 &       1.8284 \\
      & 2  2 &      -0.0484 &       1.6672 &      -0.0347 &       0.1161 &      -0.7453 &      -1.4052 \\
      & 3  2 &      -0.1349 &       2.6377 &      -0.0518 &       0.1747 &       1.5855 &      -0.2960 \\
      & 3  3 &       0.0016 &       2.0213 &      -0.0789 &       0.2560 &      -1.1539 &      -1.0416 \\    
    \hline
    \hline
    4  & 2  1 &      -0.0464 &       0.1260 &       0.4288 &       0.2772 &       1.0397 &       1.9334 \\
       & 2  2 &      -0.0396 &       1.5639 &       0.0004 &       0.1342 &      -0.5509 &      -1.1731 \\      
       & 3  2 &      -0.1360 &       2.3134 &       0.0559 &       0.2795 &       1.9825 &      -0.0350 \\
       & 3  3 &       0.0079 &       1.8294 &       0.0243 &       0.2807 &      -0.9226 &      -0.5157 \\
    \hline
    \hline
    6 & 2  1 &      -0.0323 &      -0.0701 &       0.5183 &       0.2770 &       1.2750 &       2.0649 \\
      & 2  2 &      -0.0229 &       1.4177 &       0.0397 &       0.1498 &      -0.4375 &      -0.9124 \\      
      & 3  2 &      -0.1114 &       1.7472 &       0.2487 &       0.3207 &       2.2001 &       1.5262 \\
      & 3  3 &       0.0296 &       1.5816 &       0.1347 &       0.3014 &      -0.7664 &      -0.1969 \\
    \hline
  \end{tabular}
  \end{ruledtabular}  
\end{table*}

When completed (by iteration), the above two steps completely define an NR-completed EOB model up to merger. The EOB waveform 
is then extended through merger and ringdown by attaching QNMs at the end of the inspiral-plus-plunge 
waveform, i.e. at the EOB time $\teobLR$ (which corresponds to the NR time  $\tnrLR$).
 This extension does not require the extraction of further 
NR information, but only requires to choose, by trial and error, reasonably good values of the 
number of QNM modes $N$, and of the total width of the matching comb $\Delta^{\rm match}$ around  
$\teobLR$. As already said, we use $N=5$ and  $\Delta^{\rm match}=0.7 M$.

\begin{figure}[t]
 \includegraphics[width=0.45\textwidth]{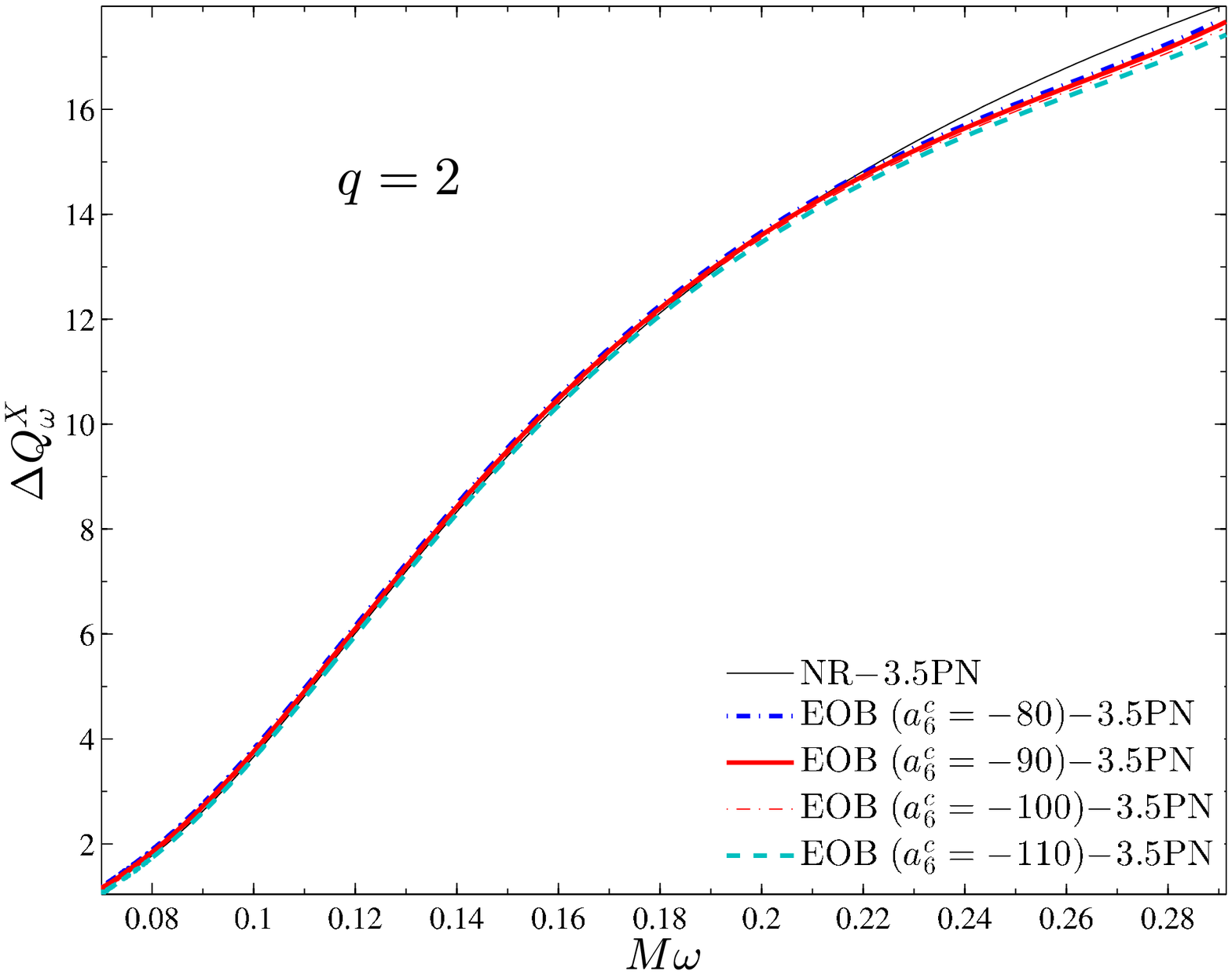} \\
\vspace{5mm}
 \includegraphics[width=0.47\textwidth]{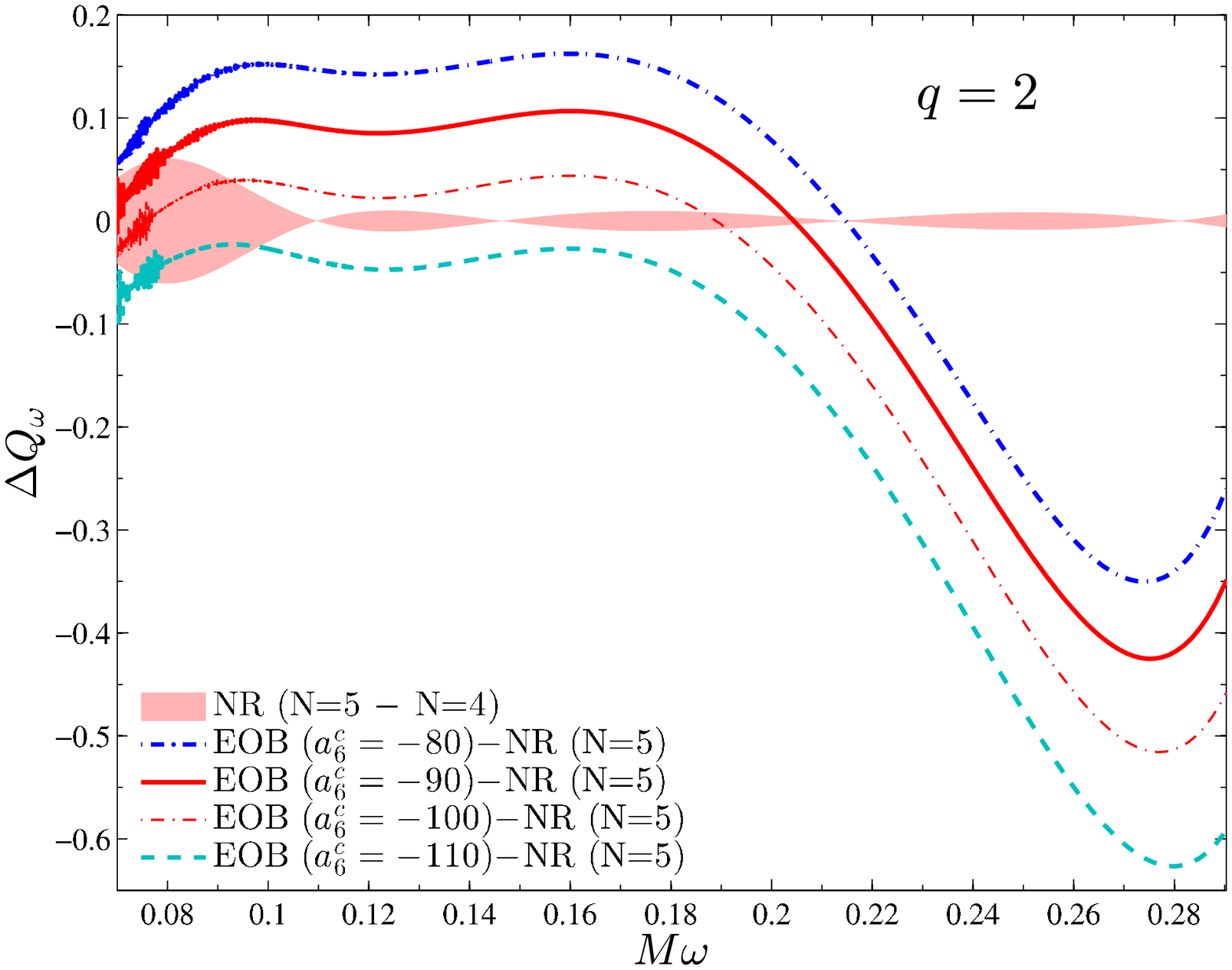}
    \caption{\label{fig:Delta_Qomg_q2}(color online) Using the $Q_\omega(\omega)$ 
      diagnostics to constrain the good values 
    of $a^c_6(\nu)$. The figure refers to the  case $q=2$, $\nu=2/9$. 
    Top panel: difference between either $Q_\omega^{\rm EOB}$ 
    or $Q_\omega^{\rm NR}$ and the 3.5PN-accurate, Taylor-expanded $Q_\omega^{\rm 3.5PN}$ given by Eq.~\eqref{eq:Qomg_PN}.
    Bottom panel: the lines show the differences $\Delta Q_\omega=Q_\omega^{\rm EOB}-Q_\omega^{\rm NR, N=5}$ for different 
    values of $a_6^c$. The shaded region exhibits the difference  $\Delta Q_\omega=Q_\omega^{\rm NR, N=5}-Q_\omega^{\rm NR, N=4}$
    where $N=4,5$ labels two different resolutions, respectively medium and
high, of the NR data~\cite{Buchman:2012dw}. See text for further details.}
\end{figure}

\begin{figure}[t]
 \includegraphics[width=0.45\textwidth]{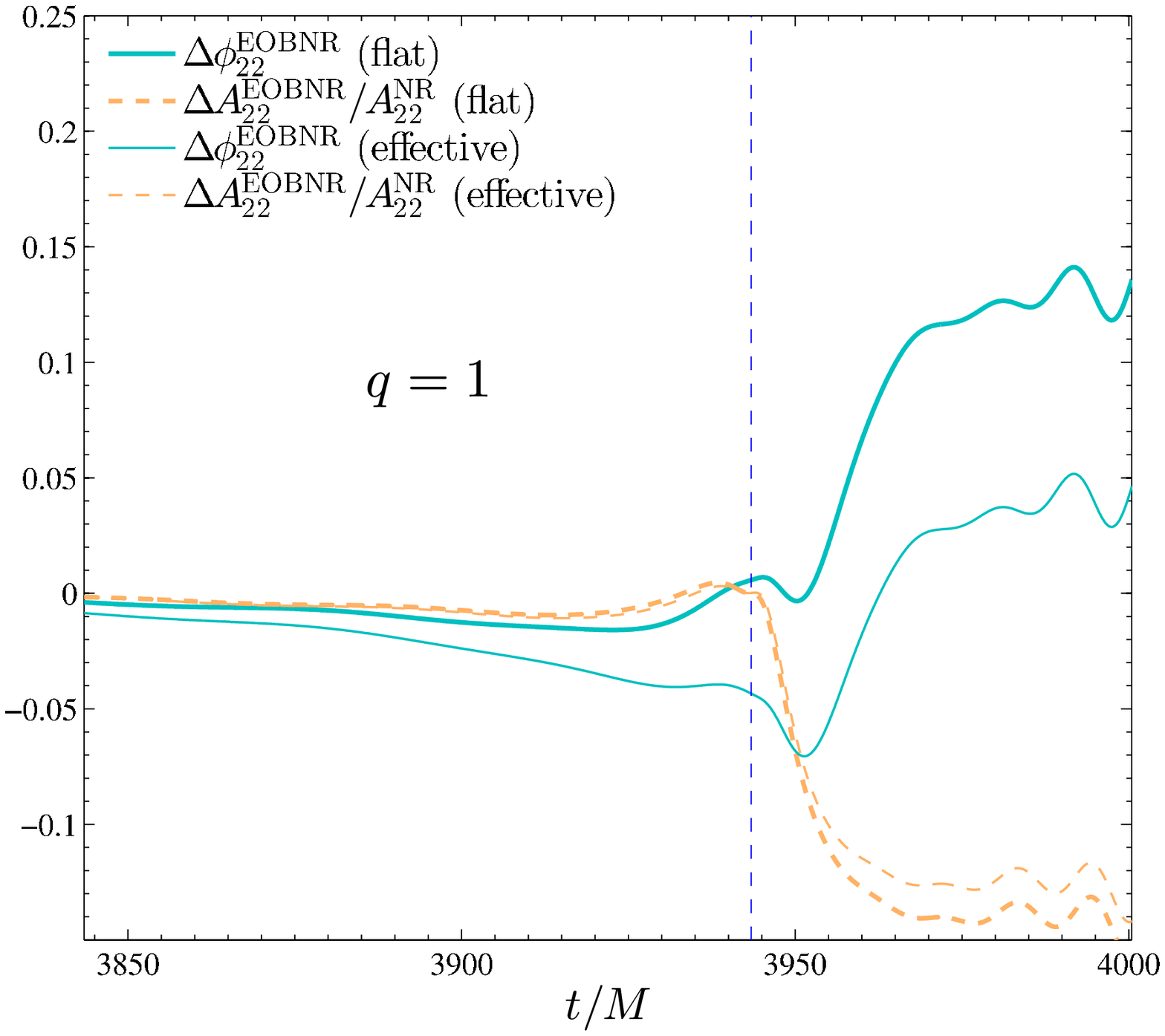}\\
 \vspace{5mm}
\includegraphics[width=0.47\textwidth]{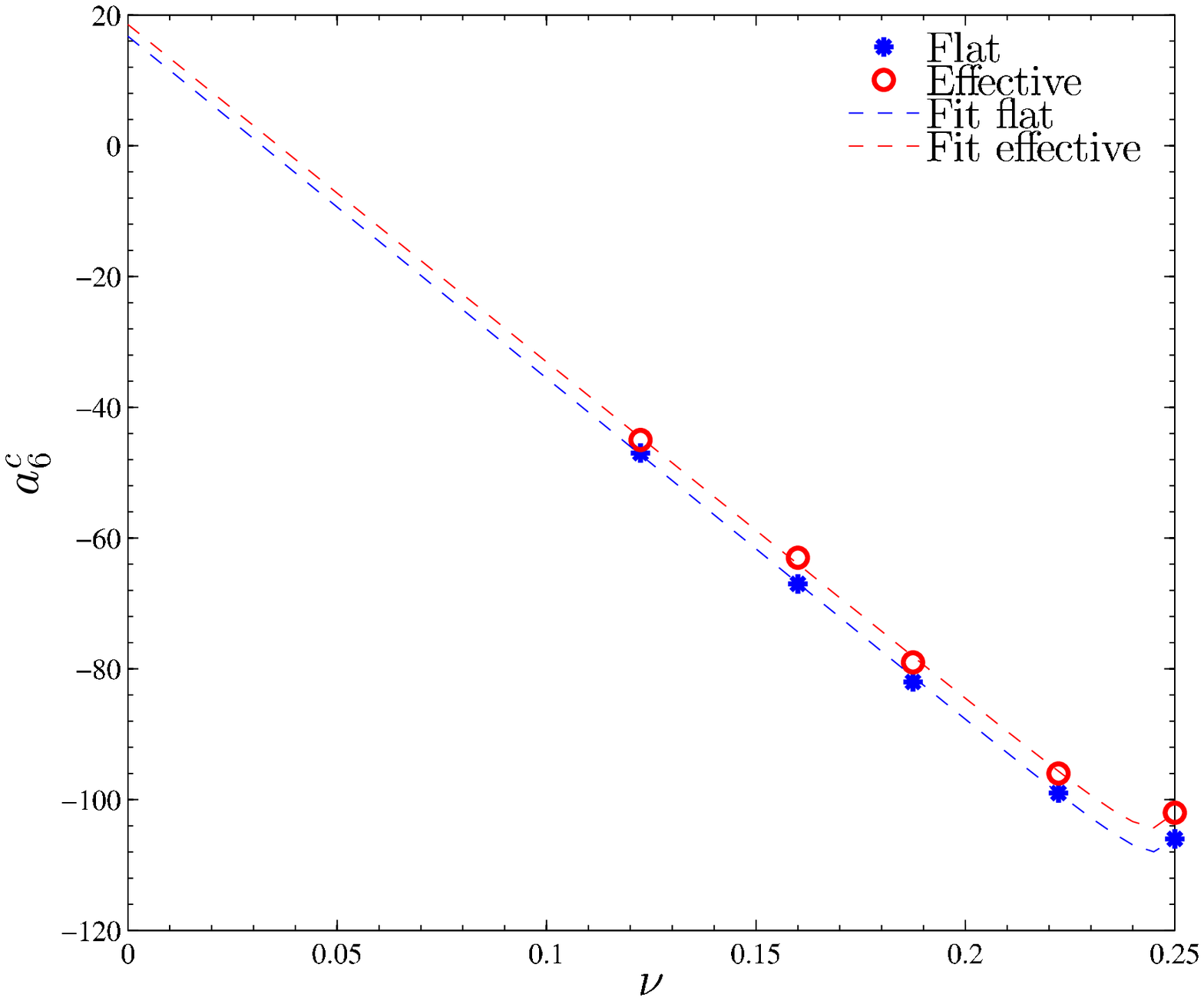}
    \caption{ \label{fig:dphi_flat_vs_effective} (color online) Top panel: Illustrating the meaning of 
    ``flat'' and ``effective'' EOB/NR phase differences around merger for $q=1$ ($\nu=0.25)$. 
     The flat phase difference is obtained here with $a_6^c(0.25)=-105.719 $ from Eq.~\eqref{eq:a6c_nu_flat}, 
     while the effective one uses $a_6^c(0.25)=-101.876$, from Eq.~\eqref{eq:a6c_nu}. 
     Bottom panel: ``flat" and ``effective" best values of $a_6^c$ and their analytical fits (dashed lines).} 
\end{figure}

\begin{figure*}[t]
 \includegraphics[width=0.45\textwidth]{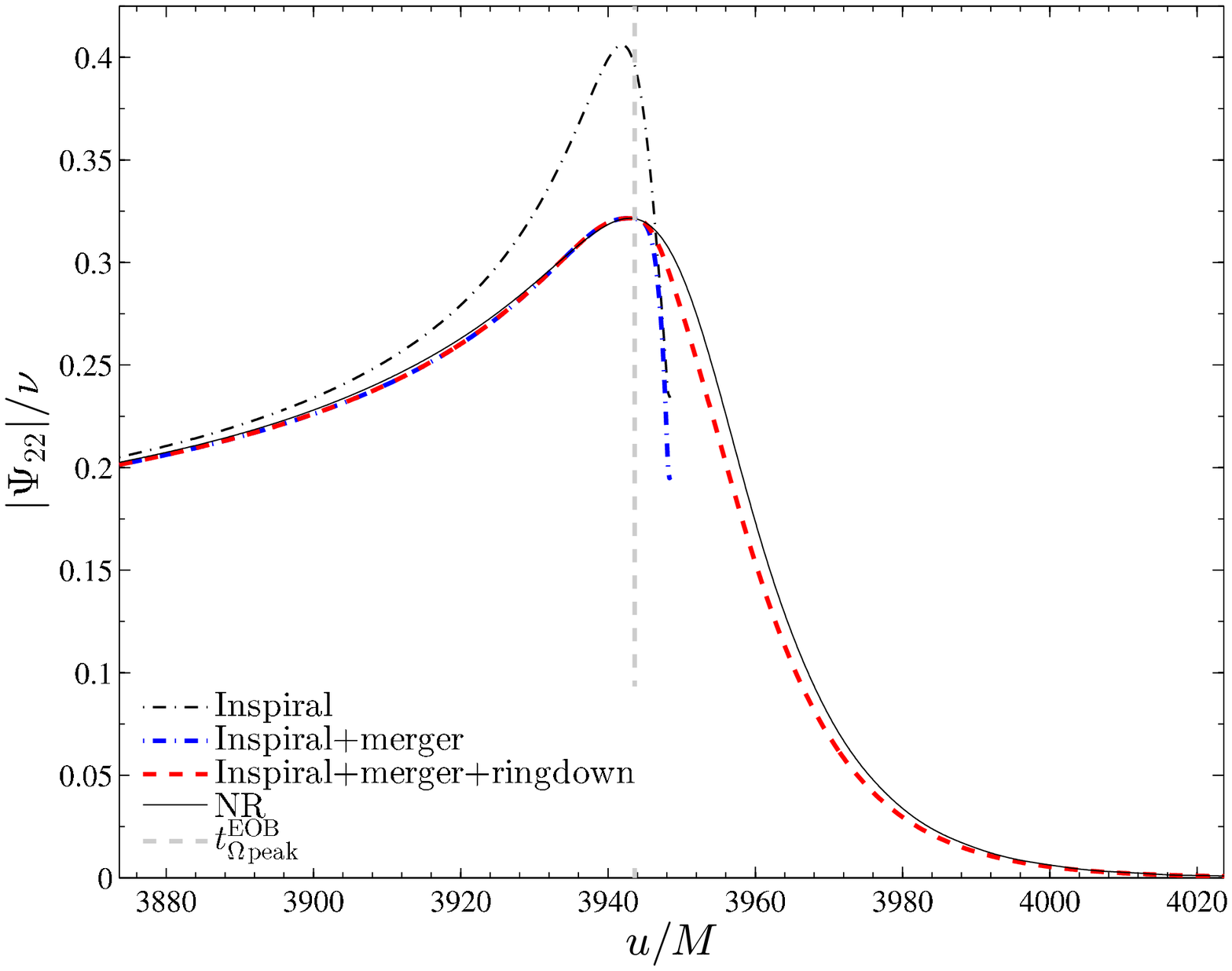}
\hspace{5mm}
 \includegraphics[width=0.45\textwidth]{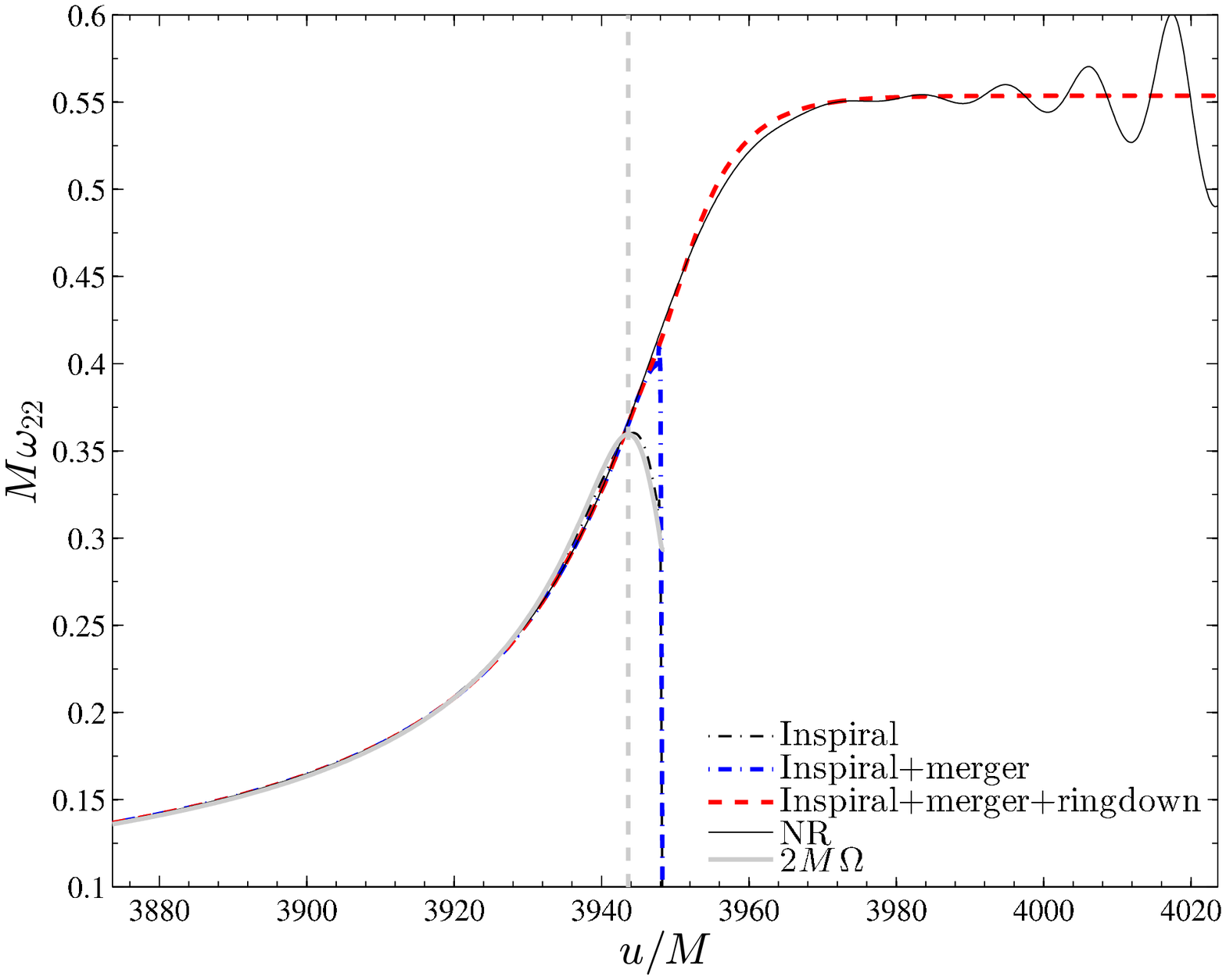}
    \caption{ \label{fig:effect_of_nqc} (color online) Illustrating the effect of the NQC factor on the ``bare'', 
    inspiral EOB waveform (equal-mass case): modulus (left panel) and frequency (right panel).} 
\end{figure*}

\begin{figure*}[t]
  \begin{center}
    \includegraphics[width=0.41\textwidth]{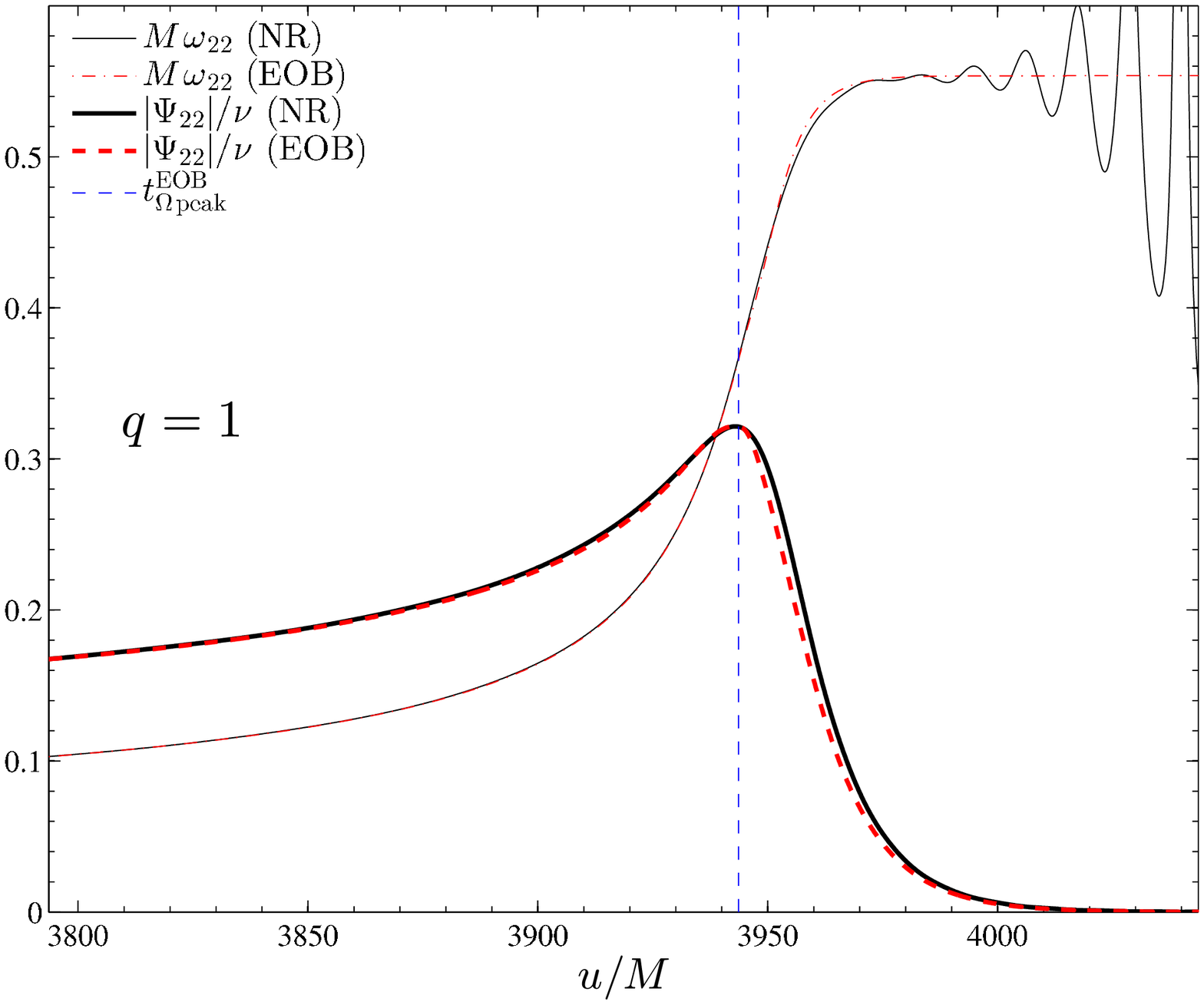}
    \hspace{5mm}
    \includegraphics[width=0.41\textwidth]{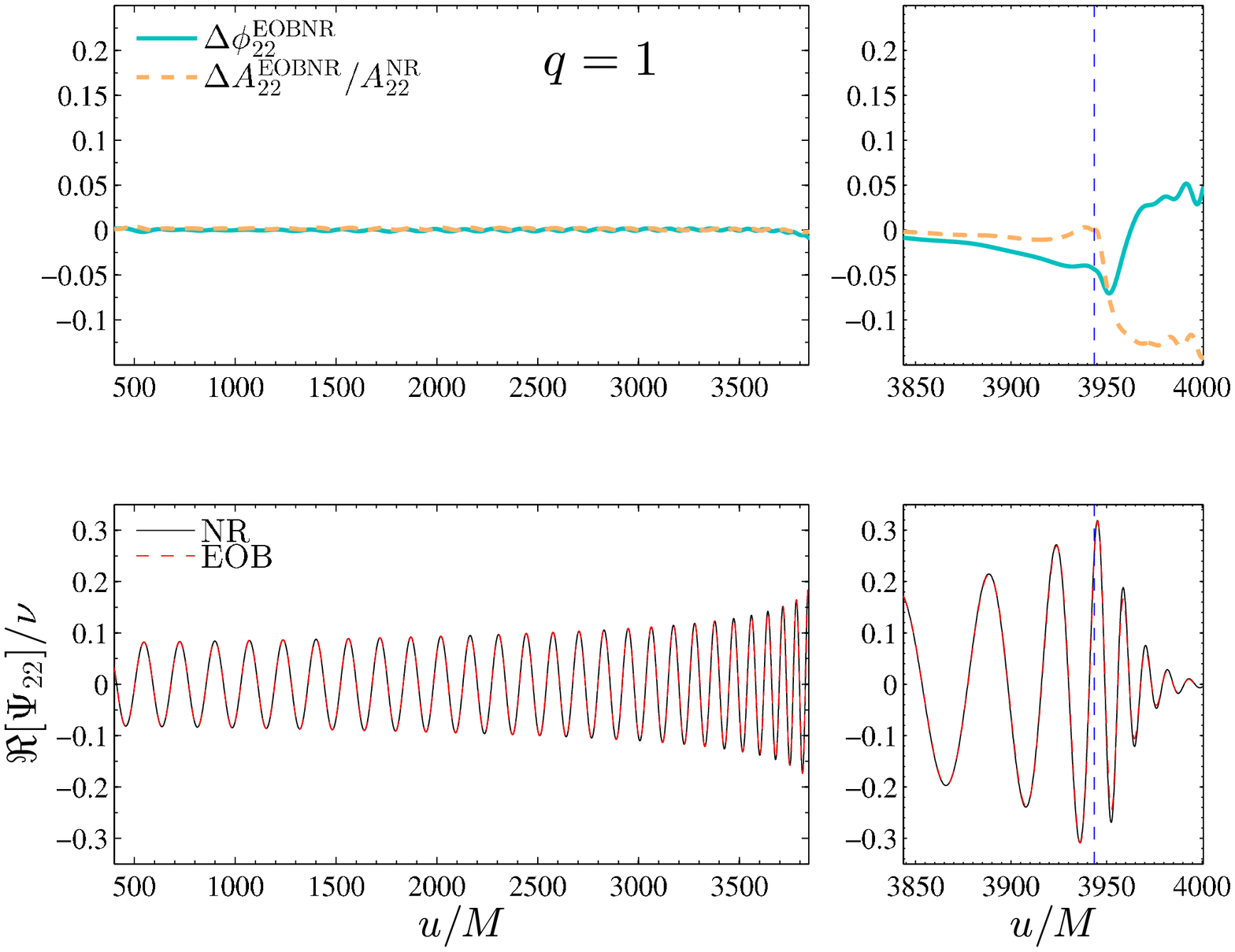} \\
    \vspace{5mm}
    \includegraphics[width=0.41\textwidth]{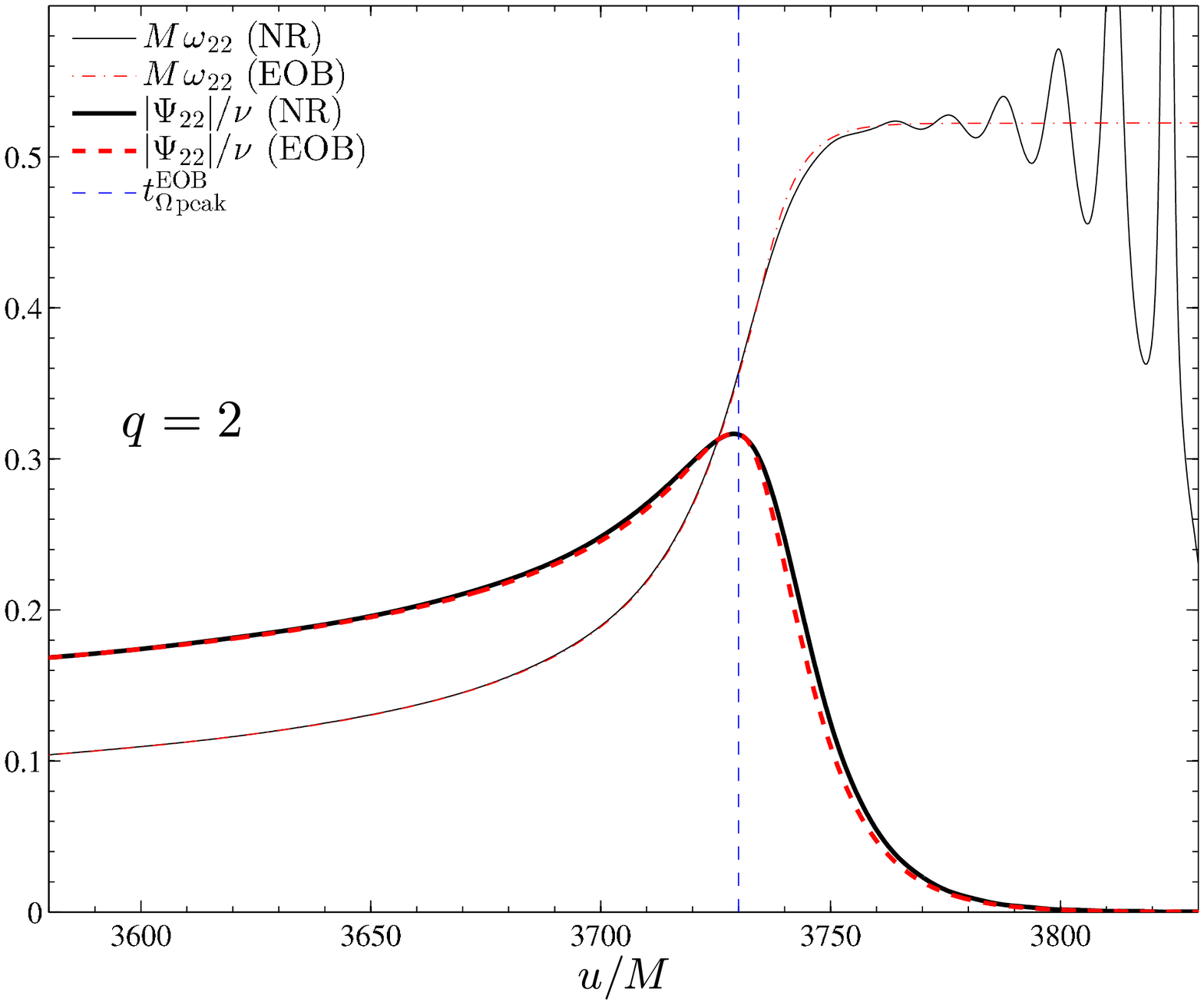}
    \hspace{5mm}
    \includegraphics[width=0.41\textwidth]{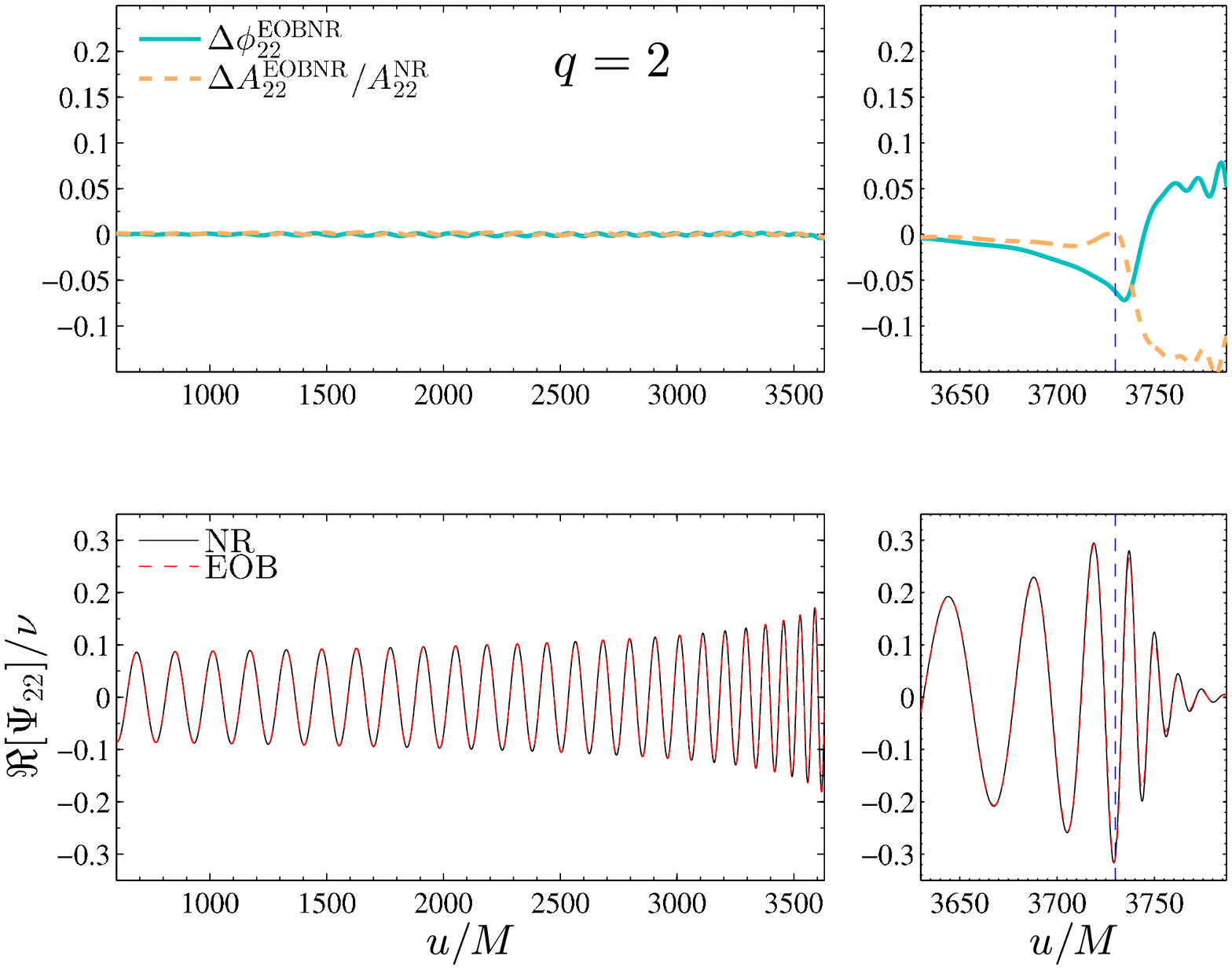}\\
    \caption{ \label{fig:phasing_q1q2} 
      (color online) Comparison between EOB and NR (Zerilli-normalized) waveforms for mass ratios $q=1,\,2$. 
      Left panels: amplitude and frequency. In the right panels, each subplot shows the phase and amplitude 
      differences between the EOB and NR waveform (top) and the real part of $\Psi_{22}$ (bottom). The time axis is 
the NR one: $u= t^{\rm NR}$. The EOB waveform has been time- and phase-shifted so as to minimize the EOB $-$ NR
phase difference for frequencies $M \omega < 0.1$.
The vertical dashed lines 
  mark the $\teobLR$ crossing time.} 
  \end{center}
\end{figure*}

\begin{figure*}[t]
  \begin{center}
    \includegraphics[width=0.41\textwidth]{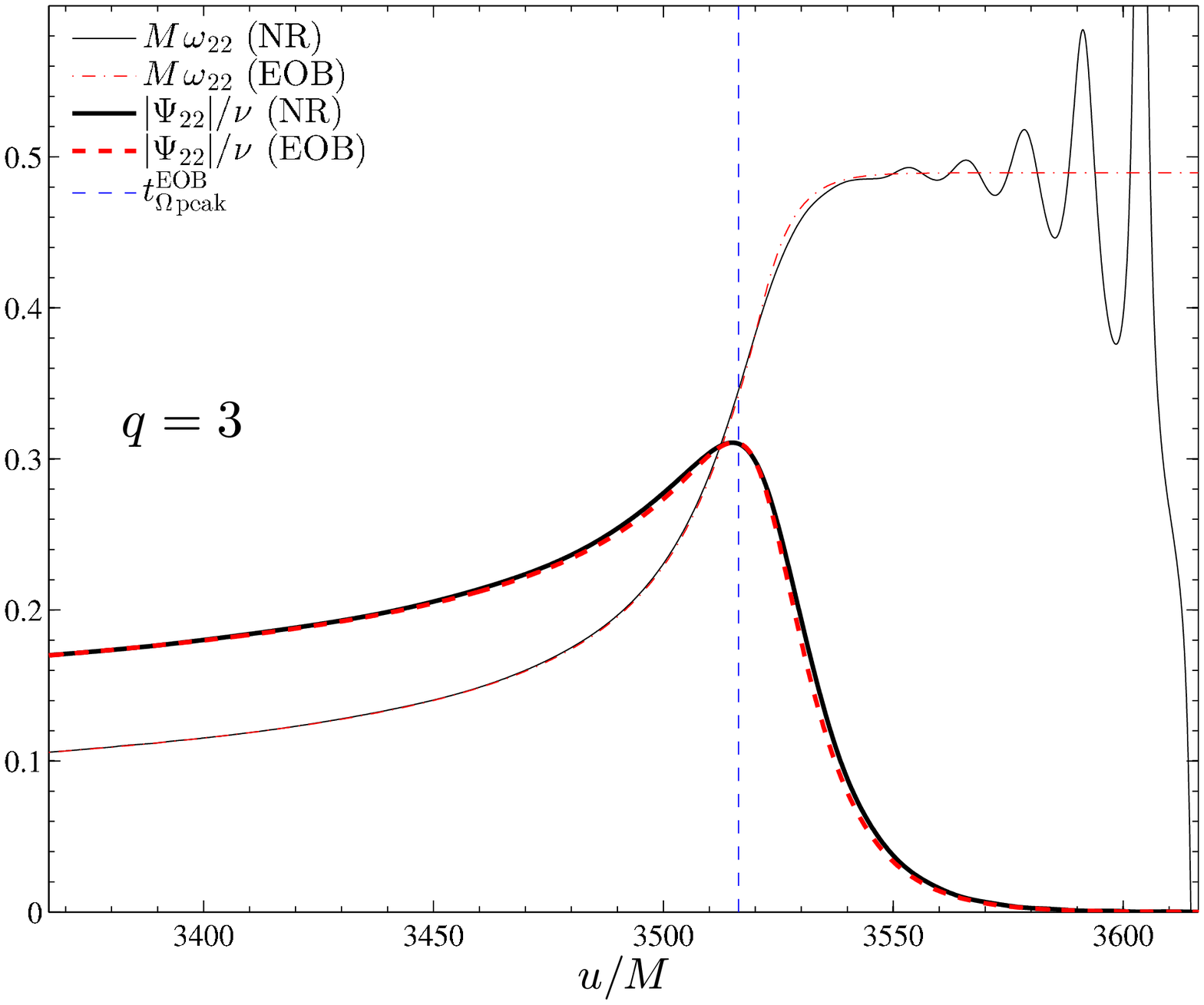}
    \hspace{5mm}
    \includegraphics[width=0.41\textwidth]{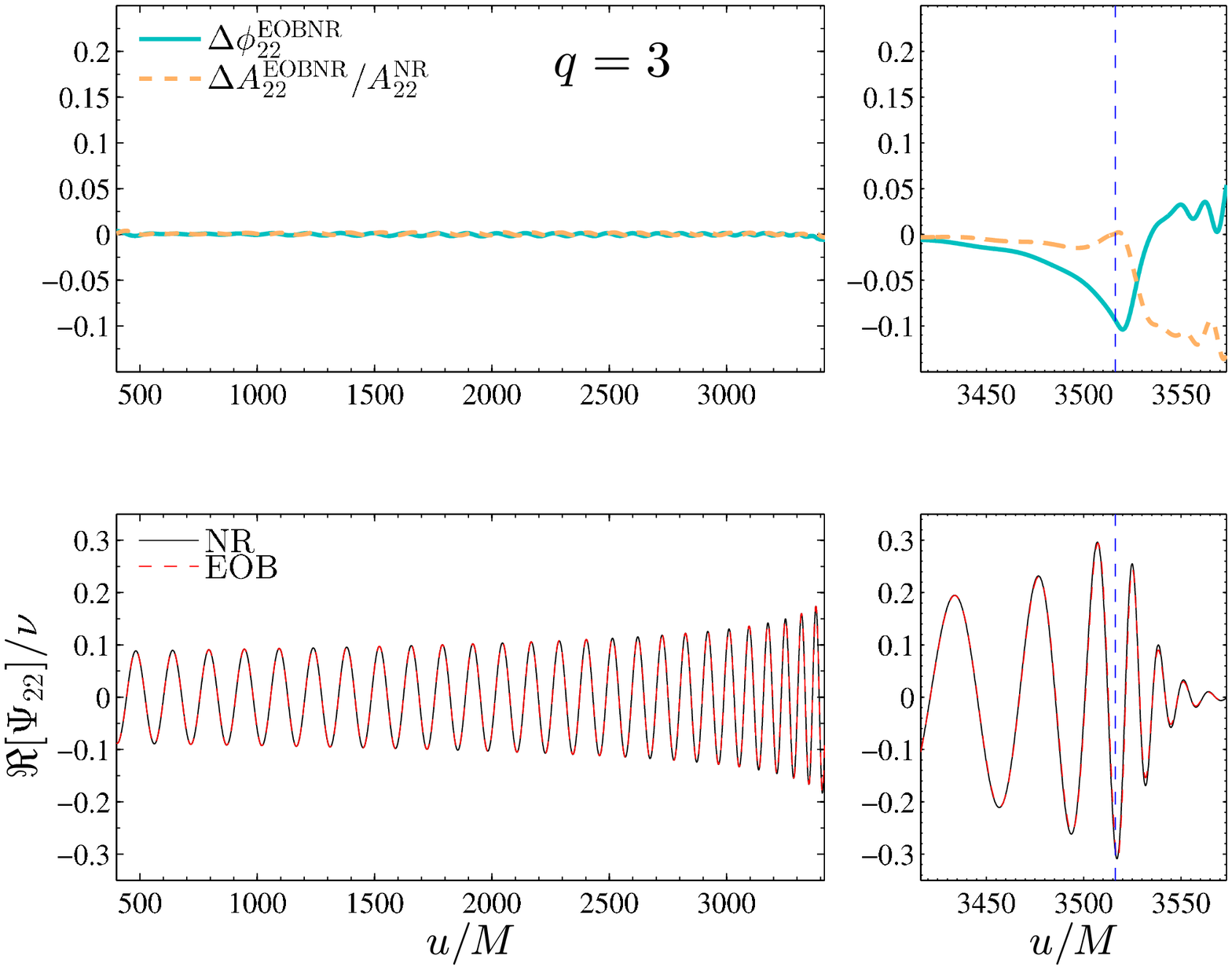}\\
    \vspace{2.0mm}
    \includegraphics[width=0.41\textwidth]{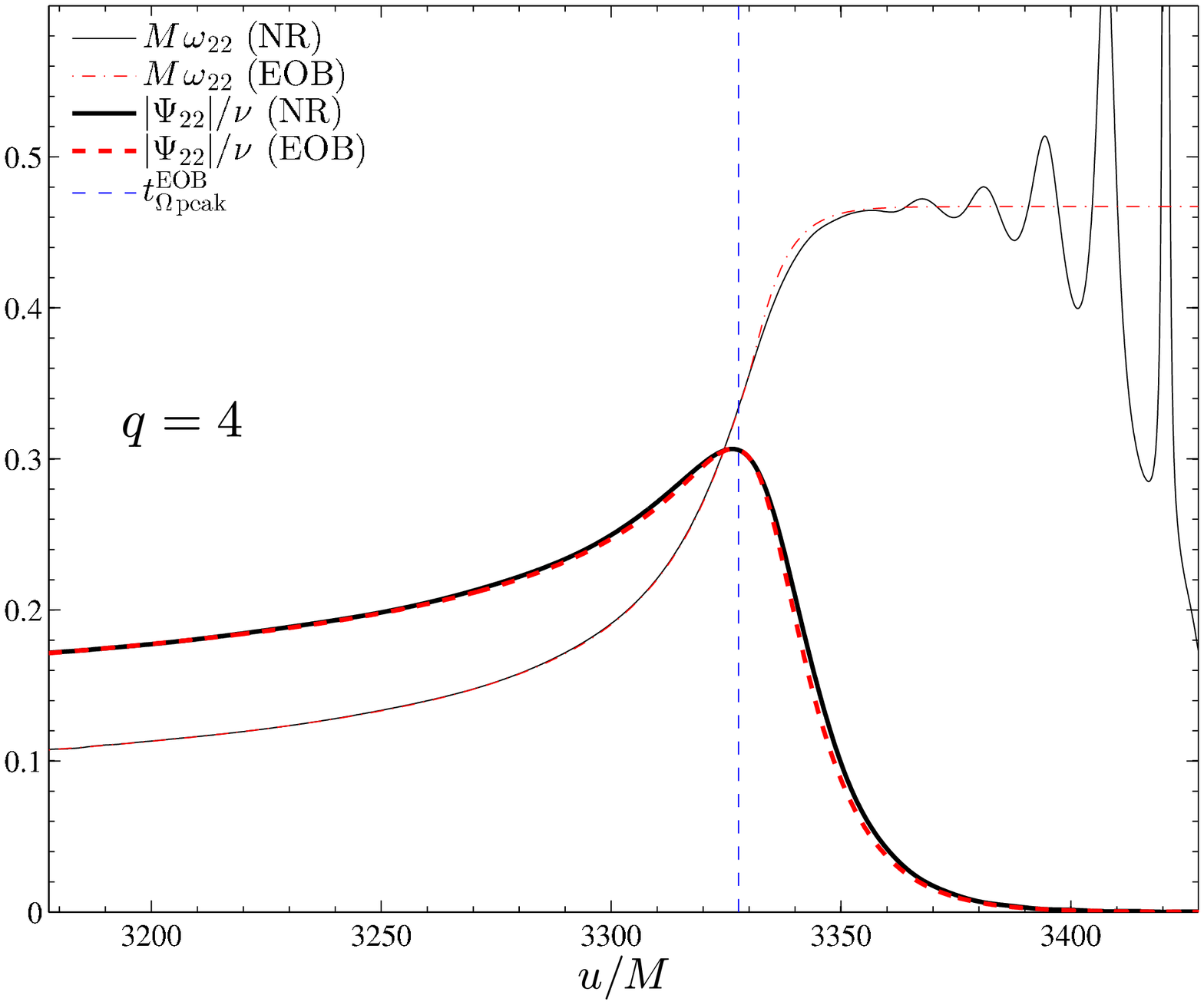}
    \hspace{5mm}
    \includegraphics[width=0.41\textwidth]{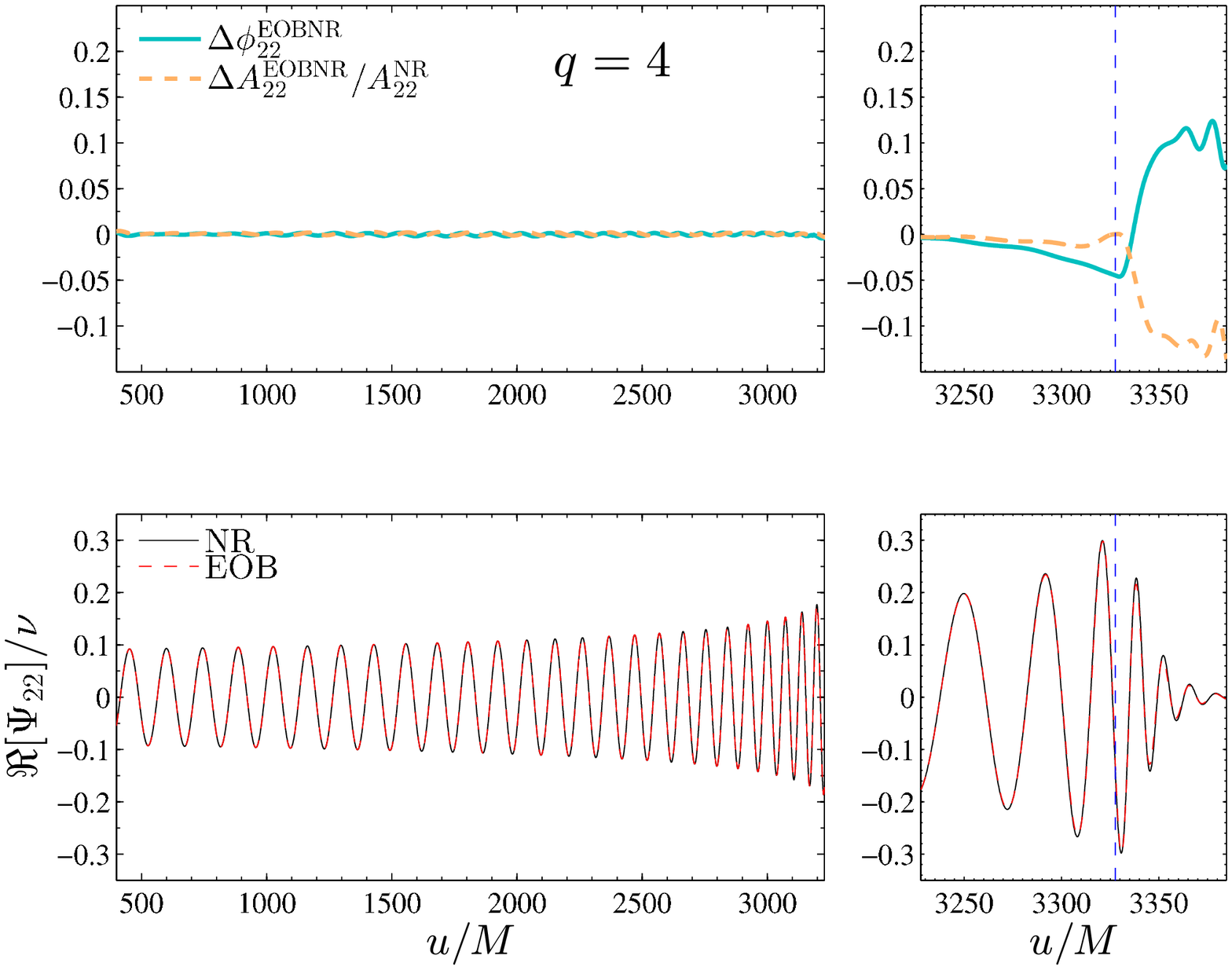}\\  
    \vspace{2.0mm}
    \includegraphics[width=0.41\textwidth]{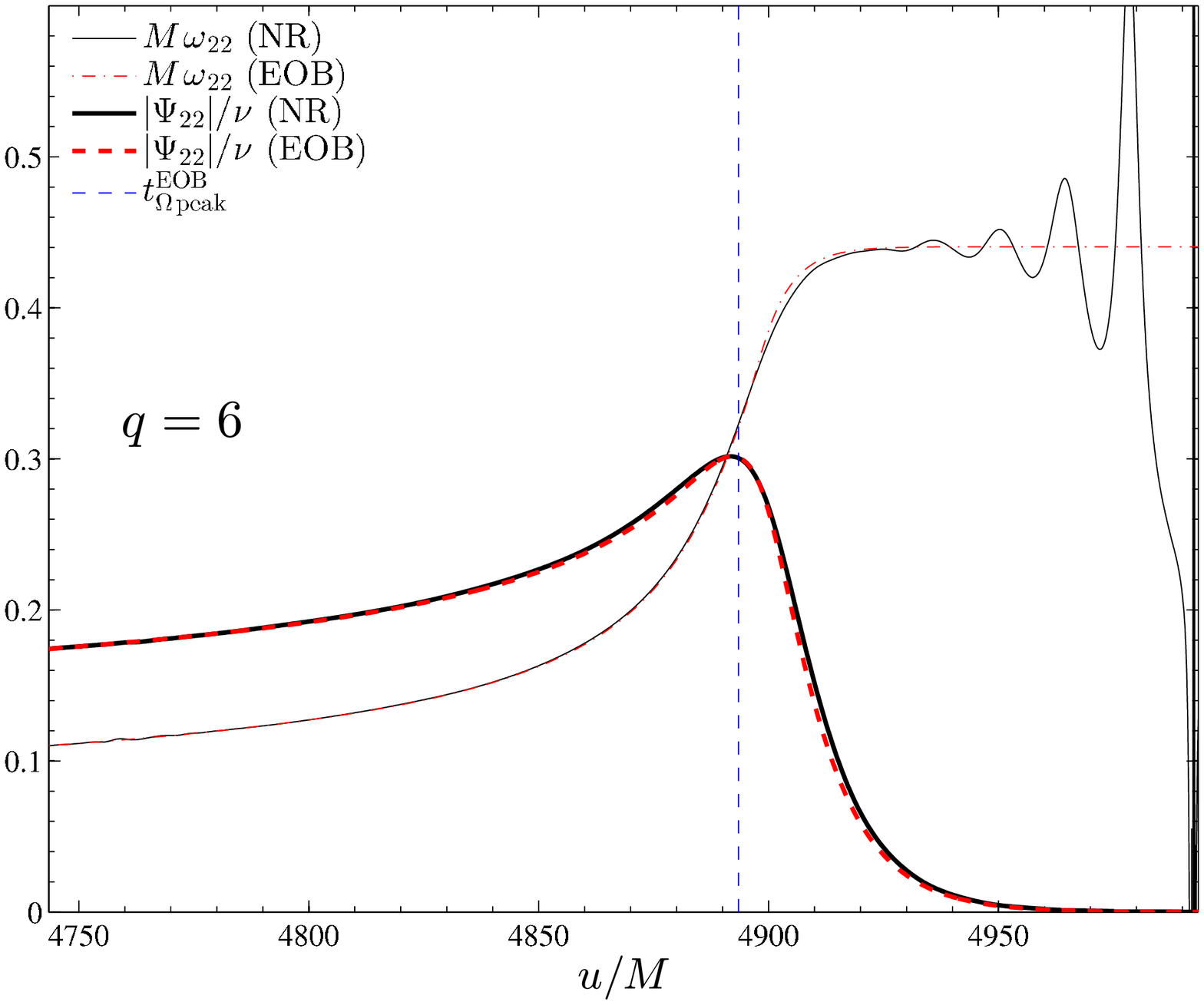}
    \hspace{5mm}
    \includegraphics[width=0.41\textwidth]{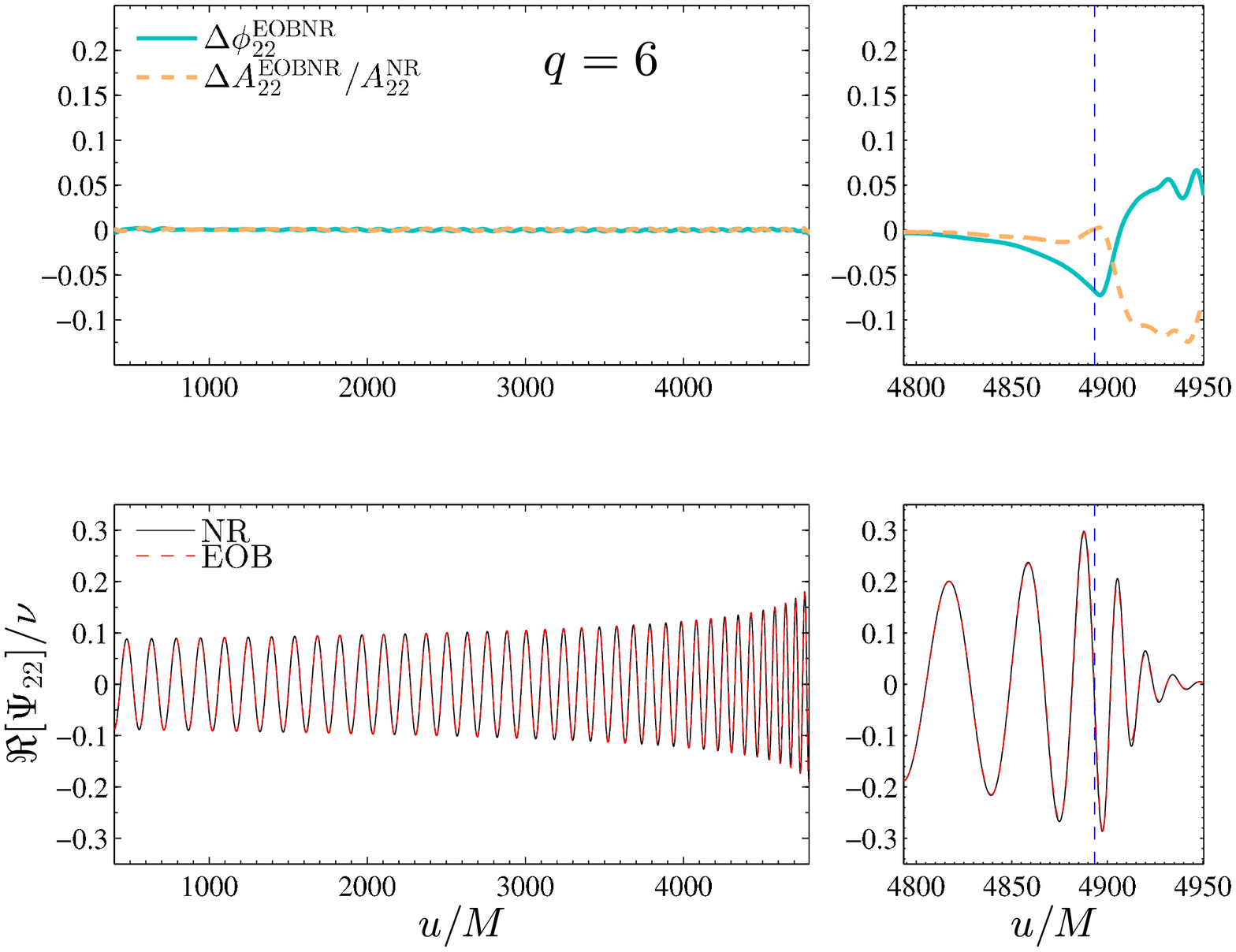}
    \caption{ \label{fig:phasing_q3q4} 
(color online) Comparison between EOB and NR (Zerilli-normalized) waveforms for mass ratios $q=3,\,4,\,6$. 
      Left panels: amplitude and frequency. In the right panels, each subplot shows the phase and amplitude 
      differences between the EOB and NR waveform (top) and the real part of $\Psi_{22}$ (bottom). The vertical dashed lines 
  mark the $\teobLR$ crossing time.}  
  \end{center}
\end{figure*}

\subsection{Determining  $\tnrLR(\nu)$}

We started by applying this iterative procedure to the equal-mass case 
$q=1$ (i.e. $\nu=0.25$). After trial and error,
we concluded that, for $q=1$,  the coefficient $f(\nu)$ in Eq.~\eqref{eq:NR_lr_position}
could be taken to have the value $f(0.25)=1/6$. In other words, when $q=1$,  $\tnrLR$
can be taken to be rather close to the peak of the $A_{22}$ modulus, 
as was indeed assumed in all previous EOB works. By contrast, when considering 
larger mass ratios, we found more and more advantageous to increase the value of 
$f(\nu)$, up to values of order of the test-mass value discussed above, $f(0) = 2/3$, 
for large mass ratios. Then, as a simplifying choice, we decided to assume for the $\nu$ 
dependence of $f(\nu)$  a simple  {\it linear behavior} between the two extreme values for $\nu=0$ and $\nu=0.25$,
in the form
\be
f(\nu) = f(0.25) + \left(f(0)-f(0.25)\right)(1-4\nu);
\ee
which yields, when using  $f(0.25)=1/6$ and $f(0)=2/3$, the explicit expression
\be
\label{eq:f_nu_choice}
f(\nu) = \frac{2}{3}-2\nu \, .
\ee
Having so chosen $\tnrLR(\nu)$,  we measure, for each $(\ell,m)$, on the NR mutipolar 
waveform the vector $(A_\lm^{\rm NR},\dot{A}_\lm^{\rm NR},\ddot{A}_\lm^{\rm NR},\omega_\lm^{\rm NR},\dot{\omega}_\lm^{\rm NR},\ddot{\omega}_\lm^{\rm NR})$ 
at $\tnrLR(\nu)$. Then, for any value of $a^c_6$, we first compute the EOB dynamics, 
then we solve the linear system given by Eqs.~\eqref{comp_1}-\eqref{comp_6} 
to obtain the NQC parameters $(a_i^\lm,b_i^\lm)$; and finally we iterate the procedure 
until  $(a_i^\lm,b_i^\lm)$ converge at the fourth digit.

\subsection{Determining $a_6^c(\nu)$}

At this stage, the only freedom left in the model is the value of $a_6^c(\nu)$. 
Let us now explain how we investigated the phase difference $\Delta \phi^{\rm EOB NR}(a_6^c(\nu))$
and used it to determine  $a_6^c(\nu)$.  Actually, we used a two-pronged approach towards studying  
$\Delta \phi^{\rm EOB NR}$.  We first studied the $Q_\omega(\omega)$ function defined by the NR data, 
and compared it to the EOB-predicted one. Then, in a second step, we considered the time-domain 
phase difference $\Delta \phi^{\rm EOB NR}(t)$.

Let us start by explaining how we used the
$Q_\omega(\omega)$ diagnostics to constrain the possible good values of $a_6^c(\nu)$.
Since, as we explained above, we could extract from NR data a rather accurate estimate of $Q_\omega^{\rm NR}(\omega)$,
we compared it to the value   $Q_\omega^{\rm EOB}(\omega; a_6^c(\nu))$ predicted, for each value of  $a_6^c(\nu)$,
by EOB theory. Such a comparison (in the $q=2$ case) is illustrated in 
Fig.~\ref{fig:Delta_Qomg_q2}. The top panel of this figure shows the EOB$-$PN and NR$-$PN 
differences $\Delta Q_\omega^X\equiv Q_\omega^{\rm X}-Q_\omega^{\rm 3.5PN}$, where
$X$ labels either EOB (for the three indicated values of $a_c^6$) or NR, 
and $Q_\omega^{\rm 3.5PN}$ is the 3.5PN-accurate, Taylor-expanded 
expression given by Eq.~\eqref{eq:Qomg_PN}.
Note first that the black solid line, corresponding to NR$-$PN, shows that the current best PN
knowledge of the intrinsic phasing function, $Q_\omega^{3.5 \rm PN}(\omega)$, differs from the NR result by
a large amount, reaching  $Q_\omega^{3.5 \rm PN}(\omega)- Q_\omega^{ \rm NR}(\omega) \simeq - 18$ at $M\omega_2=0.29$,
which is close to  merger. The corresponding integrated dephasing between PN and NR,
\be 
\Delta\phi^{\rm PN NR}\equiv \int_{\omega_1}^{\omega_2} d \ln \omega (Q_\omega^{3.5 \rm PN}(\omega)-  Q_\omega^{\rm NR}(\omega) )\ ,
\ee
accumulated from  $M\omega_1=0.07$ to $M \omega_2=0.29 $, is found to be equal to $- 11.72$ radians. 

By contrast to the NR $-$ PN, or EOB $-$ PN
differences displayed in the top panel of Fig.~\ref{fig:Delta_Qomg_q2}, its bottom panel displays the 
much smaller EOB$-$NR difference 
$\Delta Q_\omega\equiv Q_\omega^{\rm EOB}(\omega; a_6^c(\nu))-  Q_\omega^{\rm NR}(\omega)$ 
for five different values of $a_6^c$. In addition, the shaded region represents 
the NR$-$NR difference $\Delta Q_\omega=Q_\omega^{\rm NR, N=5}-Q_\omega^{\rm NR, N=4}$, where 
where $N=5$ (respectively $N=4$) labels the numerical waveform with 
the highest (resp. medium) resolution~\cite{Buchman:2012dw}.
The visual comparisons displayed in Fig.~\ref{fig:Delta_Qomg_q2} are made quantitative
in Table~\ref{tab:q2_find_a6}, which lists corresponding values of the EOB$-$NR 
phase difference over the frequency interval $M(\omega_1,\omega_2)=(0.07,0.29)$ obtained from 
the integral
\be 
\label{eq:Dphi_from_DQomg}
\Delta\phi\equiv \int_{\omega_1}^{\omega_2} d \ln \omega (Q_\omega^{\rm EOB}(\omega; a_6^c(\nu))-  Q_\omega^{\rm NR}(\omega)) \ .
\ee
Note that $M\omega_2=0.29$ approximately corresponds to the merger.
These phase differences indicate that a good range of values of $a_6^c(2/9)$ is roughly
between $-90$ and $-100$. Within such a range, $\Delta\phi$ remains of the order of the
NR phasing uncertainty as estimated in Ref.~\cite{Buchman:2012dw,Pan:2011gk} by comparing
the two resolutions $N=4$ and $N=5$. Note that the small phase differences corresponding
to  $-100\leq a_6^c(2/9)\leq-90$ result from a cancellation between positive and negative
contributions to the above integral. However, a look at Fig~\ref{fig:Delta_Qomg_q2} shows
that within this range of $a_6^c$ the nonzero values of $\Delta Q_\omega$ remain of 
the order $\pm 0.05$ 
for most of the integration region. Such range of values of $\Delta Q_\omega$ is comparable
to the numerical uncertainty on $Q_\omega$ (at least) during the inspiral,
as illustrated by the shaded region in the figure. Note indeed that the frequency $M\omega=0.1$
is reached only $150M$ before merger (cf. bottom left panel of Fig.~\ref{fig:phasing_q1q2}). 
Note also that the frequency interval $0.2\leq M\omega_{22}\leq 0.3$ (where the top panel of 
Fig.~\ref{fig:Delta_Qomg_q2} shows visible differences, made quantitative in the bottom panel) 
{\it only}  corresponds to the last $25M$ before merger. [The GW frequency $0.2$ approximately 
corresponds to the adiabatic LSO crossing, i.e. the end of the quasi-adiabatic inspiral].

\begin{table}[t]
  \caption{\label{tab:q2_find_a6} Mass ratio $q=2$: phase difference $\Delta\phi=\phi^{\rm EOB}-\phi^{\rm NR}$ 
    accumulated between frequencies $\omega_1=0.07$ and $\omega_2=0.29$ versus $a_6^c$ as obtained using Eq.~\eqref{eq:Dphi_from_DQomg}.} 
  \begin{center}
    \begin{ruledtabular}
      \begin{tabular}{cccccc}
        &$a_6^c$   &   $\Delta\phi$ [rad] \\
        \hline 
        &-80        & +0.0810\\
        &-90        & -0.0010 \\
        &-100       & -0.0909 \\
        &-110       & -0.1942 \\
   \end{tabular}
  \end{ruledtabular}
\end{center}
\end{table}

\begin{table}[t]
  \caption{\label{tab:best_a6c} Best values of $a_6^c$ selected according to the behavior 
  of the $\Delta\phi=\phi^{\rm EOB}-\phi^{\rm NR}$ phase difference around merger time.} 
  \begin{center}
    \begin{ruledtabular}
      \begin{tabular}{cccccc}
       $q$ & $\nu$  &$a_6^c$ (flat)   &   $a_6^c$ (effective) \\
        \hline 
       1  & 0.25        &-106      & -103\\
       2  & $0.\bar{2}$ &-99       & -96 \\
       3  & 0.1875      &-82       & -79 \\
       4  & 0.1600      &-67       & -63 \\
       6  & 0.1224      &-47       & -45 \\      
   \end{tabular}
  \end{ruledtabular}
\end{center}
\end{table}

This analysis based on the $Q_\omega$ diagnostics selects, for each value
of the mass ratio ($q=1, 2, 3, 4 , 6$), a range of good values of $a_6^c(\nu)$, 
which then needs to be confirmed and refined by directly comparing the time-domain 
phase evolution of the EOB waveform to the NR one.
We have done such an analysis by considering, for each value of $a_6^c(\nu)$ 
within the above range, the phase evolution from the beginning 
of the simulation up to merger, and also after merger, during ringdown. 
The comparison up to merger only depends on the choices of  $\tnrLR(\nu)$  and 
$a_6^c(\nu)$, while the comparison during the subsequent ringdown also depends 
on the choices made in attaching QNMs to the NQC-corrected pre-merger signal.
The time-domain phasing comparison allowed us to close up, for each value of $\nu$,
on a more precisely determined value of $a_6^c(\nu)$ (with an uncertainty of order
unity).
Actually, depending on the criterion we put on the quality of the EOB/NR phase
agreement, the resulting best values of $a_6^c(\nu)$ are slightly different.
However, in all the cases we have explored, we found that the good, $\nu$-dependent 
values of  $a_6^c$ were approximately lying along a straight line. 

We choose $a_6^c$ according to the following two criterions: on the one hand, we can
require that the time-domain phase difference (after alignment) $\Delta \phi^{\rm EOB NR}(t; a_6^c(\nu))$
remains near zero in as flat a manner as possible {\it up to merger}. In this case,
the price to pay for this is that the  subsequent, somewhat coarse QNM attachment 
defined by the current EOB prescriptions,  will cause, after merger and during ringdown, 
the EOB-NR phase difference  $\Delta \phi^{\rm EOB NR}(t; a_6^{c \, \rm flat}(\nu))$
to jump to positive values of order $\sim +0.15$ rad (more about this below).
On the other hand, one can also look for a more ``effective'' description of the phasing
where we allow  $\Delta\phi^{\rm EOBNR}$ to take slightly negative values just before merger, 
but to jump to smaller values $\sim + 0.05$~rad after merger (see more details below).
The so obtained corresponding good values 
of $a_6^c$ are listed in Table~\ref{tab:best_a6c}.
 The bottom panel of  Fig.~\ref{fig:dphi_flat_vs_effective} 
plots these values versus $\nu$. One sees that, for both the effective and 
flat cases, they approximately lie along a straight line. However, as evidenced by these plots,
a linear fit to   $a_6^c(\nu)$ does not give an accurate representation 
of the points when the $\nu=0.25$ value is taken into account.  Before discussing
a way to fit such a behavior, let us note that the top panel of 
Fig.~\ref{fig:dphi_flat_vs_effective} displays, for $q=1$, the phase
differences for the ``flat" and ``effective" values of $a_6^c(0.25)$.
The same behavior, with very similar 
phase differences, is found for all other mass ratios.   

Let us come back to the issue of constructing an analytical fit for the
behavior of the functions  $a_6^c(\nu)$ exhibited in the bottom panel of 
 Fig.~\ref{fig:dphi_flat_vs_effective}.  We checked that the use of a global linear
 fit for  the values of $a_6^c(\nu)$ 
would give unacceptably large phase differences 
( $>0.1$ rad) accumulated up to merger. This suggests the need of using a fitting
function which deviates from a linear function of $\nu$ only in a rather limited
interval  $0.\bar{2}<\nu\leq0.25$. There are many ways to construct such fits.
Here, as a first attempt (to be possibly improved in future work), we have used
the following, factorized, mostly-linear, functional form
\be
\label{a6cnu_fitting}
a_6^c(\nu) = \left[a + b(1-4\nu)\right]\tilde{s}(c;\;\nu),
\ee
where $\tilde{s}$ denotes a localized (when the parameter $c$ is much smaller than one) 
correction to the linear behavior parametrized by $a$ and $b$:
\be
\tilde{s}(c;\;\nu) \equiv \left(1+\dfrac{c}{(0.26 - \nu)^2}\right)^{1/2}.
\ee
We have determined sufficiently accurate values of the parameters $(a,b,c)$ by fitting the
the values of $a_6^c$ listed in Table~\ref{tab:best_a6c} in two steps. [For simplicity,
we fixed the location of the pole in  the function $\tilde{s}^2(\nu)$ to the 
fiducial value $\nu=0.26$.]
 First  $(a,b)$ were determined by fitting  only the $q=(2,3,4,6)$ data
 in  Table~\ref{tab:best_a6c}  to a straight line. 
The raw data were then divided by the outcome of the fit and the resulting ratios were further
fitted against the factor of Eq.~\eqref{a6cnu_fitting} so as to determine $c$. Applying
this fitting procedure, we find $(a,b,c)^{\rm flat}=(-114.006, 130.774,-1.352\times 10^{-5})$
for the flat choices of $a_6^c$ and $(a,b,c)^{\rm effective}=(-110.467, 129.022,-1.468\times 10^{-5})$ 
for the effective choices of $a_6^c$. Rounding up these numbers, we summarize our search 
of a ``flat" $a_6^c(\nu)$ by the following analytical expression
\begin{align}
\label{eq:a6c_nu_flat}
a_6^{c\,{\rm flat}}(\nu) &= \left[-114 + 131(1-4\nu)\right]\tilde{s}(-1.4\times 10^{-5};\,\nu).
\end{align}
For the effective description of the phasing we found instead
\begin{align}
\label{eq:a6c_nu}
a_6^c(\nu) &= \left[-110.5 + 129(1-4\nu)\right]\tilde{s}(-1.5\times 10^{-5};\,\nu).
\end{align}
This is one of the central results of our work, and one of the most important
pieces in the NR-completion of our EOB model. 

In conclusion, we propose to define the NR completion of our EOB model by adopting the 
analytical expressions~\eqref{eq:f_nu_choice} and~\eqref{eq:a6c_nu} for defining, respectively, 
$\tnrLR(\nu)$ and $a_6^c(\nu)$. In addition, we found that the following QNM-attachment 
choices define a reasonably accurate ringdown completion of the EOB waveform: $N=5$ 
QNM modes, and $\Delta^{\rm comb}= 0.7 M$. In the following, we shall illustrate the 
comparison of the EOB multipolar waveform defined by these choices to the corresponding 
NR multipolar waveform.

\begin{figure*}[t]
  \begin{center}   
    \includegraphics[width=0.45\textwidth]{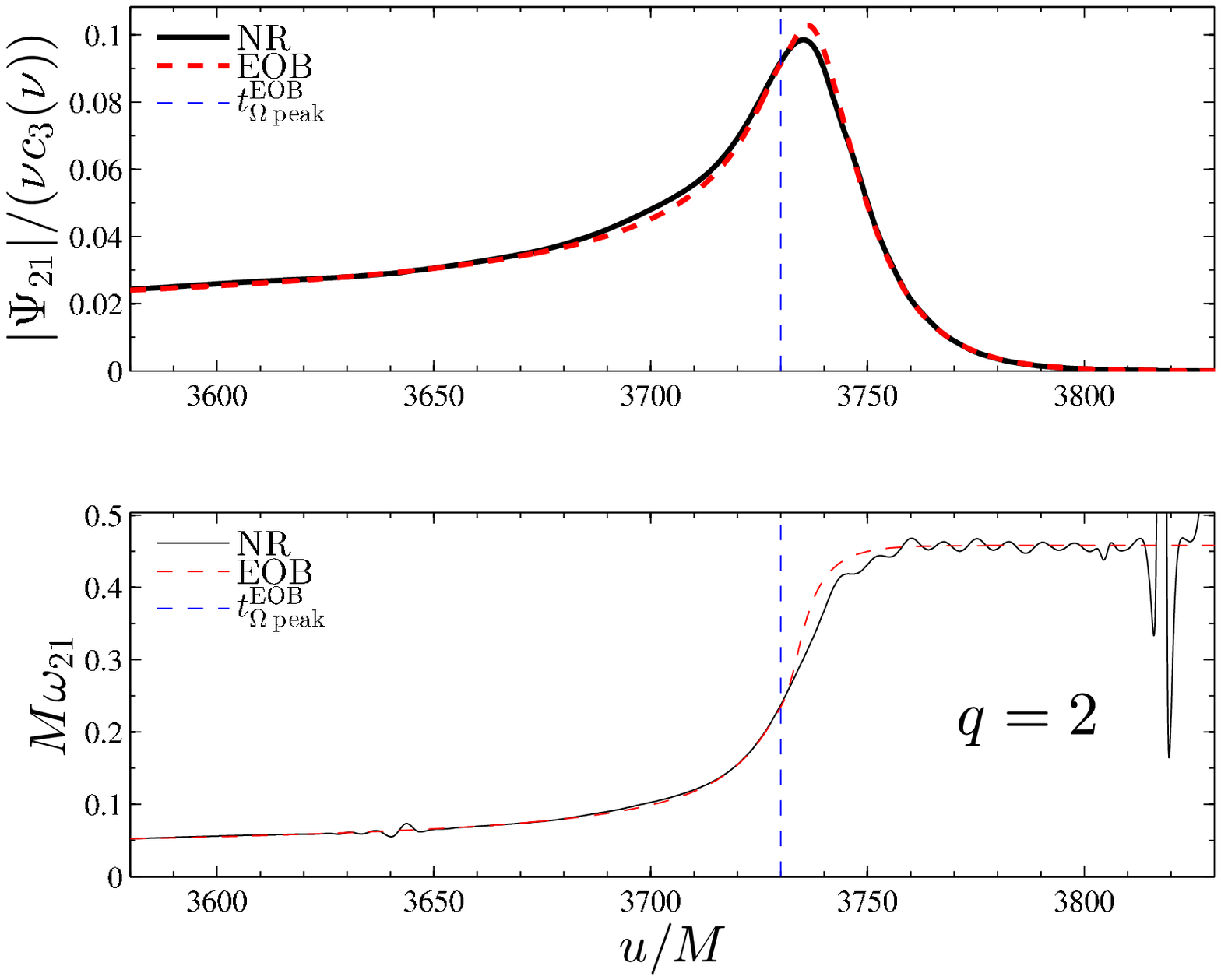}
    \hspace{5mm}
    \includegraphics[width=0.45\textwidth]{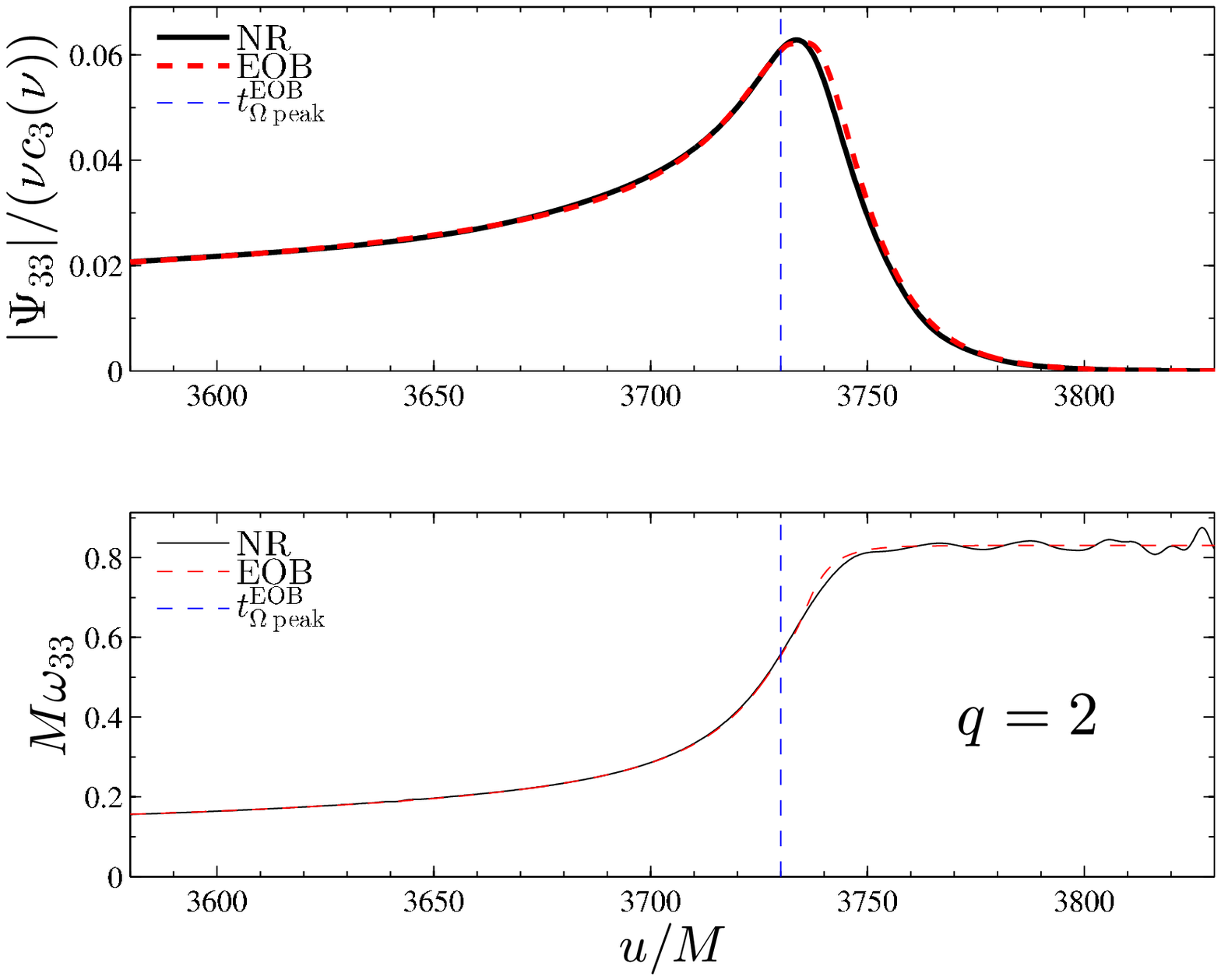} \\
    \vspace{5mm}
    \includegraphics[width=0.45\textwidth]{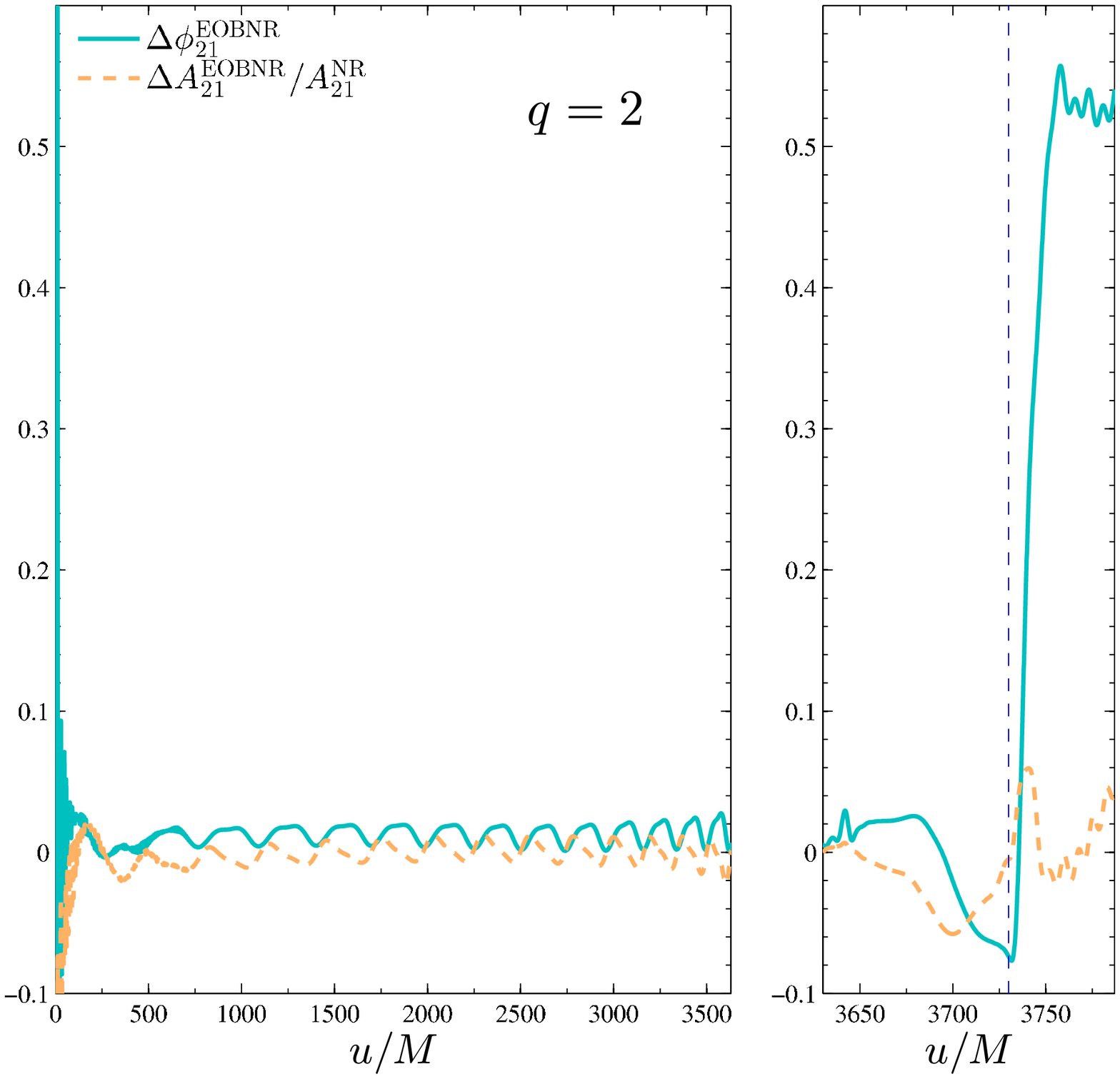} 
    \hspace{5mm}
    \includegraphics[width=0.45\textwidth]{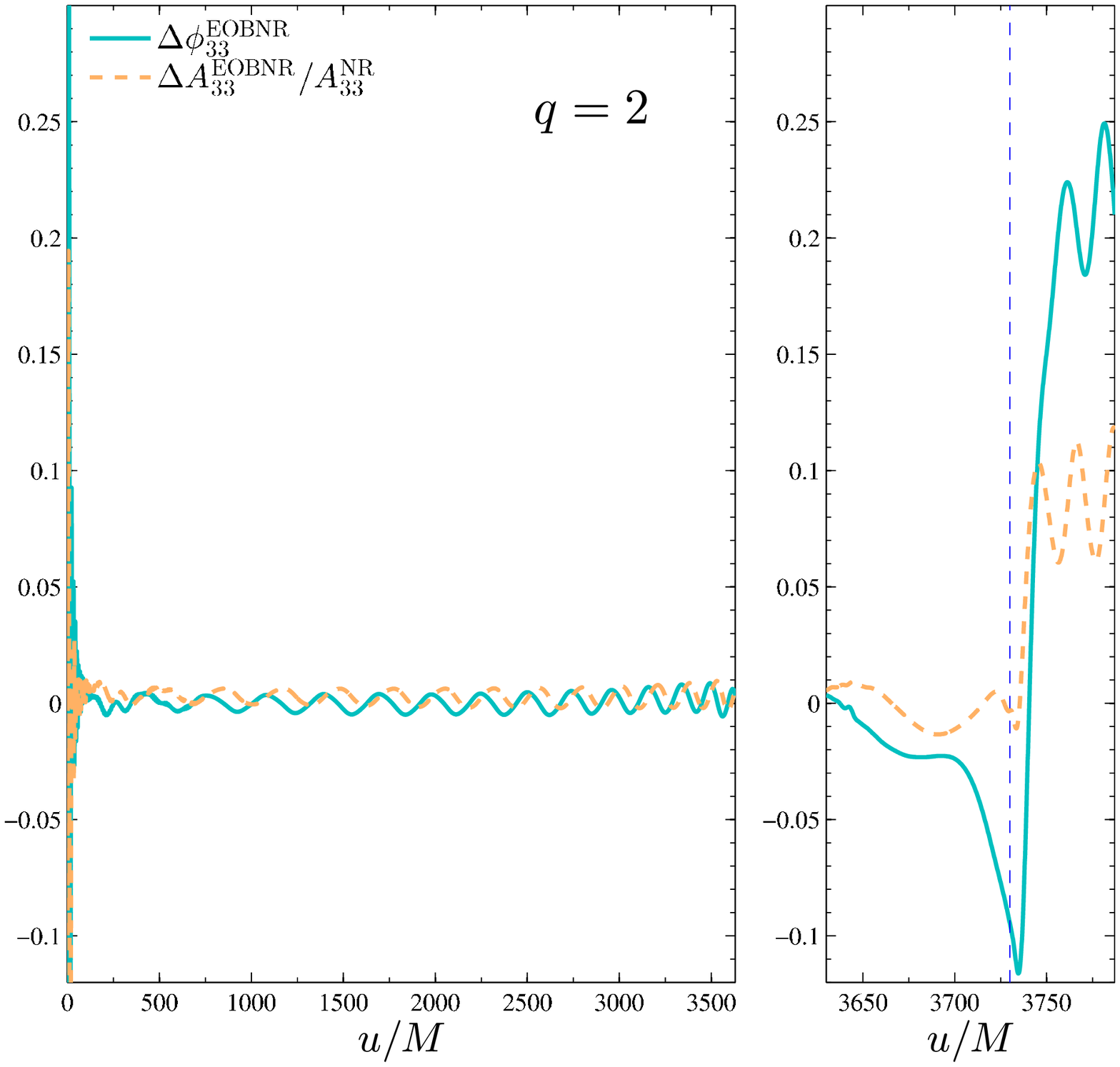} 
    \caption{ \label{fig:phasing_higher_q2} (color online) Subdominant multipoles, $\ell=2$, $m=1$ (left panels) 
      and $\ell=m=3$ (right panels) Comparison between EOB model and NR (Zerilli-normalized) waveform for mass ratio $q=2$. 
      Top:  amplitudes and frequencies. Bottom panels: amplitude and phase differences. The vertical dashed lines 
  mark the $\teobLR$ crossing time.}
  \end{center}
\end{figure*}

\begin{figure*}[t]
  \begin{center}   
    \includegraphics[width=0.45\textwidth]{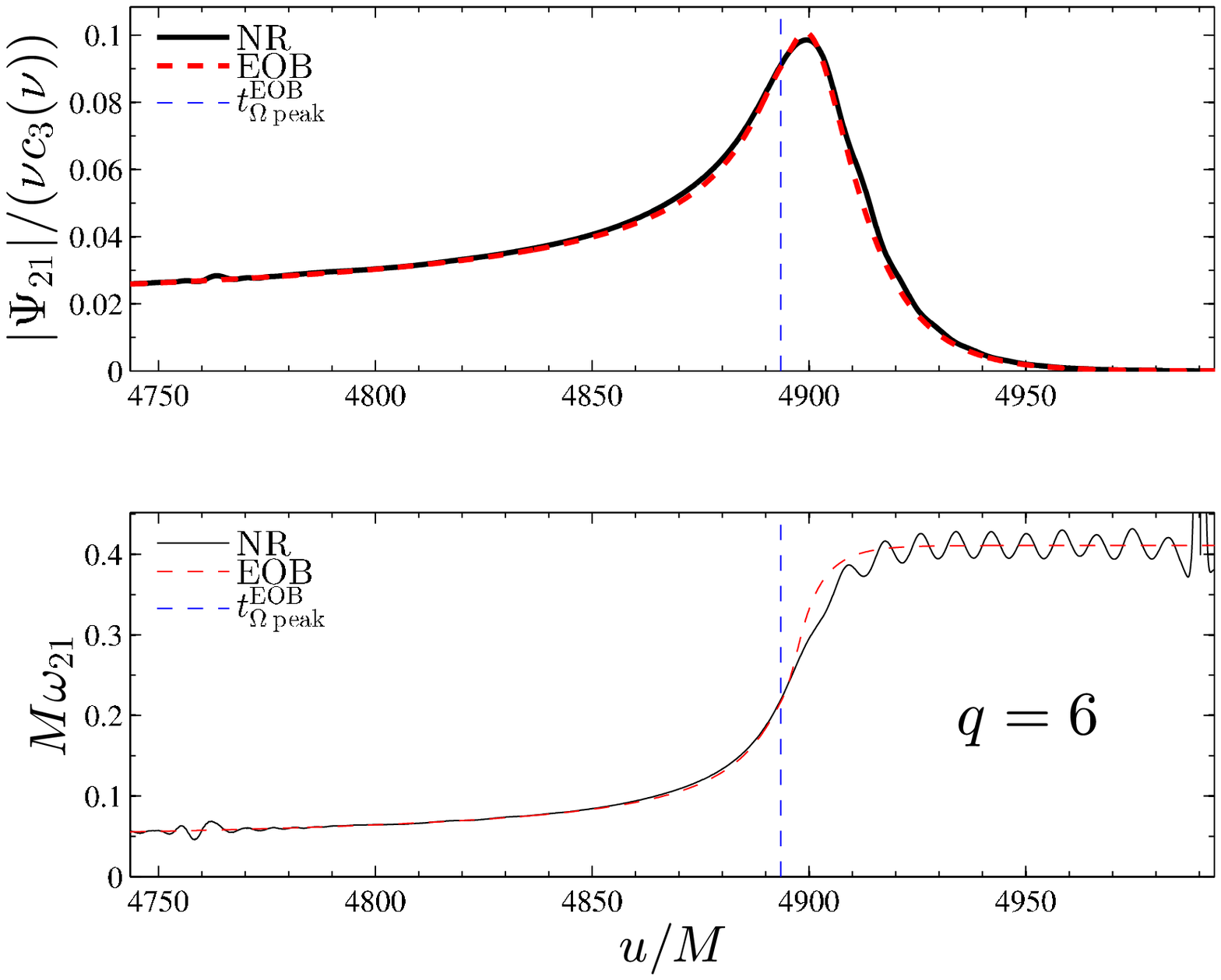}
    \hspace{5mm}
    \includegraphics[width=0.45\textwidth]{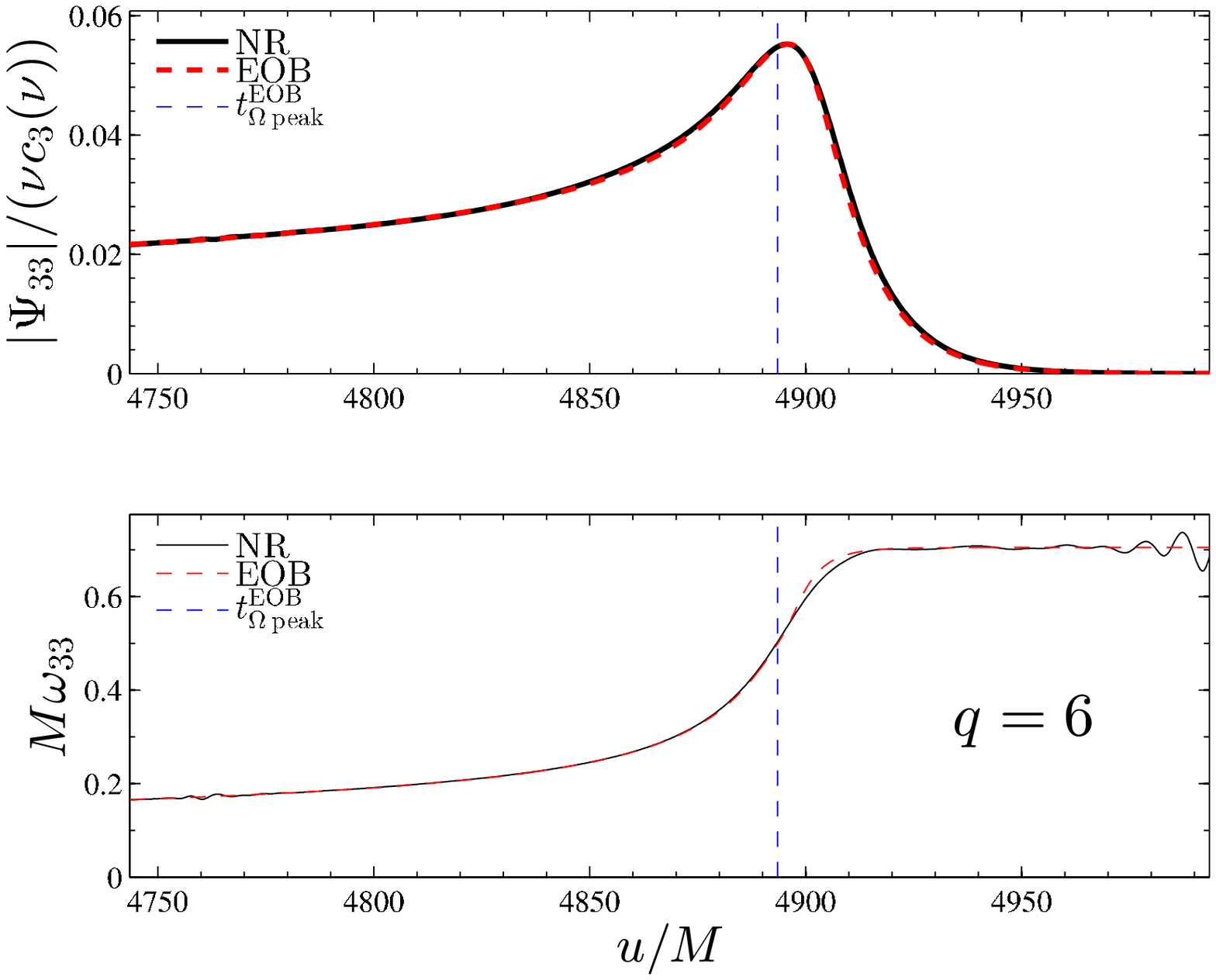} \\
    \vspace{5mm}
    \includegraphics[width=0.45\textwidth]{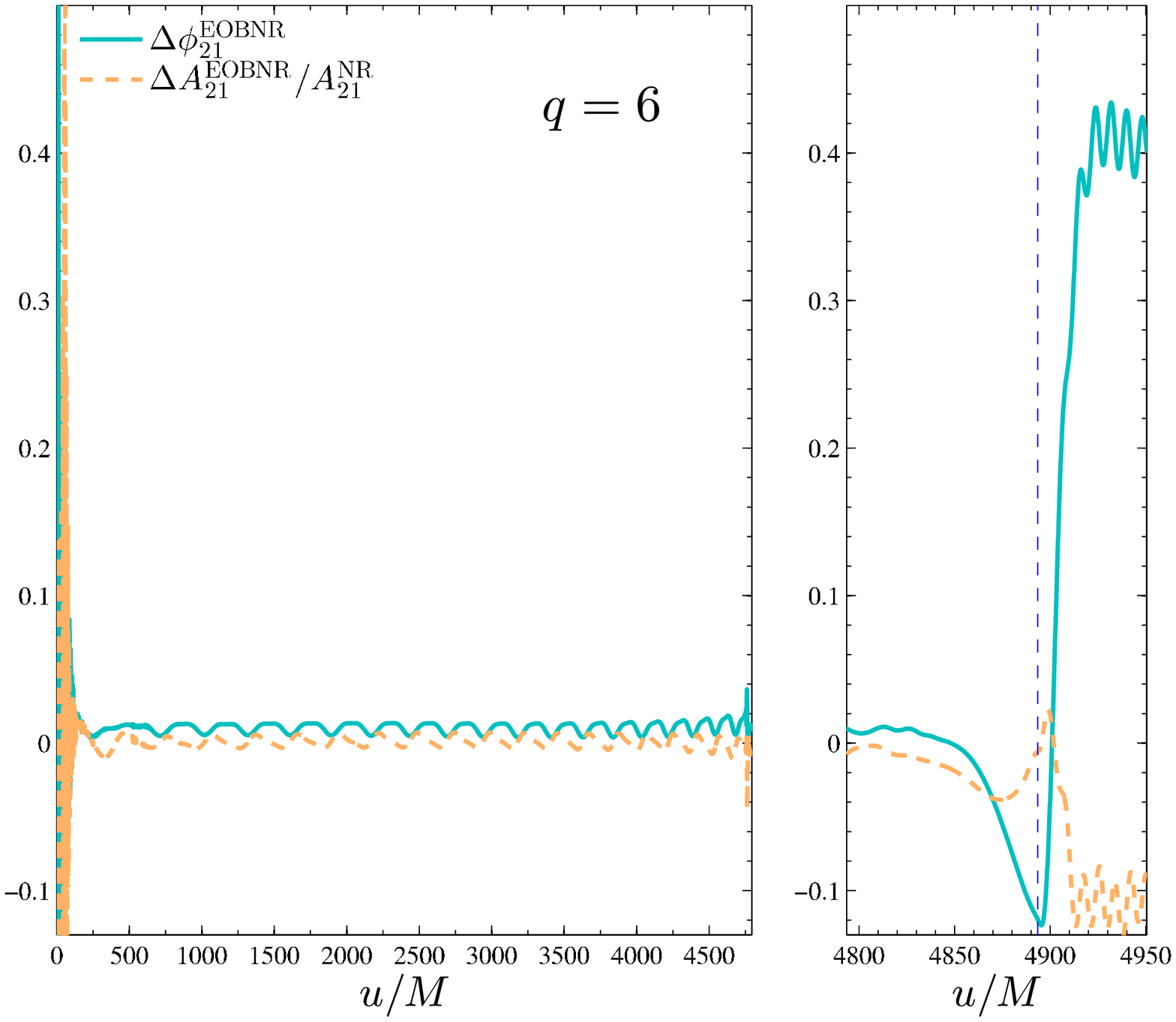} 
    \hspace{5mm}
    \includegraphics[width=0.45\textwidth]{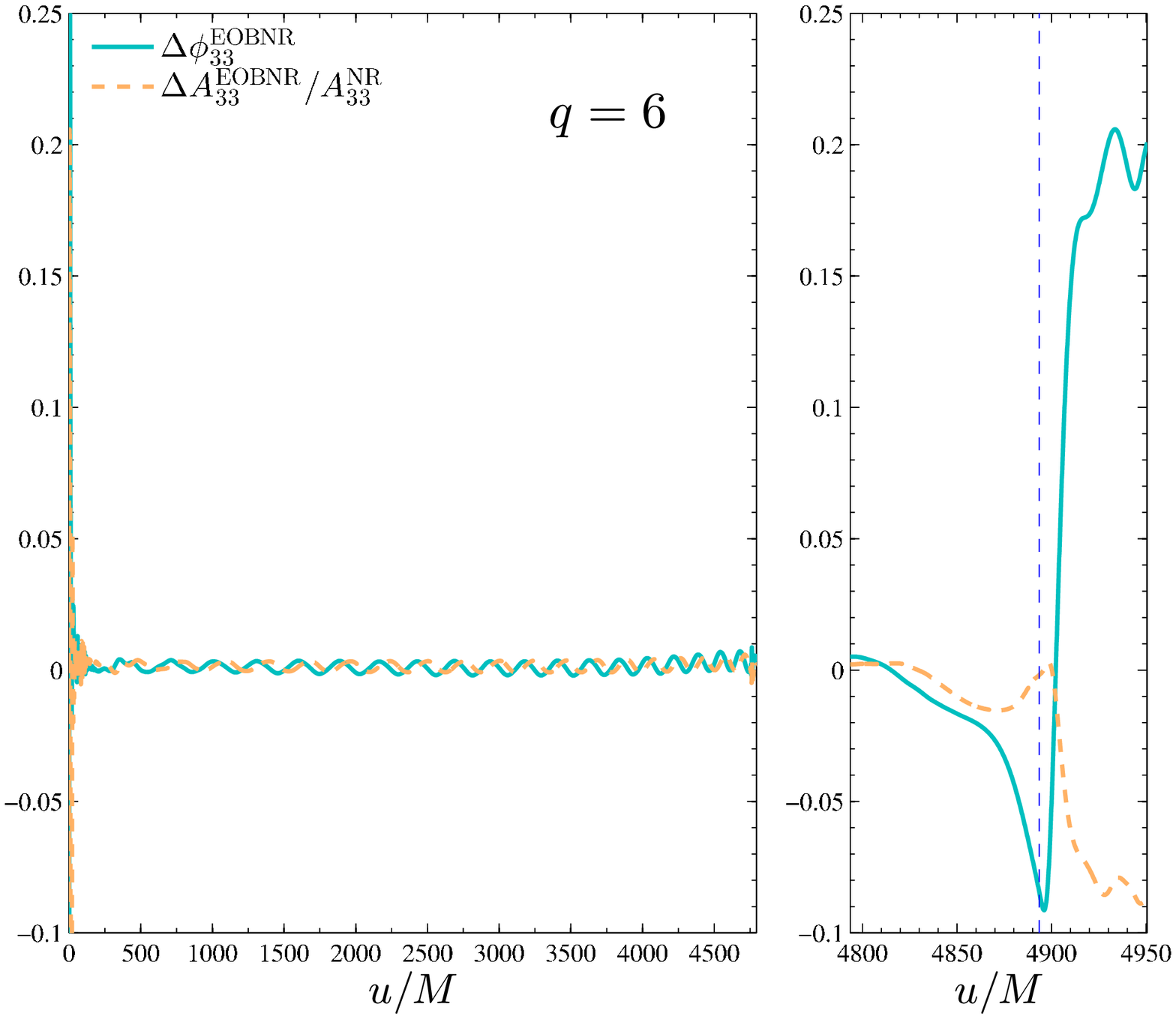} 
 \caption{ \label{fig:phasing_higher_q6} (color online) Subdominant multipoles, $\ell=2$, $m=1$ (left panels) 
      and $\ell=m=3$ (right panels) Comparison between EOB model and NR (Zerilli-normalized) waveform for mass ratio $q=6$. 
      Top: amplitudes and frequencies. Bottom panels: amplitude and phase differences. The vertical dashed lines 
  mark the $\teobLR$ crossing time.}
  \end{center}
\end{figure*}

\begin{figure}[t]
  \begin{center}   
    \includegraphics[width=0.45\textwidth]{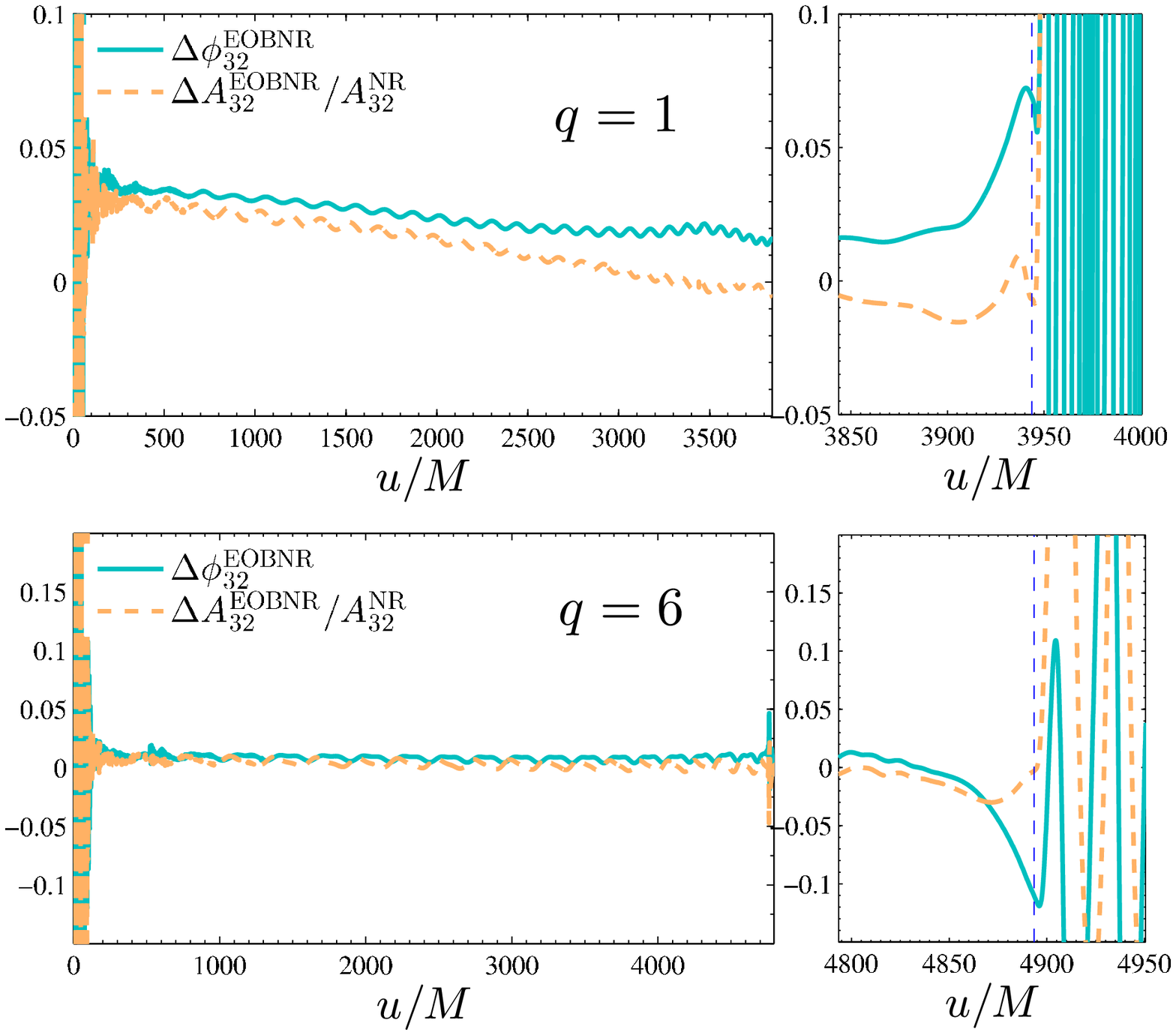}
 \caption{ \label{fig:multipole_l3m2} (color online) Subdominant multipole $\ell=3$, $m=2$: phase 
  difference for $q=1$ (top panel) and $q=6$ (bottom panel). The vertical dashed lines 
  mark the $\teobLR$ crossing time.}
  \end{center}
\end{figure}

\subsection{Values of the NQC parameters  $(a_i^\lm,b_i^\lm)$}

Before doing so, let us recall that, for each mass ratio, we must determine (by iteration) the NQC parameters $(a_i,b_i)$ 
defined by the above choices (using given NR data). 
In Table~\ref{tab:nqc_ab}  we list, for the mass ratios $q=(1,2,3,4,6)$ and for multipoles (2,2), (2,1), (3,3), (3,2),
the values of the $(a_i^\lm,b_i^\lm)$'s that define the NQC corrections to the bare inspiral-plus-plunge EOB waveform.
[When $q=1$ there are no entries for $\lm=(2,1)$ and $(3,3)$, because these modes are identically zero 
in this case for symmetry reasons.]  We will discuss below the issue of replacing the information 
contained in this table by $\nu$-dependent fitting formulas.

\subsection{Effect of the NQC factor on the EOB waveform}
\label{sec:nqc_study}

Let us first illustrate how the NQC factor modifies the purely inspiral 
EOB waveform. The $q=1$ case is considered in Fig.~\ref{fig:effect_of_nqc}:
modulus (left panel) and frequency (right panel). Similar results
are obtained for any other mass ratio (see also Ref.~\cite{Bernuzzi:2010xj}
for the test-mass limit). We show together: (i) the purely inspiral waveform,
i.e., Eq.~\eqref{eq:hlm} {\it without} the NQC factor $\hat{h}_\lm^{\rm NQC}$ 
(dash-dotted, thin line, black online); (ii) the 
inspiral+merger waveform, including the NQC factor (dash-dotted and thick 
line, blue online); (iii) the extended EOB waveform, including the ringdown
part (thick, solid line, red online); and the NR waveform (thin, solid line,
black online).
As noted already in Ref.~\cite{Bernuzzi:2010xj} the most striking feature
of this plot is that the pure inspiral EOB waveform modulus peaks (after 
alignment as explained in Sec.~\ref{sec:phasing})
just $\sim 1.4M$ before the peak of the NR modulus. 
On the other hand, its amplitude is about $20\%$ larger than the NR 
one\footnote{Such a behavior follows from our use of $x=v_\varphi^2$ 
as argument in $\rho_{22}(x)$. As noted in Fig.~2 of Ref.~\cite{Pan:2011gk}, 
the different choice $x\to \Omega^{2/3}$ (which is however not physically justified
during the plunge), makes the EOB waveform peak considerably earlier 
(by $6.2M$) than NR, but with an amplitude much closer to 
the NR one ($\approx -0.23\%$ smaller).}.
Note that the largish difference in amplitude is very effectively
corrected by the NQC factor. In order to reduce the amplitude and 
displace it to the right we need a NQC factor that, near merger, is 
smaller than one and growing. This is what $n_2$ succeeds 
in doing thanks  to its shape, as illustrated in Appendix~\ref{sec:ddotr}.
This explains why the values of the NQC parameter $a_2^{22}$ 
are the dominant ones, see Table~\ref{tab:nqc_ab}.
By contrast, if one has to increase the amplitude and displace
it to the right (as was needed in Ref.~\cite{Pan:2011gk} because
of the use of the argument $\Omega^{2/3}$ in $\rho_{22}(x)$), 
one needs a NQC factor which, near merger, is larger than one 
and growing, as, for instance, our $n_1$, Eq.~\eqref{eq:n1_nqc}.

\subsection{Comparison between the $ \ell=m=2$  NR and EOB  waveforms}
\label{sec:phasing}

Let us now present the results of the comparison between the dominant quadrupolar ($(\ell,m) = (2,2) $)
NR waveform, and the corresponding NR-completed EOB waveform introduced in this work.
For each mass ratio among $q=(1,2,3,4,6)$ ,  Figs.~\ref{fig:phasing_q1q2}-\ref{fig:phasing_q3q4}  compare the EOB
and NR modulus and frequency (left panels), the real parts of the waveforms (right panels, bottom) and
also show  the phase difference $\Delta\phi^{\rm EOBNR}\equiv \phi^{\rm EOB}-\phi^{\rm NR}$ 
and the relative amplitude difference \hbox{$\Delta A^{\rm EOBNR}/A^{\rm NR}\equiv   (A^{\rm EOB}-A^{\rm NR})/A^{\rm NR}$}
(right panels, top).
The vertical dashed line present in all panels marks the location of the peak of the EOB orbital
frequency, $\teobLR$.
These time-domain comparisons are done by suitably determining a relative time and phase shift 
between the two phases $\phi_{22}^{\rm NR}(t^{\rm NR})$ and  $\phi_{22}^{\rm EOB}(t^{\rm EOB})$.
These shifts are estimated by minimizing  the time integral of the square of the phase difference on 
a time interval corresponding to a given frequency interval $[M\omega_L,M\omega_R]$. Following 
Refs.~\cite{Boyle:2008ge,Pan:2011gk} , we perform this waveform alignment  on the long inspiral phase. 
Note that, in doing so, we do not enforce the constraint
that $\tnrLR$ corresponds to $\teobLR$. However, the EOB/NR agreement is so good up to merger
 that such an early-inspiral alignment succeeds in realizing, a posteriori, a near coincidence between
$\tnrLR$ and $\teobLR$. For instance, we find that, for $q=1$, $\tnrLR - \teobLR \simeq -0.13M$ .
The right limit of the frequency for each mass ratio is $M\omega_R = 0.1$. 
The left bounds are $M\omega_L=(0.035,0.035,0.035,0.044,0.045)$.

These figures indicate an excellent EOB/NR agreement in  phasing and in  modulus from the early inspiral up to merger.
The remaining disagreements are well within the nominal error bar on numerical data. Actually, the only estimate of the numerical error on
the phasing of these numerical data that is available in the literature is a rather conservative one that is done 
by taking the difference between the highest and the medium resolution. This procedure gives uncertainties that are 
very small during the inspiral phase ($<0.01$ rad) and small, though not negligible, in the late plunge phase up to
merger ($\sim~0.1-0.3$ rad, depending on the mass ratio)~\cite{Buchman:2012dw}. 
A less conservative NR error estimate might be smaller by (at least) a factor two\footnote{We thank Harald Pfeiffer 
and Luisa Buchman for informing us of this more realistic estimate of the NR errors.}.
Keeping this in mind, it is remarkable that our EOB model, with the very simple law for $ a_6^c(\nu)$ given in
Eq.~\eqref{eq:a6c_nu} is able to reproduce all numerical data within $\lesssim 0.06$ radians at merger.

Let us also emphasize the very good agreement between the moduli before and at merger  (see the top-right inset in 
the right-panels of Figs.~\ref{fig:phasing_q1q2}-\ref{fig:phasing_q3q4}), though they exhibit a visible difference
during the subsequent ringdown. The good agreement before merger  is an  improvement with respect to previous 
works~\cite{Damour:2009kr, Pan:2009wj,Pan:2011gk}  that is due to a combination of effects coming both from 
the use of an improved analytical EOB model, from a new choice of the basis of NQC functions $n_i$, and from
the choice of an NQC determination point which differs from the maximum of the amplitude. [ Note that such an
agreement before merger is also comparable to the one obtained by Taracchini et al.~\cite{Taracchini:2012ig} 
with an EOB model that is rather different from the one discussed here].
Let us also note that, as already mentioned, we have, on purpose, chosen effective values 
of $a_6^c(\nu)$ causing the phase difference $\Delta\phi^{\rm EOBNR}$ to dip towards negative values 
$\sim - 0.05$ rad just before merger, before jumping towards positive values of order $+0.05$ 
or $+0.1$~rad during ringdown. Such a behavior ensures a good average phase agreement during
the entire process. Had we instead chosen the slightly different ``flat'' values 
of  $a_6^c(\nu)$, Eq.~\eqref{eq:a6c_nu_flat}, they would have led 
to a near perfect phase agreement up to merger. However, the price for 
doing so would then have been the presence of a larger global 
phase disagreement (of order $\sim + 0.15$ rad), due to a positive 
jump in $\Delta\phi^{\rm EOBNR}$ after merger, and during ringdown. 
We note that such a positive jump $\sim + 0.15$ rad in $\Delta\phi^{\rm EOBNR}$ 
is consistent with the study, done in Ref.~\cite{Pan:2011gk}, of the {\it intrinsic} error in $\Delta\phi^{\rm EOBNR}$ 
coming from the procedure of QNM attachment itself. 
This indicates that more work should be devoted towards improving the current
 EOB technique for attaching QNMs onto the inspiral-plus-plunge waveform.

\subsection{Subdominant multipoles}

Up to now, our study has only considered the dominant quadrupolar $\ell=m=2$ waveform.
Let us now compare some of the subdominant multipolar waveforms. We consider here the 
$\ell=2$, $m=1$ and $\ell=m=3$ subdominant waveforms,
for the two mass ratios $q=2$ and $q=6$ (similar results were obtained for  $q=3$ and $q=4$).  
We limit ourselves to such a partial
comparison here to show the capability of the EOB model, as it was defined above, to get 
the main characteristics of the subdominant multipoles, without introducing ad hoc modifications, 
or tuning further parameters. At the end of this section we will also mention some results for
the $\ell=3$, $m=2$ multipole.

In Figs.~\ref{fig:phasing_higher_q2}-\ref{fig:phasing_higher_q6} we compare,   for the two mass ratios $q=2$ and $q=6$, the NR and EOB 
frequency and modulus for the two subdominant multipoles $\ell=2$, $m=1$ and $\ell=m=3$ (top panels) as well as the phase and 
amplitude differences (bottom panels). We use the same matching interval as 
for the $\ell=m=2$ mode, i.e. $\Delta^{\rm match} = 0.7 M$, and the same number of QNM modes, i.e. $N=5$. 
Note the good agreement of the moduli in all cases, both up to merger, and during ringdown
[In the $A_{21}$, $q=2$ case the multiple crossings between the NR and EOB moduli may be due to
inaccuracies in the NR waveform.] Note also the good agreement, {\it up to merger}, of the frequencies,
in all cases, and the good agreement of the frequency of the $(3,3)$ mode after merger, and during
ringdown. The only case which is slightly less successful is the discrepancy between the EOB frequency
and the NR frequency in the $\ell=2$, $m=1$ case for both mass
ratios (compare with Ref.~\cite{Pan:2011gk}, but note we have not introduced
here any ad hoc treatment of the the $\ell=2$, 
$m=1$ case.)
Namely, the EOB frequency of the $(2,1)$ mode shoots up, just after
merger , a bit faster than its NR counterpart. In turn, such a frequency difference builds up a phase difference
after merger. This is illustrated in the bottom panels of the figure, which shows the phase differences 
$\Delta\phi^{\rm EOBNR}_{21}$ (left) and $\Delta\phi^{\rm EOBNR}_{33}$ (right) as  functions of time during the entire simulation. 
Note that the dephasing is remarkably small up to merger for {\it both} multipoles, and then
accumulates a dephasing $\Delta\phi^{\rm EOBNR}_{21}\sim 0.5$~rad (and $\Delta\phi^{\rm EOBNR}_{33}\sim 0.15$~rad) 
during the ringdown.

Let us emphasize that  the phase difference $\Delta\phi^{\rm EOBNR}_{21}(t)$ plotted in the bottom panels 
of Figs.~\ref{fig:phasing_higher_q2}-\ref{fig:phasing_higher_q6} has been computed 
{\it without introducing any new arbitrariness}, neither in time, nor in phase, in comparing the two phase evolutions.
Indeed, the least-squares alignment procedure of the NR and EOB dominant $(2,2)$ waveforms has determined
both a shift in time, say $\tau_{2 2}$, and a phase shift, say $\alpha_{2 2}$, connecting them. The time shift  $\tau_{2 2}$
determines the (a priori unknown)  connection between the two time variables $t^{\rm NR}$ and  $t^{\rm EOB}$, 
and should therefore be used in comparing the time evolutions of all the other physical quantities,
and in particular the subdominant multipoles.  The case of the phase shift   $\alpha_{2 2}$ is similar, but with
a difference. Indeed, in our case (with a common, preferred $z$ axis given by the total angular momentum
of the sytem) the only a priori unknown angular difference between NR and EOB is a rotational shift,
by some angle $\beta$,  connecting the NR basis of tensorial spherical harmonics to the corresponding EOB basis.
This common angle $\beta$ then introduces a phase shift in all the various  $\lm$ multipoles simply given by
\be
\alpha_\lm = m \, \beta,
\ee
independently of $\ell$. As this result applies in particular to $\alpha_{2 2}$ (which is determined modulo $2 \pi$
by the alignment of the $(2,2)$ waveforms), we see that the phase shifts in the subdominant multipoles
are determined to be
\be
\alpha_\lm = \frac{m}{2} \, \alpha_{22} \, \, \text{modulo } \,  m \pi
\ee
In addition to this phase shift, there might be extra phase shifts due to the use of different conventions
in defining the phase of the tensorial spherical harmonics. Such phase conventions differ at most by multiples
of $\pi/2$, corresponding to powers of $i$. In other words, we can always write that
$\alpha_\lm = \frac{m}{2} \, \alpha_{22} $ modulo $\pi/2$, which is sufficient for unambiguously computing
$\Delta\phi^{\rm EOBNR}_\lm$ for all subdominant multipoles. 
This absence of phase-shift ambiguity in $\Delta\phi^{\rm EOBNR}_\lm$ makes it all the more remarkable
that, in the $(2,1)$ case, the phase difference  $\Delta\phi^{\rm EOBNR}_{21}$ plotted in 
Fig.~\ref{fig:phasing_higher_q2} (for $q=2$) and Fig.~\ref{fig:phasing_higher_q6} (for $q=6$) 
stays very small up to merger.

Let us finally comment on Fig.~\ref{fig:multipole_l3m2}, were we show the phase difference one gets 
for the $\ell=3$, $m=2$ multipole, for the two representative cases $q=1$ (top panel) and $q=6$ (bottom panel).
The figure, again, illustrates a rather good consistency between EOB and NR up to merger. The differences
after merger are mostly due to our simplified description of the ringdown (see Appendix~A of Ref.~\cite{Pan:2011gk}
for a detailed analysis of the structure of the $(3,2)$ ringdown waveform). 

We leave to future work a more detailed analysis of the subdominant multipoles, and the investigation of 
possible ways of improving their EOB representation, in case the slight dephasing exhibited in 
Figs.~\ref{fig:phasing_higher_q2}-\ref{fig:phasing_higher_q6} for the $(\ell,m) = (2,1)$ multipole
happens to significantly degrade the faithfulness of the complete EOB waveform (summed over 
all multipoles).

\section{Structure of the EOBNR radial potential $A(u)$  and its connection with other results}
\label{sec:A_structure}

\begin{figure}[t]
  \begin{center}   
    \includegraphics[width=0.45\textwidth]{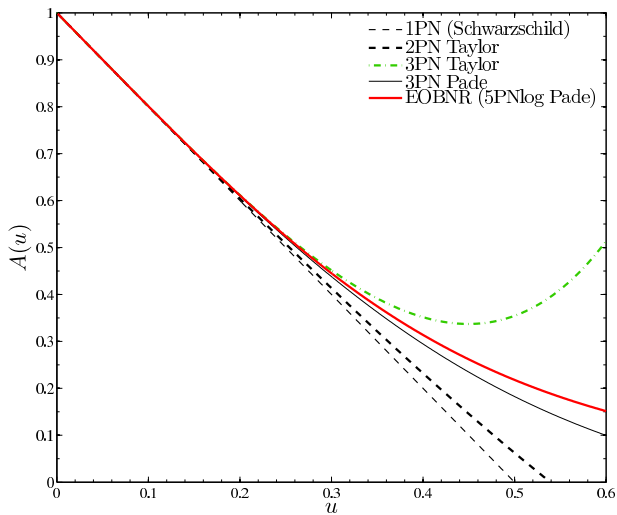}
    \caption{ \label{fig:Aplot} (color online) Contrasting various estimates of the  $A(u;\,\nu)$ function. 
      in the equal-mass case, $\nu=0.25$. The plot shows the 1PN, 2PN and 3PN Taylor-expanded versions of 
      $A(u;\,0.25)$; its 3PN-accurate Pad\'e resummed form as well as the EOBNR one (5PN accurate with 
      logarithmic terms and $a_5^c=23.5$, $a_6^c(0.25)=-101.876$, as per Eq.~\eqref{eq:a6c_nu}).}
  \end{center}
\end{figure}

One of the most important nonperturbative dynamical knowledge acquired in this work by
comparing EOB predictions to the Caltech-Cornell-CITA simulations concerns the
function $A(u ; \nu)$.  We recall that $A(u;\nu)$ is the main radial potential of the EOB Hamiltonian,
and represents the time-time component of the effective EOB metric: $A(u;\nu) = - g^{\rm eff}_{00}(R)$.
In the test-mass limit, $\nu \to 0$, the effective metric is the Schwarzschild metric,  so that
$\lim_{\nu \to 0}  A(u;\nu)= 1-2u \equiv 1 - 2 GM/(R c^2)$.  We saw above that NR data selected,
in the strong-field domain, an $A$ function given by  Eq. \eqref{Alogs} with 
 $a_5^c=23.5$  and $a_6^c(\nu)=[-110.5 + 129(1-4\nu)]\tilde{s}\left(-1.5\times 10^{-5};\,\nu\right)$).  
Let us now  discuss some properties of this NR-informed EOB potential 
(or simply EOBNR potential) and its connection with other relevant results.

\subsection{Global shape of $ A^{\rm EOBNR}(u;\nu)$ as a function of $u$, and comparison with previous purely analytical estimates }

As a first orientation, we contrast in Fig. \ref{fig:Aplot} various estimates of the function  $A(u;\nu)$
in the equal-mass case, i.e. $\nu=0.25$.  Our NR-informed estimate
(5PN-log--Pad\'e resummed and with $a_5^c=23.5$  and $a_6^c(\nu)=[-110.5 + 129(1-4\nu)]\tilde{s}\left(-1.5\times 10^{-5};\,\nu\right)$) 
is shown as a thick solid line (red online), i.e. the second line from the top.  
The dashed bottom line represents the 1PN-accurate estimate of $A$, which happens to  
coincide with the simple Schwarzschild-metric result $A^{\rm Schw}(u)=1-2u$. 
[Indeed, in  Eq. \eqref{ATaylor} there are no terms of order $u^2$ corresponding to the 1PN level.]
The thicker dashed line just above this 1PN estimate represents the
Taylor-expanded  2PN estimate,  i.e. 
Eq. \eqref{ATaylor} taken up to the term $\O(u^3)$ included.
The upper dashed line 
represents the Taylor-expanded 3PN estimate of   $A(u;\nu)$, as given by Eq. \eqref{ATaylor}  up to the term $\O(u^4)$ 
included. Finally,  the thin solid line (black online) just below the NR-completed 5PNlog Pad\'e curve is the Pad\'e-resummed estimate of the
analytically known 3PN result, which was proposed by Damour, Jaranowski and Sch\"afer \cite{Damour:2000we}
in 2000, i.e. five years before NR simulations started yielding information about the strong-field
dynamics of binary black holes.  It is remarkable that the latter simple 3PN-Pad\'e estimate is rather close to the
best current NR-informed estimate: (i) it is numerically quite close to it if one considers values $ u \lesssim 0.3$ which are already
beyond the last stable orbit, and therefore are crossed during the plunge; and (ii) even in the very strong field domain
$0.3 \lesssim u \lesssim 0.6$ (where the merger occurs) the  3PN-Pad\'e estimate is a much better approximation
to   $ A^{\rm EOBNR}(u;\nu)$ than any of its standard PN approximants.
 This closeness explains the
success of the simple Pad\'eed 3PN $A$ function in agreeing with several recent NR studies of
dynamical aspects of close black hole binaries \cite{LeTiec:2011bk,Damour:2011fu}, and confirms the 
effectiveness of using Pad\'e approximants to improve the strong-field behavior of Taylor approximants.

\subsection{Detailed study of the $\nu$-dependence of  $ A^{\rm EOBNR}(u;\nu)$}

\begin{figure}[t]
  \begin{center}   
    \includegraphics[width=0.5\textwidth]{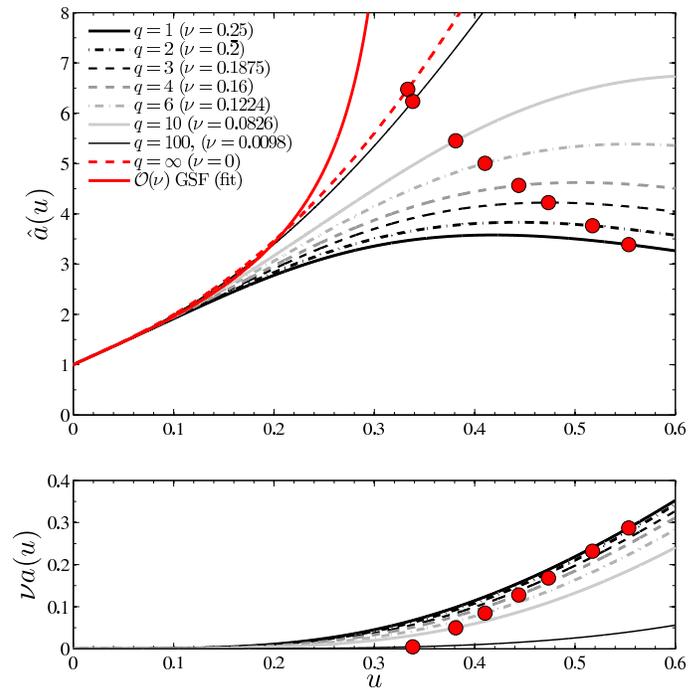}
    \caption{ \label{fig:hata} (color online) Top panel: Behavior of the EOBNR $\hat{a}(u)$ function defined 
      in Eq.~\eqref{eq:hat_u} with $a_5^c=23.5$ and $a_6^c(\nu)=[-110.5+129(1-4\nu)]\tilde{s}\left(-1.5\times 10^{-5};\, \nu\right)$. 
      The red line shows the $\nu=0$ function as obtained from the fit of GSF data~\cite{Akcay:2012ea}. 
      Bottom panel: the difference $A^{\rm EOBNR}(u;\nu)-A^{\rm Schw}(u)$ with $A^{\rm Schw}(u)=1-2u$.
      For each value of $\nu$, the marker indicates the EOB-defined adiabatic light ring location.}
  \end{center}
\end{figure}

The  comparison of the previous subsection has indicated that an accurate description of the gravitational wave emission of coalescing
binary black holes  requires a very precise determination of the shape of   $A(u;\nu)$ in the very strong-field
domain  $ u \gtrsim 0.3$  (i.e. $ R \lesssim 3 GM/c^2$).  Let us zoom on the detailed shape 
of the $A$ function in the strong-field domain by focusing on the properties of 
the associated $a$ function, defined by writing
\be
A(u;\nu) \equiv 1 -2u + \nu a(u;\nu).
\ee
The Taylor expansion of this small-$a$ function starts as
\be
a(u;\nu) =2 u^3 + \left(\dfrac{94}{3}-\dfrac{41}{32}\pi^2\right) u^4 + O(u^5 \ln u).
\ee
Note that the $\nu$ dependence of $a(u;\nu)$ is only contained in the  $ O(u^5 \ln u)$  remainder term.
In order to zoom on the $\nu$ dependence of $a(u;\nu)$ it is then useful, following Ref.~\cite{Akcay:2012ea},
to normalize the $a$ function by its LO PN behavior, $a^{2PN}(u;\nu) =2 u^3 $, i.e. to consider
 the $\hat{a}(u;\nu)$ function defined as
\be
\label{eq:hat_u}
\hat{a}(u;\nu) \equiv \dfrac{a(u;\nu)}{ 2 u^3} \equiv \dfrac{A(u;\nu)-(1-2 u)}{2\nu u^3}.
\ee 
In the upper panel of Fig.~\ref{fig:hata},  we plot the values of the EOBNR $\hat{a}(u;\nu)$ 
functions for the values of $\nu$ corresponding to the five mass ratios we used in our EOB/NR comparisons
above, namely $q=1,2,3,4,6$, as well as the EOBNR predicted $\hat{a}$ curves corresponding to $q=10$, to $q=100$
and also to $q=\infty$, i.e. to the $\nu =q/(q+1)^2 \to 0$ limit of  $\hat{a}^{\rm EOBNR}(u;\nu)$. 
The (red online) round markers on the curves indicate the EOB-defined, light-ring locations, 
i.e. the solutions of the equation $(u^2 A(u))'=0$ (see Table~\ref{tab:LR} for the precise
numbers). In addition, we have also indicated the recently derived (GSF-computed) ``exact'' value of the limit
$\lim_{\nu \to 0} \hat{a}(u;\nu)$ \cite{Akcay:2012ea}  (using their best analytical fit).
  In the bottom panel of   Fig.~\ref{fig:hata} we plot the corresponding values of the products 
  $\nu {a}^{\rm EOBNR}(u;\nu) = 2 \nu u^3 \hat{a}^{\rm EOBNR}(u;\nu)$, i.e. the corresponding differences 
  of  ${A}^{\rm EOBNR}(u;\nu)$ away from its test-mass limit, i.e. 
  ${A}^{\rm EOBNR}(u;\nu) -  A^{\rm Schw}(u)$, where
  $A^{\rm Schw}(u)=1-2u= \lim_{\nu \to 0} A^{\rm EOBNR}(u;\nu)$.
This shows again how the physics of the GW emission by coalescing  black hole binaries depends on
fine features in the $A$ potential. Note  how, as $\nu$ decreases,   $\hat{a}(u;\nu)$  monotononically
increases, in a way which is qualitatively compatible with the shape of the limiting
GSF result  $ \hat{a}(u;0) =\lim_{\nu \to 0} \hat{a}(u;\nu)$. [The latter limiting GSF shape has
a singularity at $u=1/3$, which is probably smoothed out by higher-order corrections in $\nu$ around $\nu=0$.
See  \cite{Akcay:2012ea} for a detailed discussion of the origin of this singularity, and its probable
fictitious character.]  Though the $\nu \to 0$ limit of  $\hat{a}^{\rm EOBNR}(u;\nu)$ 
(which is a polynomial in $u$,
with logarithmic coefficients) does not coincide with the exact $\O(\nu)$ GSF result, 
it stays quite close to it up to $ u\lesssim 0.2$. It is interesting in this respect 
to point out that the $\nu\to 0$ limit of our NR fitted $a_6^c(\nu)$, 
given by Eq.~\eqref{a6_final}, is $a_6^c(0) = +18.4979\approx + 18.5$. 
This is completely different from the true Taylor 
value $a_6^{c \, \rm Taylor}(0)= -131.72(1)$~\cite{Barausse:2011dq}. However, it has the same 
sign and order of magnitude as the effective value obtained
above, in  Eq.~\eqref{eq:a6c_from_GSFLSO}, by requiring compatibility with the GSF determination of LSO precession 
for $\nu\to 0$. This shows a reasonable compatibility 
between two effective determinations of $a_6^c(0)$ in the strong field regime.

Note also, on the bottom panel, 
how the behavior of the corresponding contribution to the $A$ potential,
i.e. the product $\nu a(u;\nu)$, seems to tend continuously (though maybe not uniformly) towards zero
as $\nu \to 0$.  This bottom panel suggests that the $q=10$ case should be thought of as belonging
to the class of the  normal comparable-mass cases $q=\O(1)$. 
One needs $q$'s of order at least $\O(100)$
to belong to the class of extreme-mass-ratio binaries. The EOBNR potential derived here has anyway
been tuned to the physics of comparable-mass binaries with $1\leq q \leq 6$. As we knew 
(from Ref.~\cite{Akcay:2012ea}) that the $\nu \to 0$ limit of
the (exact) $A$ potential was (probably) mildly singular, and as we are mainly interested in
describing the physics of comparable-mass systems, we did not attempt to incorporate in the $A$ function
too much of the information contained in its $\nu \to 0$, GSF limit.  In our work above, we only incorporated
some information about the $\nu \to 0$ limit of the 4PN coefficient $\lim_{\nu \to 0}  a_5^c(\nu)$.
But, as we shall discuss next, this was mainly done as a practical way of  reducing the number
of unknowns to be fitted to NR data.

\begin{table}[t]
  \caption{\label{tab:LR} EOB-defined adiabatic light-ring (LR) and last-stable-orbit (LSO) 
    locations for $a_5^c=23.5$ and $a_6^c=[-110.5+129(1-4\nu)]\tilde{s}\left(-1.5\times 10^{-5};\,\nu\right)$.} 
  \begin{center}
    \begin{ruledtabular}
      \begin{tabular}{cccccc}
        $q$   &   $\nu$  &    $r_{\rm LR}$   &   $u_{\rm LR}$ &$r_{\rm LSO}$   &   $u_{\rm LSO}$   \\
        \hline 
        1        & 0.25              &   1.8067   &  0.5535    &  4.5108 & 0.2217 \\
        2        & $0.\bar{2}$       &   1.9324   &  0.5175    &  4.6964 & 0.2129 \\
        3        & 0.1875            &   2.1119   &  0.4735    &  4.9226 & 0.2031 \\
        4        & 0.1600            &   2.5223   &  0.4440    &  5.0962 & 0.1962 \\
        6        & 0.1224            &   2.4366   &  0.4104    &  5.3235 & 0.1878 \\
        10       & 0.0826            &   2.6240   &  0.3811    &  5.5529 & 0.1801 \\
        $\infty$ & 0                 &   3.0000   & $0.\bar{3}$ & 6.000  & $0.1\bar{6}$
   \end{tabular}
  \end{ruledtabular}
\end{center}
\end{table}

\subsection{On the ``equivalence classes'' of the $A(u)$ potential}

\begin{figure}[t]
  \begin{center}   
    \includegraphics[width=0.45\textwidth]{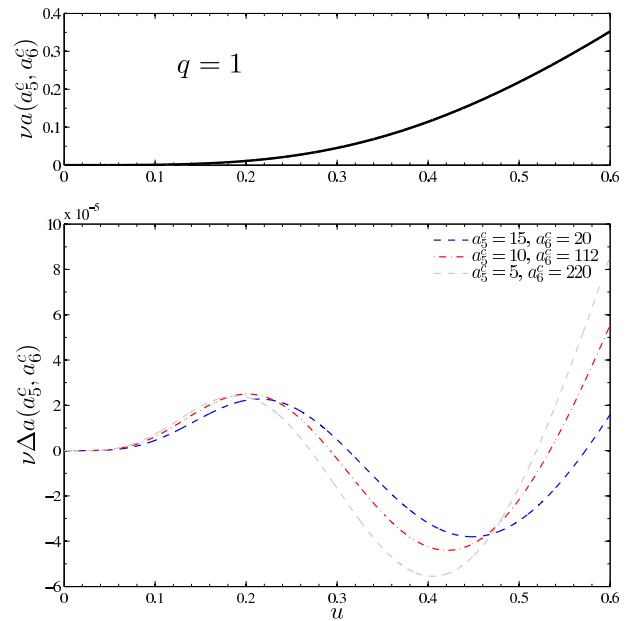}
    \caption{ \label{fig:hata_eq_class} (color online) Elements of the equivalence class of $\hat{a}(a_5^c,a_6^c)$ 
    functions for $q=1$. The bottom panel shows the fractional difference with our favorite 
    choice $a_5^c=23.5$, $a_6^c(0.25)=-101.876$.}
  \end{center}
\end{figure}

 References~\cite{Damour:2009kr,Damour:2009ic}  found, for the $q=1$ case,
  that there was a strong degeneracy between the two parameters
  entering a 5PN-accurate Pad\'e representation of the $A$ function, say $a_5^c$ and $a_6^c$.
This  was confirmed for other values of $q$ in Ref.~\cite{Pan:2011gk}.
This finding leads to the idea that the good values
of $a_5^c$ and $a_6^c$ can be organized in ``equivalence classes''  of quasi-interchangeable values of the
pairs   $(a_5^c, a_6^c)$.  An explicit way of constructing these equivalence classes was indicated in~\cite{Damour:2009sm}:
it consists in defining the equivalence class of some given pair  $(a_5^{c \, (0)}, a_6^{c \, (0)})$ 
as the set of  pairs  $(a_5^{c }, a_6^{c })$ such that the $u$-derivative $A'(u;\nu;a_5^c,a_6^c)$ of the $A$ function,
evaluated at some fiducial strong-field point, say $u_b$ (the value $u_b\simeq 0.215$ was suggested there), takes the same value at  $(a_5^{c }, a_6^{c })$ and at 
$(a_5^{c \, (0)}, a_6^{c \, (0)})$. In equations
\be
 A'(u_b;\nu;a_5^c,a_6^c)=  A'(u_b;\nu;a_5^{c \, (0)},a_6^{c \, (0)}) \ ,
\ee
or, equivalently, 
\be
a'(u_b;\nu;a_5^c,a_6^c)=  a'(u_b;\nu;a_5^{c \, (0)},a_6^{c \, (0)}) \ .
\ee
When working, as we do
here, with the normalized function   $\hat{a}(u;\nu)$, we could alternatively define these
equivalence classes as level sets (in the space of pairs  $(a_5^{c }, a_6^{c })$) of   $\hat{a}'(u_b;\nu; a_5^c,a_6^c)$,
or even, simply,  of  $\hat{a}(u_b;\nu; a_5^c,a_6^c)$. Evidently, all those possible ``definitions''  lead (when one
changes the fiducial value $u_b$, and/or the considered function $a'$, $\hat{a}'$, $a$, etc.)
to different equivalence classes. However, because of the properties of the $A$ function, one checks that,
as long as one bases one's definition on the value of $A$ or some related function in the strong-field region, this leads,
to a good approximation,  to a numerically rather well-defined equivalence class of $(a_5^{c }, a_6^{c })$ pairs.
This is illustrated in  Fig.~\ref{fig:hata_eq_class}.  This figure shows (for the case $q=1$) that our NR-tuned
preferred values  $(a_5^{c \, (0)}, a_6^{c \, (0)}) = (23.5, -101.876)$ define a   $\nu a(u)$ function which can be
very nearly reproduced by using other pairs of  $(a_5^{c }, a_6^{c })$ values, namely $(0,220)$, $(5,125)$, or $(10,40)$.
The upper panel shows together, versus $u$, the functions $\nu a(u; \nu; a_5^{c }, a_6^{c })$,  for $q=1$, i.e. $\nu=0.25$,
and for the four different
pairs of parameters values $(a_5^{c }, a_6^{c }) = {(23.5,-101.876), (10,40), (5,125), (0,220)}$. The upper panel illustrates that these
five different functions are indistinguishable by eye.
The bottom panel of the figure zooms on the differences away from our standard choice  
$(a_5^{c \, (0)}, a_6^{c \, (0)}) = (23.5, -101.876)$, i.e. it plots $\nu \Delta a(u; \nu; a_5^{c }, a_6^{c }) \equiv A(u;\nu;a_5^c,a_6^c)- A(u;\nu;23.5,-101.876) $.
For any choice of the parameters, these differences are of the order $10^{-4}$. 
Note that we have not used, here, any precise, level-set type, criterion for selecting the pairs
equivalent to our preferred value, but we have selected them by simple trial and error, until we could reduce
the (maximum) difference to the smallest level we could find .  This smallest level was $O(10^{-4})$. 
The reason why such a level of deviation is small enough for our purpose can be seen by
turning back to our analysis above,
when we were fixing  the fiducial value $a_5^{c }=23.5$, and then tuning the value of  $a_6^c$ for the
EOB phasing to best agree with the NR one. In that case, as is clear from the number of digits we were giving 
 in Table~\ref{tab:best_a6c} above for  $a_6^c$ (before fitting them), we found that the ``good'' values of $a_6^c$ were determined, roughly,
within an uncertainty $\delta  a_6^c  = \O(1)$.  Such an uncertainty on the good value of  $a_6^c$ (for the fixed $a_5^{c }=
23.5$)  entails a corresponding uncertainty on the value of the function $A(u; a_5^c,a_6^c)$ of
order $\delta A(u; a_5^c,a_6^c) \sim \partial A(u; a_5^c,a_6^c)/ \partial a_6^c$. The latter 
quantity   is found to increase with $u$,
and to reach a value of order $ 0.8 \times 10^{-4}$ when $u$ takes the light-ring value $u_{\rm LR} \simeq 0.55$
(for $q=1$). In conclusion, a possible variation in the $A(u)$ function of $L_{\infty}$ norm $\sim 10^{-4}$, 
for $0\leq u \leq u_{\rm LR}$, is a reasonable way of  defining the equivalence class of $A(u)$, and  Fig.~\ref{fig:hata_eq_class}
shows that one can indeed, starting from the (analytically fitted) 
values $(a_5^{c \, (0)}, a_6^{c \, (0)}) = (23.5, -101.876)$, find
a (relatively thin) strip of values of   $(a_5^{c }, a_6^{c })$  along which the  5PN Pad\'ed
function $A^{\rm EOBNR}(u;a_5^{c }, a_6^{c };\nu)$  stays within
such an equivalence class.

Though here we focus only on the $q=1$ case, similar classes of equivalence of $\hat{a}$ 
functions exist for {\it any} mass ratio. In summary, this exercise confirms that we 
were justified in a priori fixing the value of $a_5^c$. Finally, the important  fact is 
that  NR data allow one to directly determine the $A(u;\nu)$ function itself,  
essentially independently of the chosen ``representative'' $(a_5^{c }, a_6^{c })$ within
some equivalence strip in the  $(a_5^{c }, a_6^{c })$ plane.  This determination
of the $ A^{\rm EOBNR}(u;\nu)$ function is exemplified on Fig.~\ref{fig:hata} 
(keeping in mind the invisible deviations plotted in the upper panel of  Fig.~\ref{fig:hata_eq_class}).

\subsection{Comparison between the present  determination of $A^{\rm EOBNR}(u;\nu)$ (5PN with logs), with previous
estimates (5PN without logs). }  

The present work is the first EOB work to include logarithmic terms in a comparison with NR data.
Let us now compare our final NR-aided determination of such an $A$ function (with logarithmic terms)
to  the 5PN-accurate $A$ functions ({\it without} logarithmic terms) used in previous EOB
works~\cite{Damour:2009kr,Damour:2009ic,Pan:2011gk,Bernuzzi:2012ci,Barausse:2009xi,Taracchini:2012ig}.
In particular, Ref.~\cite{Damour:2009kr}, using a 5PN-accurate $A(a_5^c,a_6^c;\,\nu)$ 
function (without logs), exploited a previous version of the 
$q=1$  Caltech-Cornell-CITA numerical waveform  to
find a banana-like region of good  values in the $(a_5^c,a_6^c)$ plane such that the
phase difference between EOB and NR waveform through inspiral, plunge and merger was $<0.02$ rad.
The  values $a_5^c=-6.37$ and $a_6^c=50$ lie in the middle of this good region and have been used extensively
in subsequent EOB work~\cite{Damour:2011fu,Damour:2012yf,Bernuzzi:2012ci,Bernuzzi:2012ku}. 
[By contrast Ref.~\cite{Damour:2009kr} actually used the values  $a_5^c=0$ and $a_6^c=-20$ which
lie on the boundary of the good region.]
The analog, banana-shaped equivalence classes in the $(a_5^c,a_6^c)$ plane corresponding
to other values of $q$ were then first investigated in Ref.~\cite{Pan:2011gk}. [The latter reference basically used
 the same conceptual structure as Ref.~\cite{Damour:2009kr} with some technical differences.] 
 Ref.~\cite{Pan:2011gk} found a very good agreement between EOB and NR waveforms with 
 an $A$ function defined by the following choices 
\begin{align}
\label{eq:pan_et_al}
a_5(\nu) &= -5.828 - 143.5\nu + 447 \nu^2,\;\quad a_6      &= 184 \ .
\end{align} 
More recently, Barausse and Buonanno ~\cite{Barausse:2009xi} introduced a differently resummed $A$
function, which is 3PN accurate and does not contain the 4PN and 5PN logarithmic contributions used in the present work.
Their resummation does not rely on a Pad\'e approximant, but imposes by hand the presence 
of a horizon, by factoring out of $A(u)$ a binomial of the form $1 - 2 (1-K(\nu) \nu) u + a^2 (1-K(\nu) \nu)^2 u^2$ .
[Here, $a$ is a Kerr-like spin parameter, which vanishes in the non-spinning case considered here.] The flexibility
parameter $K(\nu)$, which effectively parametrizes 4PN and higher contributions, was then calibrated 
in Ref. ~\cite{Taracchini:2012ig} against Caltech-Cornell-CITA  non-spinning waveforms 
(for $q=1,\,2,\,3,\,4,\,6$), with the result:
\be
K(\nu) = 1.447 -  1.715 \nu - 3.246 \nu^2 \, .
\ee

\begin{figure}[t]
  \begin{center}   
    \includegraphics[width=0.45\textwidth]{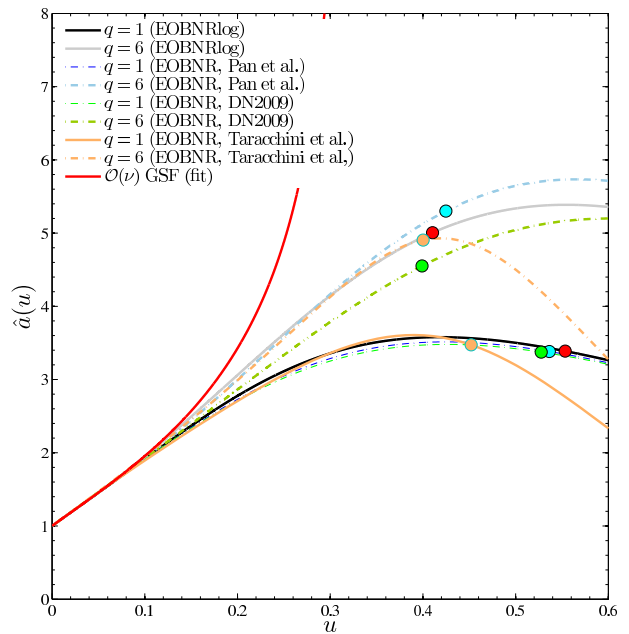}
    \caption{ \label{fig:hata_all_EOB} (color online) Comparing $\hat{a}$ functions for 
    different, 5PN-accurate, EOBNR-completed models. The markers indicate the location of the
     EOB-defined adiabatic light-ring for each curve. See text for explanations.}
  \end{center}
\end{figure}

In Fig.~\ref{fig:hata_all_EOB} we consider the two mass ratios $q=1$ and $q=6$ and for each mass ratio
we  compare four different $\hat{a}(u)$ curves, namely: (i) the log-containing 5PN-accurate one determined
in this work (``EOBNRlog'' with $a_5^c=+23.5$, and $a_6^c(\nu)=[-110.5+129(1-4\nu)]\tilde{s}\left(-1.5\times 10^{-5};\,\nu\right)$);
(ii) the log-less 5PN-accurate one of~\cite{Damour:2009kr} (with $a_5^c=-6.3$ and $a_6^c=50$); 
(iii) the log-less 5PN-accurate one of ~\cite{Pan:2011gk}, see Eq.~\eqref{eq:pan_et_al};
and (iv) the (log-less) 3PN-accurate Barausse-Buonanno~\cite{Barausse:2009xi} one, $\hat{a}^{\rm BB}(u)$, for the
value of the  adjustable parameter, $K(\nu)$, cited above ~\cite{Taracchini:2012ig}.

The figure shows that while the first three different analytical descriptions seem to be visually close
for the equal-mass case, $q=1$, they exhibit visible differences in the $q=6$ case. 
However, we have seen above that only differences of order $10^{-4}$ in the $A$ function 
can be considered as being negligibly small. When computing the differences 
$\Delta A^X(u;\,\nu)\equiv A^X-A^{\rm EOBNRlog}$ for the two labels $X=\text{DN2009}$, 
Ref.~\cite{Damour:2009kr} and $X=\text{Pan et al}$, Ref.~\cite{Pan:2011gk}, one finds 
that, for $q=1$, $\Delta A^X(u)$ is a monotonically decreasing function of $u$
which reaches values of order $\simeq -0.004$ for  $X=\text{DN2009}$ and  $\simeq -0.0025$
for $X=\text{Pan et al}$ when $u\simeq 0.5$, i.e., close to the corresponding adiabatic light-ring position.
Such differences are therefore quite significant on the $10^{-4}$ scale of the equivalence
classes of $A$ functions exhibited in Fig.~\ref{fig:hata_eq_class}.
In the $q=6$ case the corresponding differences taken at $u\simeq 0.4$, close to the adiabatic
light-ring position, are  $\simeq -0.006$ for  $X=\text{DN2009}$ and  $\simeq +0.003$ for  
$X=\text{Pan et al.}$. Again these differences are quite significant.
Note however that for $u\leq 0.3$ the log-less model of~\cite{Pan:2011gk}, Eq.~\eqref{eq:pan_et_al}, 
(which had been tuned to the same $q=6$ NR data as ours) stays quite close to our present 
log-containing model ($\Delta A=2\times 10^{-4}$).

Let us finally discuss the comparison with the (log-less) Barausse-Buonanno $\hat{a}^{\rm BB}(u)$ function calibrated by Taracchini 
et al.~\cite{Taracchini:2012ig} (orange line online, solid for $q=1$, 
dashed for $q=6$). Figure~\ref{fig:hata_all_EOB} shows that up to its 
own light ring (marked by an orange circle on the curves) this function 
stays rather close to our EOBNRlog one with $a_5^c=+23.5$, 
and $a_6^c(\nu)=[-110.5+129(1-4\nu)]\tilde{s}\left(-1.5\times 10^{-5};\,\nu\right)$. 
The differences are however so large that $\hat{a}^{\rm BB}$ cannot be considered to be part of the  
equivalence class of EOBNRlog in the sense discussed above. More precisely, we find that the 
difference between the respective $A$ potentials varies, roughly, between $\pm 0.01$ 
for $q=1$ and between $\pm 0.005$ for $q=6$. This is two orders of magnitudes larger than 
the $10^{-4}$ level that we used above to define the equivalence class 
(see Fig.~\ref{fig:hata_eq_class} and corresponding text). 
Despite this, one finds that the adiabatic
LSO orbital frequencies predicted by the two 
potentials are very close. For $q=1$, we have 
$\Omega^{\rm EOBNRlog}_{\rm LSO}=0.0993$ and $\Omega^{\rm BB}_{\rm LSO}=0.1010$
(i.e. a $\sim 2\%$ difference), and for $q=6$ we obtain 
$\Omega^{\rm EOBNRlog}_{\rm LSO}=0.0801$ and $\Omega^{\rm BB}_{\rm LSO}=0.0797$.
In addition, we see on the figure that the difference
$\hat{a}^{\rm BB}(u) - \hat{a}^{\rm EOBNRlog}(u)$ {\it oscillates in sign around zero},
so that the phasing defined by $\hat{a}^{\rm BB}(u)$ can be expected to agree, {\it on average},
with that defined by $\hat{a}^{\rm EOBNRlog}(u)$. We have performed a quantitative check of this expectation
by considering  the phasing during the quasi-adiabatic inspiral, which is rather directly 
related to the conservative part of the dynamics and thereby to the $A(u)$ function. 
More precisely, we computed, for each $A(u)$ potential, the adiabatic 
phasing along the sequence of EOB circular orbits. This phasing is best measured by the (adiabatic)
$ Q_\omega^{\rm adiabatic}(\omega) \equiv -5/(24\nu)x^{-1}\de_x j_0(\hat{f}(x))^{-1}$ function. 
Here, $\omega = 2 \Omega$ is the adiabatic GW frequency,  $x=\Omega^{2/3}$, $\hat{f}$ is the 
resummed, Newton-normalized, energy flux as introduced in Eq.~\eqref{eq:fhat} above, 
and $j_0$ is the angular momentum along the sequence of EOB circular orbits 
defined by Eq.~\eqref{eq:j02} for a given $A$ potential. We then focus on the 
 difference $\Delta Q_\omega(\omega)=Q_\omega^{\rm EOBNRlog}(\omega) -Q_\omega^{\rm BB}(\omega)$.

Inspection of the $\Delta Q_\omega(\omega)$ function  more or less confirms the conclusion drawn from the comparison
between the $\hat{a}(u)$ functions in Fig.~\ref{fig:hata_all_EOB}. More precisely, we find that:
for $q=1$ it basically averages around zero up to the LSO, varying between $\pm 0.5$ 
in a frequency range $\Delta\omega=(0.03,0.2)$; on the contrary, for $q=6$ the
same function is negative and monotonically decreasing over the frequency interval 
$\Delta\omega=(0.03,0.16)$, reaching the value $\sim-4.4$ at $\omega=0.16$. 
As explained in Sec.~\ref{secQomega} above where the 
$Q_\omega(\omega)$ function was introduced, the usefulness of this phasing diagnostic
is that its integral over $\ln \omega$ directly gives the GW phase as a function of frequency.
Correspondingly the integral $\Delta\phi=\int_{0.03}^{\bar{\omega}_{\rm LSO}} \Delta Q_\omega d\ln(\omega)$
yields the relative dephasing (here estimated in the adiabatic approximation) between the waveforms corresponding
to the choice of two different $A$ potentials, which is accumulated between the initial frequency 
$ \omega=0.03$ and the average LSO frequencies, say  $\bar{\omega}_{\rm LSO}=0.2$ for $q=1$,  
and $\bar{\omega}_{\rm LSO}=0.16$ for $q=6$. We obtained $\Delta\phi=0.62$ rad for $q=1$ and
$\Delta\phi=2.66$~rad  for $q=6$. This result shows that the difference between the BB and EOBNRlog $A$ functions
entails, {\it when considered by itself},  a corresponding difference in the phasing (up to the LSO) that 
can be as large as$\sim 3$~rad depending on the mass ratio considered. However, the model of Ref.~\cite{Taracchini:2012ig}, 
that is based on the $A^{\rm BB}$ function, succeeded (like our EOBNRlog model) in getting an agreement 
with the NR waveform at the level of a $\%$ of a radian. This means that the $A$-dependent
intrinsic difference in the (adiabatic) phasing that we are pointing out
here can be (and has been)  effectively compensated by other adjustable elements entering the model
of Ref.~\cite{Taracchini:2012ig} (notably parameters entering the
 radiation reaction, such as the argument of the $\rho_\lm$'s, the number of 
multipoles in the flux, a different NQC basis, the tuning of $\rho_{22}^{(4)}(\nu)$, etc.).

The conclusions of this comparative analysis of various EOB $A(u)$ functions are two sided.
On the one hand, if we insist on trying to determine the $A$ function with the utmost
accuracy needed to stay within an all-purposes equivalence class of $A$ functions, our results above
show that the introduction of logarithmic contributions in the 
$A$ function {\it cannot be reabsorbed}  by tuning log-less versions of the EOB $A$ potential. 
As we know, from analytical PN work, that these
logarithmic contributions do exist, we conclude that it is necessary 
to include them, and therefore to prefer the type of improved EOB model 
presented in this work to previous  log-less versions 
of the EOB Hamiltonian. On the other hand, if we are ready to neglect the need of reaching
 an ideal all-purposes accuracy in the determination of the  $A$ function,
 the overall conclusion of the comparison done in Fig.~\ref{fig:hata_all_EOB} is that accurate NR
data (here the Caltech-Cornell-CITA ones) do constrain so much the value of the EOB $A(u)$ potential 
(at least up to $u \sim 0.5$) that various ways of parametrizing the shape of the $A(u)$ potentials
lead to final results that are rather close to each other. This comforts us in showing how the EOB formalism
is able to extract from NR data reliable information about the strong-field dynamics of binary black holes.

\section{Extension of the model by analytic  continuation in $\nu$}
\label{sec:fit_nu}
\begin{table}[t]
  \caption{\label{tab:nufits_A0_lr} Fits of Zerilli-normalized multipolar 
    quantities (amplitude, frequencies and derivatives)
    extracted at  $\tnrLR(\nu)$ as function of $\nu$. Each
    quantity is fitted to a quadratic polynomial of the form $f_\lm(\nu)=c_2^\lm\nu^2+ c_1^\lm\nu+c_0^\lm$. 
    For the amplitude and its derivatives the full leading-order 
    $\nu$-dependence ($\nu c_{\ell+\epsilon}(\nu)$, see Eq.~\eqref{eq:clm_LO}) 
    is factorized before fitting.}  
  \centering  
  \begin{ruledtabular}  
  \begin{tabular}{c|c|ccc}        
    \hline
    & $\ell\, m$ & $c_2^\lm$ & $c_1^\lm$ & $c_0^\lm$ \\
    \hline
    \multirow{4}{*}{$\dfrac{A_\lm}{\nu c_{\ell+\epsilon}(\nu)}$}
    &2\, 1& 1.8020$\times 10^{-1}$& -5.3482$\times 10^{-2}$& 9.4465$\times 10^{-2}$ \\	 
    &2\, 2& 3.6836$\times 10^{-1}$&  2.3213$\times 10^{-2}$& 2.9281$\times 10^{-1}$ \\	 
    &3\, 2& 2.3484$\times 10^{-1}$& -5.1891$\times 10^{-2}$& 1.5969$\times 10^{-2}$ \\	 
    &3\, 3& 1.5774$\times 10^{-1}$&  7.1170$\times 10^{-3}$& 5.1385$\times 10^{-2}$ \\	 
    \hline
    \multirow{4}{*}{$\dfrac{\dot{A}_\lm}{ \nu c_{\ell+\epsilon}(\nu)}$}
    &2\, 1& 1.3075$\times 10^{-2}$& -5.3660$\times 10^{-3}$&  2.7088$\times 10^{-3}$ \\	 
    &2\, 2& 6.2259$\times 10^{-3}$&  2.8059$\times 10^{-3}$& -1.5658$\times 10^{-3}$ \\	 
    &3\, 2& 2.7001$\times 10^{-2}$& -6.8708$\times 10^{-3}$&  5.0927$\times 10^{-4}$ \\	 
    &3\, 3& 1.2320$\times 10^{-2}$&  7.6133$\times 10^{-4}$&  1.5238$\times 10^{-4}$ \\	 
    \hline
    \multirow{4}{*}{$\dfrac{\ddot{A}_\lm}{\nu c_{\ell+\epsilon}(\nu)}$}
    &2\, 1&  5.2570$\times 10^{-4}$& -4.9124$\times 10^{-4}$& -1.1183$\times 10^{-4}$ \\	 
    &2\, 2&  1.4031$\times 10^{-3}$& -1.0071$\times 10^{-3}$& -7.4628$\times 10^{-4}$ \\	 
    &3\, 2&  4.9252$\times 10^{-3}$& -1.2516$\times 10^{-3}$& -3.1190$\times 10^{-6}$ \\	 
    &3\, 3& -3.5470$\times 10^{-4}$&  8.2613$\times 10^{-5}$& -1.3908$\times 10^{-4}$ \\	 
    \hline
    \multirow{4}{*}{$\omega_\lm$}
    &2\, 1& -7.1306$\times 10^{-3}$& 1.8015$\times 10^{-1}$& 1.9488$\times 10^{-1}$ \\	 
    &2\, 2&  3.1848$\times 10^{-1}$& 2.2996$\times 10^{-1}$& 2.8788$\times 10^{-1}$ \\	 
    &3\, 2& -2.3137               & 5.3441$\times 10^{-1}$& 3.5026$\times 10^{-1}$ \\	 
    &3\, 3&  3.7872$\times 10^{-1}$& 4.1589$\times 10^{-1}$& 4.4262$\times 10^{-1}$ \\	 
    \hline
    \multirow{4}{*}{$\dot{\omega}_\lm$}
    &2\, 1& -5.4429$\times 10^{-2}$&  2.3401$\times 10^{-2}$& 8.6489$\times 10^{-3}$ \\
    &2\, 2&  2.6909$\times 10^{-2}$&  1.3939$\times 10^{-2}$& 6.3061$\times 10^{-3}$ \\
    &3\, 2& -3.3131$\times 10^{-1}$&  5.7770$\times 10^{-2}$& 1.3219$\times 10^{-2}$ \\
    &3\, 3&  1.9620$\times 10^{-2}$&  2.6984$\times 10^{-2}$& 1.0610$\times 10^{-2}$ \\
    \hline
    \multirow{4}{*}{$\ddot{\omega}_\lm$}
    &2\, 1&  3.1509$\times 10^{-2}$& -5.7895$\times 10^{-3}$& 9.1507$\times 10^{-4}$ \\
    &2\, 2&  2.2304$\times 10^{-3}$&  3.2830$\times 10^{-4}$& 9.6664$\times 10^{-5}$ \\
    &3\, 2& -1.5297$\times 10^{-2}$&  2.7862$\times 10^{-6}$& 7.3264$\times 10^{-4}$ \\
    &3\, 3&  1.6612$\times 10^{-2}$& -2.0232$\times 10^{-3}$& 3.0898$\times 10^{-4}$ \\
    \hline
  \end{tabular}
  \end{ruledtabular}  
\end{table}

\begin{table}[t]
  \caption{\label{tab:nufits_aibi} Fits of the  NQC parameters $(a_i^\lm,b_i^\lm)$ 
    considered in this work as function of $\nu$. Each quantity is fitted to 
    a quadratic polynomial of the form $f_\lm(\nu)=c_2^\lm\nu^2+ c_1^\lm\nu+c_0^\lm$.}  
  \centering  
  \begin{ruledtabular}  
  \begin{tabular}{c|c|cccc}  
 \hline
    & $\ell \, m$ & $c_2^\lm$ & $c_1^\lm$ & $c_0^\lm$& \\
\hline
\multirow{2}{*}{$a_1^\lm(\nu)$}
    & 2\, 1 &  0.9150                & -0.6522     &  0.0340&\\
    & 2\, 2 &  2.1601                & -1.0937     &  0.0793&\\
    & 3\, 2 &  31.671                & -10.310     &  0.6844& \\
    & 3\, 3 &  2.6793                & -1.2792     &  0.1456& \\
\hline
\multirow{2}{*}{$a_2^\lm(\nu)$}
    & 2\, 1 &   1.9035               &  4.9785            &  -0.7106&\\ 
    & 2\, 2 &  -10.807               &  7.1420            &   0.7035&\\
    & 3\, 2 &  -132.73               &  55.153            &  -3.0449&\\
    & 3\, 3 &  -12.932               &  10.634            &   0.4704&\\
\hline
\multirow{2}{*}{$a_3^\lm(\nu)$}
    & 2\, 1 &   2.6950                &  -3.1603  &  0.8650&\\
    & 2\, 2 &  -2.7666                &  -0.1769  &  0.1012&\\
    & 3\, 2 &  -19.734                &   1.8299  &  0.3031&\\
    & 3\, 3 &  -0.8932                &  -2.9229  &  0.5078&\\
\hline          
\multirow{2}{*}{$b_1^\lm(\nu)$}
    & 2\, 1 & -2.2480              &   0.5304   &  0.2466&\\
    & 2\, 2 & -0.8568              &  -0.2417   &  0.1929&\\
    & 3\, 2 & -1.3497              &  -1.8083   &  0.5735&\\
    & 3\, 3 & -1.6468              &  -0.1611   &  0.3464&\\
\hline
\multirow{2}{*}{$b_2^\lm(\nu)$}
    & 2\, 1 &  51.726            & -21.689 &  3.1616 &\\
    & 2\, 2 &  9.6382            & -7.6453 &  0.3732 &\\
    & 3\, 2 & -4.4860            & -7.7968 &  3.2538 &\\
    & 3\, 3 &  6.9597            & -7.5958 &  0.0709 &\\
\hline
\multirow{2}{*}{$b_3^\lm(\nu)$}
  & 2\, 1 & -112.81  &   30.559 &  -0.0026 &\\
  & 2\, 2 & -80.991  &   17.075 &  -1.7974 &\\
  & 3\, 2 &  3.2489  &  -38.133 &   6.1648 &\\
  & 3\, 3 & -213.46  &   52.819 &  -3.4718 &\\
\hline
 \end{tabular}
  \end{ruledtabular}  
\end{table}

\begin{figure}[t]
  \begin{center}   
    \includegraphics[width=0.38\textwidth]{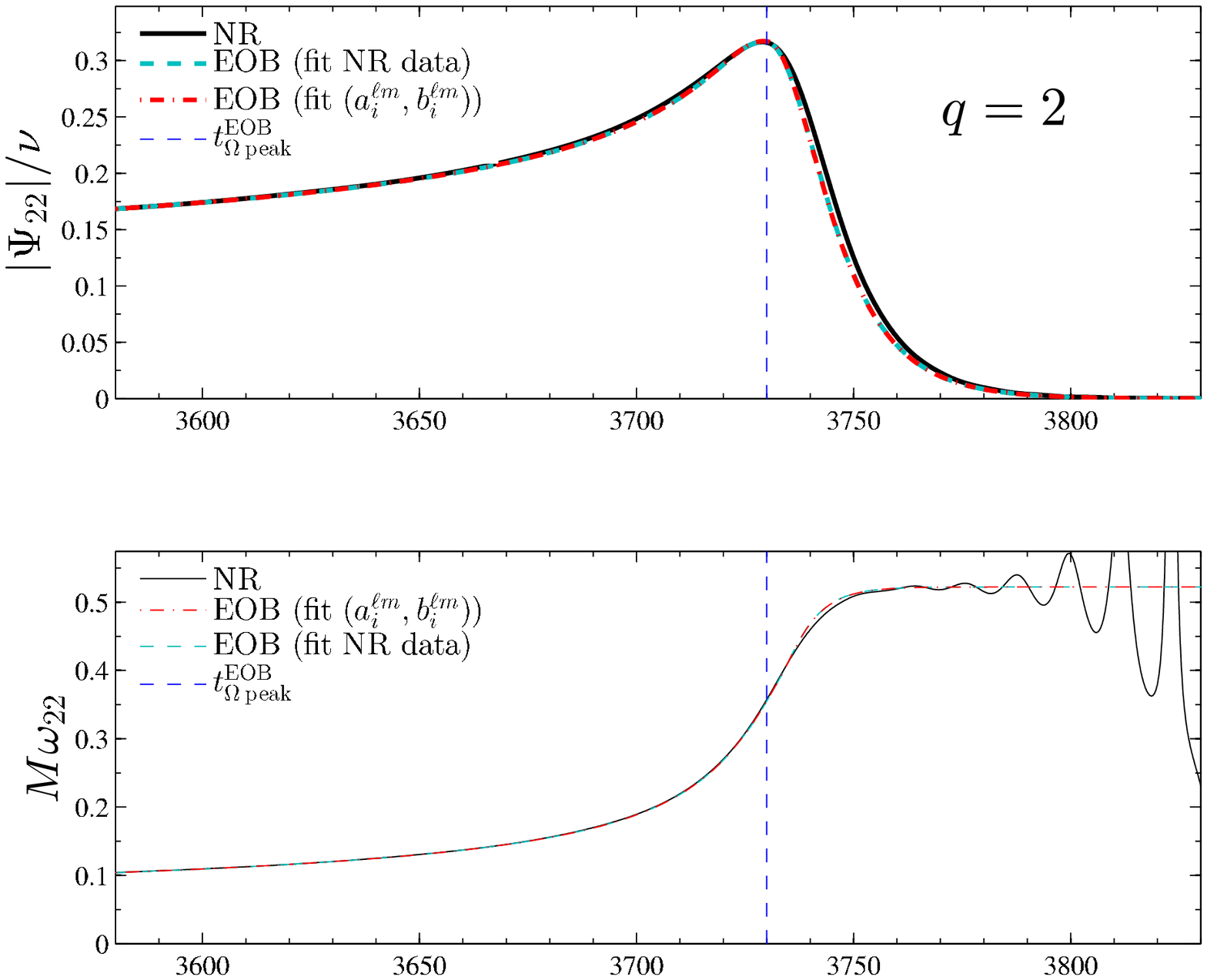}\\
\vspace{5.0mm}
    \includegraphics[width=0.38\textwidth]{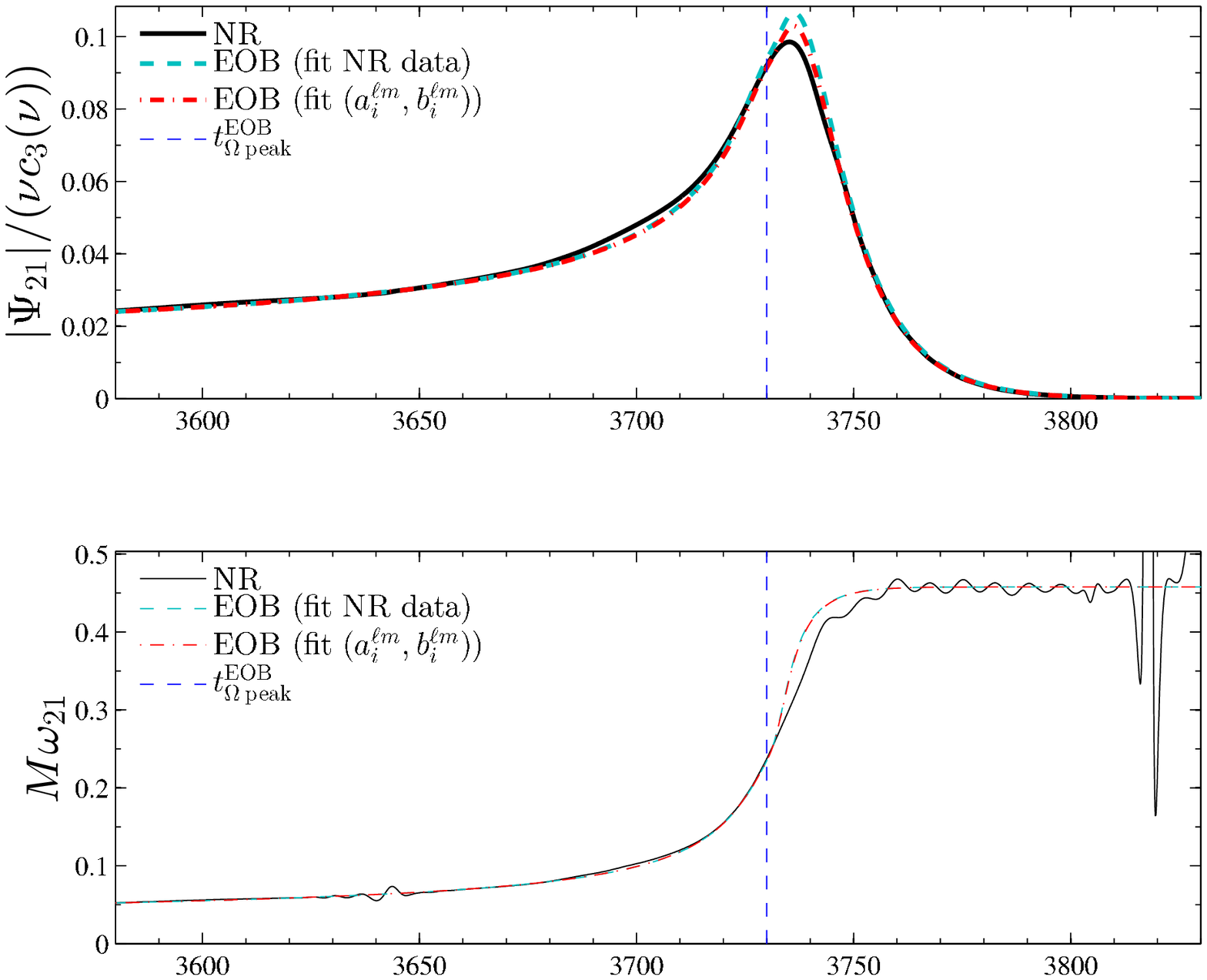}\\
\vspace{5.0mm}
    \includegraphics[width=0.38\textwidth]{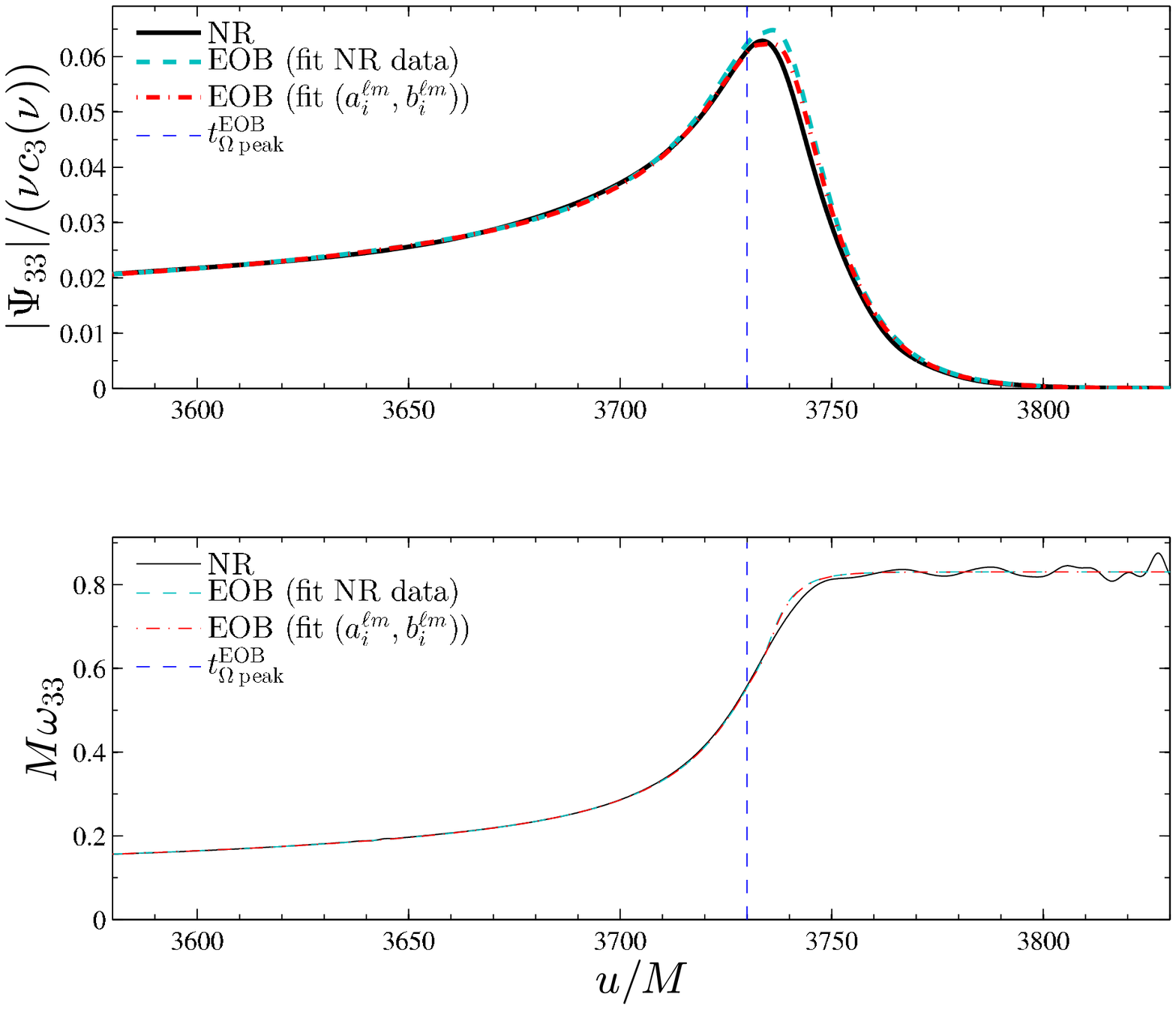}
    \caption{ \label{fig:compare_fits} (color online) Testing two possible fitting
    strategies to continuously extend the discrete sample of NQC parameters $(a_i^\lm,b_i^\lm)$ 
    to any value of $\nu$: comparison (for $q=2$) between $\ell=2$ and $\ell=3$ modulus and frequency.}
  \end{center}
\end{figure}

In the present work, we have used a discrete sample of  numerical simulations to complete 
an EOB model, notably through the use of suitable, NR-fitted NQC corrections.  
In order to be able to compute the predictions of such a NR-completed EOB model for arbitrary 
values of $\nu$, we need to fix a procedure for computing the six  NQC parameters,  $(a_i^\lm(\nu),b_i^\lm(\nu))$, 
as continuous functions of $\nu$.  [The remaining defining elements of the EOB model, 
notably $a_5^c$ and $a_6^c$ were already given as functions of $\nu$.]

One can think of  two different ways of continuously extending the definition
of the present EOB model to any value of $\nu$: first, one can interpolate the discrete sample of 
$(a_i^\lm,b_i^\lm)$ values of the NQC parameters that we obtained (from the five numerical simulations
with $q=1,2,3,4,6$) by fitting them to, say, quadratic polynomials in $\nu$; 
second, one can  instead fit  the original NR-extracted numerical values of 
$(A_\lm,\dot{A}_\lm,\ddot{A}_\lm,\omega_\lm,\dot{\omega}_\lm,\ddot{\omega}_\lm)$ to quadratic polynomials
in $\nu$, and then, for any given $\nu$, determine $(a_i^\lm(\nu),b_i^\lm(\nu))$ with
the iterative procedure described above.
We have explored in detail both procedures. The first one, i.e fitting the end parameters 
$(a_i^\lm,b_i^\lm)$ needed to compute an EOB waveform (and explicitly given 
in Table~\ref{tab:nqc_ab} for all $q$'s)
is clearly a faster way to compute, for any $\nu$, a corresponding EOB waveform. 
Indeed, this approach does not require any iteration procedure.

We found that the fitted  $(a_i^{\lm}(\nu), b_i^{\lm}(\nu)) $'s give very accurate results 
for the multipoles we have at hand, i.e. $\ell=m=2$, $\ell=2$, $m=1$, $\ell=m=3$ 
and $\ell=3$, $m=2$. This allows us to construct EOB waveforms that are as accurate 
as the ones obtained by determining $(a_i^\lm,b_i^\lm)$ by  the iterative procedure,
discussed above, that uses the actual NR data. The coefficients of these quadratic 
fits are listed in Table~\ref{tab:nufits_aibi}.

By contrast, the determination of $(a_i^\lm,b_i^\lm)$ from quadratic fits of NR data 
(given in Table~\ref{tab:nufits_A0_lr}) is equally accurate for $\ell=m=2$ waveforms,
but leads to slightly less accurate results
for the subdominant multipoles. More precisely,   this 
procedure introduces some visible, though small, differences  between the EOB and NR 
waveform modulus around the peaks of the ($\ell=2$, $m=1$ and $\ell=m=3$) waveforms. 
Note that, contrarily to the fits of the $(a_i,b_i)$ mentioned above 
(which relied only on the  $q=1,2,3,4,6$ data), we have done quadratic fits of 
$(A_\lm,\dot{A}_\lm,\ddot{A}_\lm,\omega_\lm,\dot{\omega}_\lm,\ddot{\omega}_\lm)$
to {\it six} numerical results, namely the Caltech-Cornell-CITA $q=(1,2,3,4,6)$ 
data together with the $q=\infty$ data of Ref.~\cite{Bernuzzi:2012ku}. Given 
these fits, one then needs to solve for the NQC parameters.
Actually, such a  procedure is simplified by the fact that, as we said, the quadratic fits
for  the $a_i^{22}(\nu)$'s (which are the only NQC parameters which need to be 
reinserted in the flux) can be used from the start, so that, contrary to the 
general case, one can get the needed values of the other NQC parameters in 
one go, without having to iterate the procedure. 

Figure~\ref{fig:compare_fits} illustrates the performances of the two different fitting procedures. 
The figure refers to mass ratio $q=2$ only (equivalent results are found for the other mass ratios,
with improvements for larger values of $q$) and shows the following triple comparison for 
$\ell=m=2$ (top panel) and $\ell=2$, $m=1$ (medium panel), and $\ell=m=3$ (bottom panle) between: 
(i) the NR waveform frequency and  modulus; (ii) the EOB waveform frequency and modulus obtained 
using the fits $(a_i^{\lm}(\nu),b_i^{\lm}(\nu))$; (iii) the EOB  waveform frequency and modulus 
obtained by fitting the numerical data extracted at $\tnrLR$, determining the NQC parameters in the 
usual way, but using the $a_i^{22}$ fits of Table~\ref{tab:nufits_aibi} to account for NQC corrections 
in the radiation reaction.

In conclusion, the  prescription of using the $(a_i^\lm(\nu),b_i^\lm(\nu))$ fits of 
Table~\ref{tab:nufits_aibi} a priori looks as the best (and simplest) choice to obtain the NQC 
parameters interpolating between the discrete sample of NR-computed $q$-values. Since the NR data we have at
hand are limited to the $\ell=3$ multipole, we cannot check the reliability of the procedure
also for higher values of $\ell$. We leave such an investigation to future work.

\section{Conclusions}
\label{sec:end}

We have improved the EOB description of nonspinning coalescing black hole binaries by incorporating
several recent analytical advances, namely:
\begin{itemize}
\item[(i)] 4PN and 5PN logarithmic contributions to the conservative 
    dynamics~\cite{Damour:2009sm,Blanchet:2010zd,Damourlogs,Barausse:2011dq};

\item[(ii)]  the  $\O(\nu)$ 4PN nonlogarithmic contribution to the conservative 
      dynamics ~\cite{Blanchet:2010cx,Blanchet:2010zd,LeTiec:2011dp,Barausse:2011dq};

\item[(iii)] resummed horizon-absorption contributions to angular momentum 
      loss ~\cite{Nagar:2011aa,Bernuzzi:2012ku};

\item[(iv)] the radial component of the radiation reaction force implied 
     by consistency with the azimuthal one~\cite{Bini:2012ji};

\item[(v)] an additional 3.5PN contribution to the phase of the 
    (factorized~\cite{Damour:2007xr,Damour:2007yf,Damour:2008gu}) quadrupolar waveform~\cite{Faye:2012we}.
\end{itemize}
Moreover, we have introduced new features in the EOB formalism, namely:
\begin{itemize}
\item[(a)] a Pad\'e resummation of the additional tail phases $\delta_\lm$ of the factorized EOB waveform;

\item[(b)] a new way of matching the EOB waveform to the NR one by mapping the EOB time when the orbital
  frequency reaches a maximum $\teobLR$ to a specifically chosen ($\nu$-dependent) NR time $\tnrLR(\nu)$
  around merger, Eq.~\eqref{deftnrLR}. More precisely, 
  we impose [by using six next-to-quasi-circular (NQC) parameters]
  a $C^2$ contact between the amplitudes and the frequencies of the NR and EOB waveforms at 
  the NR instant $\tnrLR(\nu)$ which corresponds to $\teobLR$.
\end{itemize}
We have extracted new information from the NR data, namely:
\begin{itemize}
\item[(c)] We showed how to extract from NR (curvature) phasing data the function $Q_\omega^{\rm NR}(\omega)\equiv \omega^2/\dot{\omega}$
which is an intrinsic measure of the phase evolution. We have given an explicit representation 
of the function $Q_\omega^{\rm NR}(\omega;\,q)$, for $q=(1,2,3,4,6)$, in terms of some fitting coefficients 
(see Eqs.~\eqref{eq:Qhat_def},\eqref{Qhat_fit} and Table~\ref{tab:Qomg_fit}).
\item[(d)] We extracted data on the NR amplitude and frequency, together with their first two derivatives,
at the specific $\nu$-dependent NR time  $\tnrLR(\nu)$, which is located a little bit after the maximum
of the quadrupolar waveform amplitude. We gave fitting formulas for the $\nu$-dependence of those
quantities for several multipoles, see Table~\ref{tab:nufits_A0_lr}.
\end{itemize}

Using such nonperturbative information from NR data, we showed how to complete the EOB model by:
\begin{enumerate}
\item Constraining the value of the main EOB radial potential, i.e. the $A(u;\,\nu)$ function; and
\item Determining the coefficients entering the NQC correction factor Eq.~\eqref{eq:hNQC}.
\end{enumerate}
Among these results, we think that the new expression of the NR-tuned $A$ function, containing logarithms,
is more refined and more accurate than its previous determinations~\cite{Damour:2009kr,Pan:2011gk,Taracchini:2012ig}.
Let us recall that, as in previous work, the $A$ function is parametrized in terms of coefficients,
here called $(a_5^c,a_6^c)$, entering a certain Pad\'e approximant, $A^{\rm Pade}(u;\,\nu;\,a_5^c,a_6^c)$, 
Eq.~\eqref{Alogs}.
Then NR data were used to constrain these parameters. We have delineated the reason why the 
two parameters  $(a_5^c,a_6^c)$ entering the Pad\'e definition of $A^{\rm Pade}(u;\,\nu;\,a_5^c,a_6^c)$ 
are degenerate by giving a definition of equivalence classes of the pairs $(a_5^c,a_6^c)$ 
in terms of some $L_\infty$ norm of the $A(u)$ function. We have determined
a good NR-tuned $A$ function by assuming a fixed value of $a_5^c$ ($a_5^c=23.5$ as suggested by
recent GSF results~\cite{Barausse:2011dq,Akcay:2012ea}), 
and by then tuning the remaining parameter $a_6^c(\nu)$. 
We found that $a_6^c(\nu)$ can be simply represented by the mostly-linear function of $\nu$ 
\be
a_6^c(\nu) = [-110.5 + 147(1-4\nu)]\left(1-\dfrac{1.5\times 10^{-5}}{(0.26-\nu)^2}\right)^{1/2},
\ee
where the last, nonlinear~\footnote{Additional NR simulations in the mass-ratio range 
$1\leq q \lesssim 2$  will be needed to probe/improve the nonlinear behavior of the 
$a_6^c(\nu)$ function there.} factor is relevant only in the range $0.\bar{2}\lesssim \nu\leq 0.25$ 
(i.e., $1\leq q \lesssim 2$). We think that the resulting function of $u$ and 
$\nu$, $A^{\rm EOBNR}(u;\,\nu)\equiv A^{\rm Pade}(u;\,\nu;\,23.5,a_6^c(\nu))$,
yields an accurate representation of the $A(u;\,\nu)$ function itself, 
independently of the way it was obtained.
Moreover, we find remarkable that the good value of $A(u;\,\nu)$ could be obtained already
by considering only the {\it inspiral} phasing (before the LSO crossing) and was then 
checked to yield (together with the NR-determined NQC corrections) 
an excellent phasing agreement {\it up to merger}.

We have presented our improved EOB model in a self contained manner so as to allow interested
readers to generate for themselves all our EOB results. We intend to make available soon a public
version of our EOB codes. In view of the new physics that we have included in our EOB model,
and of its excellent performance (obtained without introducing any ad hoc parameters) 
against the very accurate Caltech-Cornell-CITA data, we recommend to use this new EOB model (or small variations thereof)
in future EOB works (in particular in extensions to spinning and/or tidally interacting systems).

\begin{acknowledgments}
  We are grateful L.~Buchman, L.~Kidder, H.~Pfeiffer, M.~Scheel, and B.~Szil\'agyi for 
  making available to us the most recent and accurate simulations of the Caltech-Cornell-CITA
  collaboration. We are particularly indebted to H.~Pfeiffer for continuous help and informative
  communications about the waveforms, and to L.~Buchman for clarifications about the waveform error budget.
  We thank E.~Barausse, A.~Buonanno, Y.~Pan and A.~Taracchini for informative communications about 
  the implementation of their EOB model, Ref.~\cite{Pan:2011gk,Taracchini:2012ig}, as well as G.~Faye
  for checking the 3.5PN term of the $\delta_{22}$ additional phase,
  Eq.~\eqref{eq:delta22_35PN}.  
  S.B. was supported in part by DFG grant SFB/Transregio~7
  ``Gravitational Wave Astronomy''. 
  S.B. thanks IHES for hospitality during the development of part of this work.
\end{acknowledgments}

\appendix

\section{On the computation of $\ddot{r}$}
\label{sec:ddotr}
\begin{figure}[t]
 \includegraphics[width=0.45\textwidth]{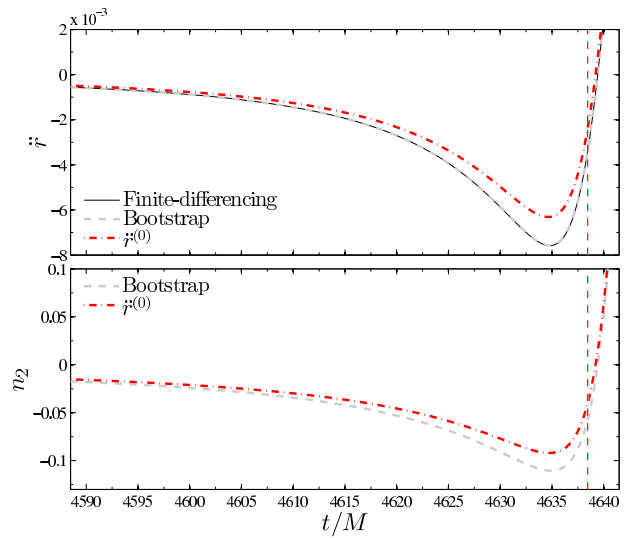}
    \caption{ \label{fig:ddotr_approx}(color online) Top: computation of $\ddot{r}$ 
    with finite differencing and analytical iterations, and comparison with $(\ddot{r})^{(0)}$. 
    Bottom: effect on the NQC basis vector $n_2$. The figure refers to $q=1$ with the choices 
    $a_5^c=23.5$ and $a_6^c(0.25)=-101.876$. See text for discussion.} 
\end{figure}
\begin{table}[t]
  \caption{\label{tab:nufits_AO_mx} Fits of Zerilli-normalized multipolar 
     quantities of the numerical waveforms (modulus, frequency 
     and their derivatives)  measured at the peak of each multipole
    (``maxima'') as function of $\nu$. Each quantity is fitted to 
    a quadratic polynomial of the form $f_\lm(\nu)=c_2^\lm\nu^2+ c_1^\lm\nu+c_0^\lm$. 
    For the modulus and its derivatives the leading order $\nu$-dependence 
    (Eq.~\ref{eq:clm_LO}) is factorized before fitting.} 
  \centering  
  \begin{ruledtabular}  
  \begin{tabular}{c|c|ccc}        
    \hline
    & $\ell\, m$ & $c_2^\lm$ & $c_1^\lm$ & $c_0^\lm$ \\
    \hline
    \multirow{4}{*}{$\dfrac{A_\lm}{\nu c_{\ell+\epsilon}(\nu)}$}
    &2\, 1 &2.9410$\times 10^{-1}$ &-1.0286$\times 10^{-1}$ &1.0691$\times 10^{-1}$ \\
    &2\, 2 &3.8132$\times 10^{-1}$ & 1.3011$\times 10^{-2}$ &2.9467$\times 10^{-1}$ \\
    &3\, 2 &3.9814$\times 10^{-1}$ &-9.2149$\times 10^{-2}$ &1.8310$\times 10^{-2}$ \\
    &3\, 3 &2.0896$\times 10^{-1}$ & 4.7198$\times 10^{-3}$ &5.1463$\times 10^{-2}$ \\
    \hline
    \multirow{4}{*}{$\dfrac{\ddot{A}_\lm}{\nu c_{\ell +\epsilon}(\nu)}$}
    &2\, 1 &-5.2646$\times 10^{-3}$ & 6.1932$\times 10^{-4}$ &-5.4059$\times 10^{-4}$ \\
    &2\, 2 & 1.5609$\times 10^{-4}$ &-1.4628$\times 10^{-3}$ &-4.8017$\times 10^{-4}$ \\
    &3\, 2 & 1.8500$\times 10^{-4}$ & 2.2093$\times 10^{-4}$ &-1.4184$\times 10^{-4}$ \\
    &3\, 3 &-3.4677$\times 10^{-3}$ &-6.6072$\times 10^{-6}$ &-1.6713$\times 10^{-4}$ \\
    \hline
    \multirow{4}{*}{$\omega_\lm$}
    &2\, 1 &5.8728$\times 10^{-1}$ &-8.3459$\times 10^{-2}$ &2.9074$\times 10^{-1}$ \\
    &2\, 2 &4.1410$\times 10^{-1}$ & 2.4377$\times 10^{-1}$ &2.7221$\times 10^{-1}$ \\
    &3\, 2 &3.6315                &-9.5776$\times 10^{-1}$ &4.5459$\times 10^{-1}$ \\
    &3\, 3 &1.0192                & 5.4557$\times 10^{-1}$ &4.5319$\times 10^{-1}$ \\    
    \hline
    \multirow{4}{*}{$\dot{\omega}_\lm$}
    &2\, 1 &-2.2041$\times 10^{-1}$ & 1.0228$\times 10^{-1}$ & 6.2835$\times 10^{-4}$ \\
    &2\, 2 & 2.8060$\times 10^{-2}$ & 1.4581$\times 10^{-2}$ & 5.8725$\times 10^{-3}$ \\
    &3\, 2 &-2.8225$\times 10^{-1}$ & 3.7702$\times 10^{-2}$ & 1.6036$\times 10^{-2}$ \\
    &3\, 3 & 2.5253$\times 10^{-2}$ & 2.7690$\times 10^{-2}$ & 1.0871$\times 10^{-2}$ \\
    \hline    
    \multirow{4}{*}{$\ddot{\omega}_\lm$}
    &2\, 1 & 7.8607$\times 10^{-3}$ & 1.5684$\times 10^{-2}$ &-3.5511$\times 10^{-3}$ \\
    &2\, 2 & 2.3604$\times 10^{-3}$ & 7.2810$\times 10^{-5}$ & 2.2436$\times 10^{-4}$ \\
    &3\, 2 &-8.7028$\times 10^{-2}$ & 1.7233$\times 10^{-2}$ & 8.6570$\times 10^{-5}$ \\
    &3\, 3 &-1.1065$\times 10^{-2}$ &-2.3899$\times 10^{-6}$ & 2.1351$\times 10^{-4}$ \\
    \hline
  \end{tabular}
  \end{ruledtabular}  
\end{table}

In the definition of the NQC correction $n_2$, Eq.~\eqref{eq:ddotr}, we used
for  the second time derivative of the relative separation $r$ the quantity
$(\ddot{r})^{(0)}$, which is the value of $\ddot{r}$ along the 
{\it conservative} dynamics, i.e. neglecting the contributions proportional to $\F$. 
This choice is made for efficiency's sake because it is faster to compute  $(\ddot{r})^{(0)}$ 
along the dynamics. In spite of the neglect of $\F$ in its computation,  
$(\ddot{r})^{(0)}$ does represent an allowed NQC correction because it vanishes 
(together with $\dot{r}$ and the exact value of $\ddot{r}$) in the circular 
limit (see below).

For completeness, let us discuss here how to compute a more exact value 
of $\ddot{r}$ along the dynamics and how the result differs from $(\ddot{r})^{(0)}$.
Let us first recall that along the EOB equations of motion $\dot{r}$ is,
at any moment, a function of the phase space variables: 
$\dot{r}=\dot{r}(r(t),p_\varphi(t),p_{r_*}(t))$. Therefore, its 
total time derivative is the sum of three partial contributions
\be
\ddot{r} = \dfrac{\de \dot{r}}{\de r}\,\dot{r} 
          + \dfrac{\de\dot{r}}{\de p_{r_*}} \dot{p}_{r_*}
          + \dfrac{\de \dot{r}}{\de p_\varphi} \dot{p}_\varphi.
\ee
Using the other EOB equations of motion, this equation reads explicitly
\be
\label{eq:ddotr_full}
\ddot{r} = \dfrac{\de \dot{r}}{\de r}\,\dot{r} 
          + \dfrac{\de\dot{r}}{\de p_{r_*}} \left(\hat{\F}_{r_*} - \dfrac{\de\hat{H}_{\rm EOB}}{\de r{_*}}\right)
          + \dfrac{\de \dot{r}}{\de p_\varphi} \hat{\F}_\varphi.
\ee
where $\de H_{\rm EOB}/\de{r_*}\equiv (A/B)^{1/2}\de H_{\rm EOB}/\de r$.

By definition, the circular dynamics limit corresponds to setting $\dot{r}=0=p_{r_*}$
and $\de H_{\rm EOB}/\de r=0$. One then sees that, along the circular dynamics, one has
also $\F_{r_*}\propto p_{r_*}=0$, and (using $\dot{r}=C(r,p_{r_*},p_{\varphi})p_{r_*}$) 
$\de \dot{r}/{\de p_\varphi}\propto p_{r_*}=0$.
As a consequence, both $\ddot{r}$ and $(\ddot{r})^{(0)}$, 
defined by setting to zero the contributions proportional to $\F$, i.e.
\be
\label{eq:ddotr0}
(\ddot{r})^{(0)} = \dfrac{\de \dot{r}}{\de r}\,\dot{r} 
          - \dfrac{\de\dot{r}}{\de p_{r_*}} \dfrac{\de\hat{H}_{\rm EOB}}{\de r{_*}},
\ee
vanish in the circular dynamics approximation.
This shows that we can use either the exact $\ddot{r}$ or its ``geodesic'' approximation
$(\ddot{r})^{(0)}$ to define the second element of the ``NQC basis'', $n_2=\ddot{r}/(r\Omega^2)$.

When using the definition $n_2=(\ddot{r})^{(0)}/(r\Omega^2)$, Eq.~\eqref{eq:ddotr0} 
allows one to compute immediately $n_2$ along the exact dynamics. By contrast, if one wished to use
the definition $n_2'=\ddot{r}/(r\Omega^2)$, a complication arises. Indeed, as contributions 
proportional to $\hat{\F}_{r_*}$ and $\hat{\F}_{\varphi}$ appear on the r.h.s. of 
Eq.~\eqref{eq:ddotr_full}, and as these contain the squared modulus of the NQC 
factor (i.e., for each multipole, a factor $|1+\sum_ja_j^{\lm}n_j|^2$) we see that
$n_2'\propto \ddot{r}$ now appears on both sides of Eq.~\eqref{eq:ddotr_full}.

Schematically, defining $\xi=(r,p_{r_*},p_{\varphi})$, Eq.~\eqref{eq:ddotr_full} has the structure
\be
\label{eq:ddotr_scheme}
\ddot{r} = a(\xi) + b(\xi)\hat{\F}_{r_*}(\xi,\ddot{r}) + c(\xi)\hat{\F}_\varphi(\xi,\ddot{r})
\ee
which only gives an {\it implicit} equation for determining the exact $\ddot{r}$ along the
dynamics.  We can however get an explicit expression for $\ddot{r}$ by an iterative procedure.
Inserting $\ddot{r}^{(0)}$ as lowest order approximation on the r.h.s. of Eq.~\eqref{eq:ddotr_scheme}
defines an improved value, say $(\ddot{r})^{(1)}$ for $\ddot{r}$, namely
\be
\ddot{r}^{(1)} = a(\xi) + b(\xi)\hat{\F}_{r_*}(\xi,\ddot{r}^{(0)}) + c(\xi)\hat{\F}_\varphi(\xi,\ddot{r}^{(0)}).
\ee
By iterating the procedure once more, we then get
\be
\label{eq:ddotr_final}
\ddot{r}^{(2)} = a(\xi) + b(\xi)\hat{\F}_{r_*}(\xi,\ddot{r}^{(1)}) + c(\xi)\hat{\F}_\varphi(\xi,\ddot{r}^{(1)}).
\ee
The result~\eqref{eq:ddotr_final} leads to a sufficiently accurate computation of $\ddot{r}$ 
up to merger, as illustrated in the top panel of Fig.~\ref{fig:ddotr_approx}.
However, the  recursive presence of the flux in this iteration substantially increases
(by approximately a factor 4) the computational time needed to produce an EOB waveform.
This is why we prefer to use $n_2=\ddot{r}^{(0)}/(r\Omega^2)$.
Anyway, as Fig.~\ref{fig:ddotr_approx} shows, $n_2$ and $n_2'$ are numerically quite 
similar. In view of the arguments above their differences are essentially absorbed in
a redefinition of the coefficients $a_i$.

\section{NQC factor determined using NR data at $\tAmax$}
\label{sec:NQC_at_max}
\begin{figure}[t]
  \begin{center}   
    \includegraphics[width=0.5\textwidth]{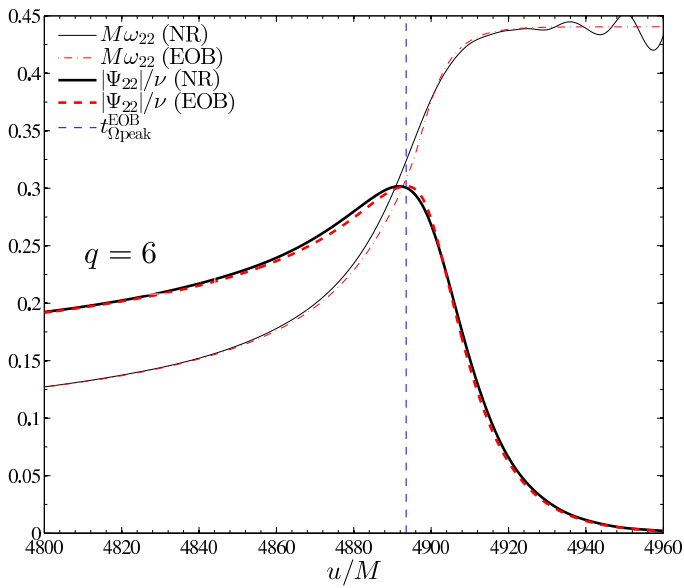}
    \caption{ \label{fig:NQC_at_maxA22} (color online) Mass ratio $q=6$: EOB waveform 
      (frequency and modulus) obtained by determining NQC corrections from NR data 
      extracted at $\tAmax$ instead of $\tnrLR$.}  
  \end{center}
\end{figure}

In the text, we argued that it was advantageous to determine NQC corrections by matching
the EOB waveform (considered at $\teobLR$) to the NR waveform considered at the time
$\tnrLR$. Let us illustrate here (see Fig.~\ref{fig:NQC_at_maxA22}) 
in the case $q=6$ the slightly different (but significantly worsened) EOB waveform 
obtained when one instead matches the $\ell=m=2$ EOB waveform (considered at time
$\teobLR$) to the NR waveform considered at the time $\tAmax$ (as was done in early EOB works). 
Figure~\ref{fig:NQC_at_maxA22} uses as before 6 NQC corrections and the 
value $a_6^c(6/49)=-44.67$. However, the NR extraction point, which is also used 
as NQC determination point, is now  $\tAmax$. 

The fits of the vector of NR quantities 
$(A_\lm^{\rm NR},\ddot{A}_\lm^{\rm NR},\omega_\lm^{\rm NR},\dot{\omega}_\lm^{\rm NR},\ddot{\omega}_\lm^{\rm NR})$
now measured at the location of the maximum of each multipole are given in Table~\ref{tab:nufits_AO_mx} 
and include, as before, the test-mass information. We checked that these fits are compatible 
with the fits given in Table~II of Ref.~\cite{Pan:2011gk}.

When comparing Fig.~\ref{fig:NQC_at_maxA22} with the bottom left panel of Fig.~\ref{fig:phasing_q3q4},
we see that, though the effect of having replaced $\tnrLR$ by $\tAmax$ is small, it leads
to visible differences. In particular, one sees that the frequency evolution near merger
was more accurately captured in Fig.~\ref{fig:phasing_q3q4} than in Fig.~\ref{fig:NQC_at_maxA22}.

\section{Effect of including NQC corrections to higher multipoles in the radiation reaction}
\label{sec:Flm_high}
In this Appendix we explore the effect of including the NQC correction factor in the higher multipole
contributions to radiation reaction, specifically in some of the main subdominant multipoles, 
$\hat{h}_{21}^{\rm NQC}$, $\hat{h}_{33}^{\rm NQC}$ and $\hat{h}_{32}^{\rm NQC}$. [By contrast in the main
text we NQC corrected only $\hat{h}_{22}^{\rm NQC}$ in the radiation reaction].
Note that with our choice $x=v_\varphi^2$ of the argument in $\rho_\lm(x)$ we need larger NQC
modulus correction factors than Ref.~\cite{Pan:2011gk} which used $x=\Omega^{2/3}$. 
Indeed as during the plunge $\Omega^{2/3}$ is larger than $v_\varphi^2$ and as the function $\rho_\lm(x)$
is a decreasing function of its argument, one has, along the EOB dynamics, 
$\left(\rho_\lm(v_\varphi^2)\right)^\ell>\left(\rho_\lm(\Omega^{2/3})\right)^\ell$. Therefore 
the inclusion of NQC corrections for higher multipoles is apriori more significant within
our EOB setup than within the one of  Ref.~\cite{Pan:2011gk}.
\begin{figure}[t]
  \begin{center}   
    \includegraphics[width=0.45\textwidth]{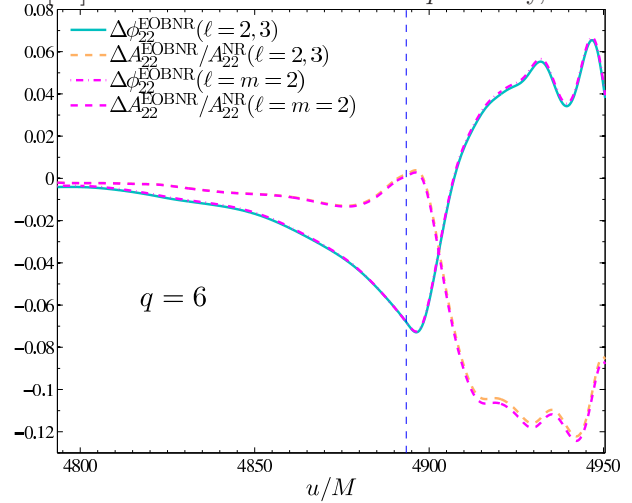}
    \caption{ \label{fig:q6_phasing_fNQC_improved}(color online) 
    Negligible effect on the phasing (and modulus), for $q=6$, of including 
    $\ell=2$, $m=1$ and $\ell=3$, $(m=2,3)$ NQC corrections to the energy 
    flux beyond the $\ell=m=2$ one.}  
  \end{center}
\end{figure}
We focus on the mass ratio $q=6$ only, because subdominant multipoles do not significantly 
contribute when $q\sim 1$. Figure~\ref{fig:q6_phasing_fNQC_improved} compares the phase 
difference and the fractional amplitude difference for two EOB models: 
one with the standard $h_{22}$--only NQC flux correction (magenta online), 
and another one which includes in addition the three subleading NQC factors 
$\hat{h}_{21}^{\rm NQC}$, $\hat{h}_{33}^{\rm NQC}$  and $\hat{h}_{32}^{\rm NQC}$.
The effect of this inclusion is totally negligible, so that it is justified
to include only the $\ell=m=2$ NQC correction to the radiation reaction.  

\section{Explicit expression of $\rho_\lm(x)$ and $\delta_\lm(y)$}
\label{sec:rholm}
In this Appendix we list the explicit expressions of the residual amplitude, $\rho_\lm(x)$, 
and phase, $\delta_\lm(y)$ corrections that we have implemented in our EOB code. They rely
on the results of Refs.~\cite{Damour:2008gu,Fujita:2010xj}. We give explicit expressions
for all multipoles up to $\ell=8$ included. Such expressions are given at the 
$3^{+2}$~PN approximation, i.e. the 3PN-accurate, $\nu\neq0$ results of Ref.~\cite{Damour:2008gu}
are {\it hybridized} with the 5PN-accurate, $\nu=0$, terms obtained in Ref.~\cite{Fujita:2010xj}.
Let us recall that we used here the following values of the arguments of these functions: $x\to v_\varphi^2$
in $\rho_\lm(x)$ and $y\to (H_{\rm EOB}\Omega)^{2/3}$.

\begin{widetext}
\begin{align}
%
\rho_{22}(x;\nu)&= 1 +\left(\frac{55 \nu }{84}-\frac{43}{42}\right) x 
+\left(\frac{19583 \nu^2}{42336}-\frac{33025 \nu
}{21168}-\frac{20555}{10584}\right) x^2 \nonumber\\
&+\left(\frac{10620745 \nu ^3}{39118464}-\frac{6292061 \nu ^2}{3259872}+\frac{41 \pi
   ^2 \nu }{192}-\frac{48993925 \nu }{9779616}-\frac{428}{105}
  \text{eulerlog}_{2}(x)+\frac{1556919113}{122245200}\right) x^3 \nonumber\\
&+\left(\frac{9202}{2205}\text{eulerlog}_2(x)-\frac{387216563023}{160190110080}\right) x^4
+\left(\frac{439877}{55566}\text{eulerlog}_{2}(x)-\frac{16094530514677}{533967033600}\right)x^5,\\
\nonumber\\
%
\rho_{21}(x;\nu)&=1+\left(\frac{23 \nu }{84}-\frac{59}{56}\right) x +\left(\frac{617 \nu ^2}{4704}-\frac{10993\nu }{14112}-\frac{47009}{56448}\right) x^2\nonumber\\
&+\left(\frac{7613184941}{2607897600}-\frac{107}{105}\text{eulerlog}_1(x)\right)x^3
+\left(\frac{6313}{5880}\text{eulerlog}_1(x)-\frac{1168617463883}{911303737344}\right)x^4\nonumber\\ 
&+ \left( - \frac {63735873771463}{16569158860800}  + \frac {5029963}{5927040} \,{\rm eulerlog}_1(x)\right)\,x^{5},\\
\nonumber\\
%
\rho_{33}(x;\nu) &= 1+\left(\frac{2 \nu }{3}-\frac{7}{6}\right) x+\left(\frac{149 \nu ^2}{330}-\frac{1861 \nu }{990}-\frac{6719}{3960}\right) x^2
+\left(\frac{3203101567}{227026800}-\frac{26}{7} \text{eulerlog}_{3}(x)\right)x^3\nonumber\\ 
& +\left(\frac{13}{3}\text{eulerlog}_{3}(x)-\frac{57566572157}{8562153600}\right)x^4
+ \left( - \frac {903823148417327}{30566888352000} + \frac {87347}{13860} \,{\rm eulerlog}_3(x)\right)\,x^{5},\\
\nonumber\\
%
\rho_{32}(x;\nu)&=1 +\frac{320 \nu ^2-1115\nu +328}{270 (3 \nu -1)}x 
+\frac{3085640 \nu ^4-20338960 \nu ^3-4725605 \nu ^2+8050045\nu -1444528}{1603800 (1-3 \nu )^2}x^2\nonumber\\
&+\left(\frac{5849948554}{940355325}-\frac{104}{63}\text{eulerlog}_2(x)\right)x^3
+ \left( - \frac {10607269449358}{3072140846775} + \frac {17056}{8505} \,{\rm eulerlog}_2(x)\right)\,x^{4},\\
\nonumber\\
%
\rho_{31}(x;\nu) &= 1+\left(-\frac{2 \nu}{9}-\frac{13}{18}\right) x 
+\left(-\frac{829 \nu ^2}{1782}-\frac{1685\nu }{1782}+\frac{101}{7128}\right) x^2
+\left(\frac{11706720301}{6129723600}-\frac{26}{63}\text{eulerlog}_1(x)\right) x^3\nonumber\\
&+\left(\frac{169}{567}
\text{eulerlog}_1(x)+\frac{2606097992581}{4854741091200}\right) x^4
+ \left( \frac {430750057673539}{297110154781440}  - \frac
{1313}{224532} \,{\rm eulerlog}_1 (x)\right)\,x^{5},\\
\nonumber\\
%
\rho_{44}(x;\nu) &=1+\frac{2625 \nu ^2-5870 \nu +1614}{1320 (3 \nu -1)}x\nonumber\\
&+\frac{1252563795 \nu^4-6733146000 \nu^3-313857376 \nu^2+2338945704 \nu -511573572}{317116800 (1-3 \nu )^2}x^2\nonumber\\
&+\left(\frac{16600939332793}{1098809712000}-\frac{12568}{3465}\text{eulerlog}_4(x)\right)x^3
+ \left( \frac {845198}{190575} \,{\rm eulerlog}_4 (x) - \frac {172066910136202271}{19426955708160000} \right)\,x^{4},\\
\nonumber\\
%
\rho_{43}(x;\nu)&=1+ \frac{160 \nu ^2-547 \nu +222}{176 (2\nu-1)}x
-\frac{6894273}{7047040}x^2
+ \left( - \frac {1571}{770} \,{\rm eulerlog}_3(x) + \frac {1664224207351}{195343948800} \right)\,x^{3}\nonumber\\
& + \left( - \frac {2465107182496333}{460490801971200}  + \frac {174381}{67760} \,{\rm eulerlog}_3(x)\right)\,x^{4},
\nonumber\\
%
\rho_{42}(x;\nu)&=1 +\frac{285 \nu ^2-3530 \nu +1146}{1320 (3 \nu -1)}x\nonumber\\
&  +\frac{-379526805 \nu ^4-3047981160 \nu ^3+1204388696 \nu ^2+295834536 \nu -114859044}{317116800 (1-3 \nu )^2}x^2\nonumber\\
&
+\left(\frac{848238724511}{219761942400}-\frac{3142}{3465}\text{eulerlog}_2(x)\right)
x^3
+ \left( \frac {300061}{381150} \,{\rm eulerlog}_2(x) - \frac {12864377174485679}{19426955708160000} \right)\,x^{4},
\end{align}
%
%
%
%
\begin{align}
%
\rho_{41}(x;\nu)&= 1 +\frac{ 288 \nu^2 -1385\nu + 602 }{528 (2 \nu -1)}x
-\frac{7775491}{21141120}x^2
+ \left( - \frac {1571}{6930} \,{\rm eulerlog}_1(x) + \frac {1227423222031}{1758095539200} \right)\,x^{3}\nonumber\\
&
+ \left( - \frac {29584392078751453}{37299754959667200}  + \frac {67553}{261360} \,{\rm eulerlog}_1(x)\right)\,x^{4},\\
%
\rho_{55}(x;\nu)&=1 + \dfrac{512 \nu ^2-1298 \nu +487}{390(2\nu-1)} x -
\dfrac{3353747}{2129400}x^2
+ \left( - \frac {1546}{429} \,{\rm eulerlog}_5(x) + \frac {190606537999247}{11957879934000} \right)\,x^{3} \nonumber\\
& + \left( - \frac {1213641959949291437}{118143853747920000} + \frac {376451}{83655} \,{\rm eulerlog}_5(x)\right)\,x^{4},\\
\nonumber\\
%
\rho_{54}(x;\nu)&= 1 + \dfrac{33320\nu^3 - 127610\nu^2 + 96019\nu -
  17448 }{13650(5\nu^2 - 5\nu + 1 )}x
- \frac {16213384}{15526875} \,x^{2} \nonumber\\
& + \left( - \frac {24736}{10725} \,{\rm eulerlog}_4(x) + \frac {6704294638171892}{653946558890625} \right)\,x^{3},\\
\nonumber\\
%
\rho_{53}(x;\nu)&= 1 + \dfrac{176 \nu ^2-850 \nu +375}{390(2\nu-1)}x   - \dfrac{410833}{709800}x^2
+ \left( - \frac {4638}{3575} \,{\rm eulerlog}_3(x) + \frac {7618462680967}{1328653326000} \right)\,x^{3} \nonumber\\
& + \left( - \frac {77082121019870543}{39381284582640000} + \frac {2319}{1859} \,{\rm eulerlog}_3(x)\right)\,x^{4},\\
\nonumber\\
%
\rho_{52}(x;\nu)&=1 + \dfrac{21980\nu^3 - 104930\nu^2 + 84679\nu - 15828}{13650(5\nu^2 -5\nu +1)}x
- \frac {7187914}{15526875} \,x^{2} \nonumber\\
& + \left( \frac {1539689950126502}{653946558890625}  - \frac
{6184}{10725} \,{\rm eulerlog}_2(x)\right)\,x^{3},\\ 
\nonumber\\
%
\rho_{51}(x;\nu)&= 1 + \dfrac{8\nu^2-626\nu+319}{390(2\nu-1)}x -
\dfrac{31877}{304200}x^2
+ \left( - \frac {1546}{10725} \,{\rm eulerlog}_1(x) + \frac
{7685351978519}{11957879934000} \right)\,x^{3} \nonumber\\ 
& + \left( - \frac {821807362819271}{10740350340720000} + \frac
{22417}{190125} \,{\rm eulerlog}_1(x)\right)\,x^{4},\\ 
\nonumber\\
%
\rho_{66}(x;\nu)&=1+\frac{273 \nu ^3-861 \nu ^2+602 \nu
  -106}{84\left(5 \nu ^2-5 \nu +1\right)}x
- \frac {1025435}{659736} \,x^{2} + \left( - \frac {3604}{1001} \,{\rm
  eulerlog}_6(x) + \frac {610931247213169}{36701493028200}
\right)\,x^{3},\\ 
\nonumber\\
%
\rho_{65}(x;\nu)&=1+\frac{220 \nu ^3-910 \nu ^2+838 \nu -185}{144 \left(3
  \nu ^2-4 \nu +1\right)}x
- \frac {59574065}{54286848} \,x^{2} + \left( - \frac {22525}{9009}
\,{\rm eulerlog}_5(x) + \frac {67397117912549267}{5798416452820992}
\right)\,x^{3},\\ 
\nonumber\\
%
\rho_{64}(x;\nu)&=1+\frac{133 \nu ^3-581 \nu ^2+462 \nu -86}{84
  \left(5 \nu ^2-5 \nu +1\right)}x
- \frac {476887}{659736} \,x^{2} + \left( - \frac {14416}{9009} \,{\rm
  eulerlog}_4(x) + \frac {180067034480351}{24467662018800}
\right)\,x^{3},\\ 
\nonumber\\
%
\rho_{63}(x;\nu)&=1+\frac{156 \nu ^3-750 \nu ^2+742 \nu -169}{144 \left(3
  \nu ^2-4 \nu +1\right)}x
- \frac {152153941}{271434240} \,x^{2} + \left( - \frac {901}{1001}
\,{\rm eulerlog}_3(x) + \frac {116042497264681103}{28992082264104960}
\right)\,x^{3},\\ 
\nonumber\\
%
\rho_{62}(x;\nu)&=1+\frac{49 \nu ^3-413 \nu ^2+378 \nu
  -74}{84\left(5 \nu ^2-5 \nu +1\right)}x
- \frac {817991}{3298680} \,x^{2} + \left( - \frac {3604}{9009} \,{\rm
  eulerlog}_2(x) + \frac {812992177581}{453104852200}
\right)\,x^{3},\\ 
\nonumber\\
%
\rho_{61}(x;\nu)&=1+\frac{124 \nu ^3-670 \nu ^2+694 \nu -161}{144 \left(3 \nu ^2-4 \nu +1\right)}x
- \frac {79192261}{271434240} \,x^{2} + \left( - \frac {901}{9009}
\,{\rm eulerlog}_1(x) + \frac
  {6277796663889319}{28992082264104960}\right)\,x^{3}, 
\end{align}
%
%
%
%
\begin{align}
%
\rho_{77}(x;\nu)&=1+\frac{1380 \nu ^3-4963 \nu ^2+4246 \nu -906}{714
  \left(3 \nu ^2-4 \nu +1\right)}x
- \frac {32358125}{20986602} \,x^{2} 
+ \left( -  \frac {11948}{3315} \,{\rm eulerlog}_7(x) + \frac
  {66555794049401803}{3856993267327200} \right)\,x^{3},\\ 
\nonumber\\
%
\rho_{76}(x;\nu)&=1+
\frac{6104 \nu ^4-29351 \nu ^3+37828 \nu ^2-16185 \nu +2144}{1666 \left(7 \nu
  ^3-14 \nu ^2+7 \nu -1\right)}x 
- \frac {195441224}{171390583} \,x^{2},\\
\nonumber\\
%
\rho_{75}(x;\nu)&=1+\frac{804 \nu ^3-3523 \nu ^2+3382 \nu
  -762}{714 \left(3 \nu ^2-4 \nu +1\right)}x
- \frac {17354227}{20986602} \,x^{2} + \left( - \frac {59740}{32487}
  \,{\rm eulerlog}_5(x) + \frac {192862646381533}{22039961527584}
  \right)\,x^{3},\\ 
\nonumber\\
%
\rho_{74}(x;\nu)&=1+\frac{41076 \nu ^4-217959 \nu ^3+298872 \nu
  ^2-131805 \nu +17756}{14994 \left(7 \nu ^3-14 \nu ^2+7 \nu
  -1\right)}x 
- \frac {2995755988}{4627545741} \,x^{2},\\
\nonumber\\
%
\rho_{73}(x;\nu)&=1+\frac{420 \nu ^3-2563 \nu ^2+2806 \nu
  -666}{714 \left(3 \nu ^2-4 \nu +1\right)}x
- \frac {7804375}{20986602} \,x^{2} + \left( 
- \frac {35844}{54145} \,{\rm eulerlog}_3(x) + 
\frac {1321461327981547}{428554807480800} \right)\,x^{3},\\ 
\nonumber\\
%
\rho_{72}(x;\nu)&=1+\frac{32760 \nu ^4-190239 \nu ^3+273924 \nu ^2-123489
  \nu +16832}{14994 \left(7 \nu ^3-14 \nu ^2+7 \nu -1\right)}x 
- \frac {1625746984}{4627545741} \,x^{2},\\
\nonumber\\
%
\rho_{71}(x;\nu)&=1+\frac{228 \nu ^3-2083 \nu ^2+2518 \nu
  -618}{714 \left(3 \nu ^2-4 \nu +1\right)}x
- \frac {1055091}{6995534} \,x^{2} + \left( - \frac {11948}{162435}
  \,{\rm eulerlog}_1(x) + \frac {142228318411021}{550999038189600}
  \right)\,x^{3}, \\
%
%
\rho_{88}(x;\nu)&=1+\frac{12243 \nu^4-53445 \nu^3+64659 \nu^2-26778 \nu
  +3482}{2736 \left(7 \nu^3-14 \nu^2+7\nu -1\right)}x
- 1.5337092502821381\, x^2,\\
\nonumber\\
\rho_{87}(x;\nu)&=1+\frac{38920 \nu ^4-207550 \nu ^3+309498 \nu ^2-154099
  \nu +23478}{18240 \left(4 \nu ^3-10 \nu ^2+6 \nu -1\right)}x
 - 1.175404252991305\, x^2,\\
\nonumber\\
\rho_{86}(x;\nu)&= 1 +\frac{2653 \nu^4-13055 \nu^3+17269 \nu^2-7498 \nu
  +1002}{912\left(7 \nu^3-14\nu^2+7\nu-1\right)}x
- 0.9061610303170207\, x^2,\\
\nonumber\\
\rho_{85}(x;\nu)&=1+\frac{6056 \nu ^4-34598 \nu ^3+54642 \nu ^2-28055 \nu
  +4350}{3648 \left(4 \nu ^3-10 \nu ^2+6 \nu -1\right)}x
- 0.7220789990670207\, x^2,\\
\nonumber\\
\rho_{84}(x;\nu)&=1+\frac{4899 \nu ^4-28965 \nu ^3+42627 \nu ^2-19434 \nu
  +2666}{2736 \left(7 \nu ^3-14 \nu ^2+7 \nu -1\right)}x
- 0.47652059150068155\, x^2
\nonumber\\
\rho_{83}(x;\nu)&=1+\frac{24520 \nu ^4-149950 \nu ^3+249018 \nu ^2-131059
  \nu +20598}{18240 \left(4 \nu ^3-10 \nu ^2+6 \nu -1\right)}x
- 0.4196774909106648\, x^2,\\
\nonumber\\
\rho_{82}(x;\nu)&=1+\frac{3063 \nu^4-22845 \nu^3+37119 \nu ^2-17598 \nu
+2462}{2736 \left(7 \nu^3-14 \nu^2+7 \nu -1\right)}x
- 0.2261796441029474\, x^2\\
\nonumber\\
\rho_{81}(x;\nu)&=1+\frac{21640 \nu ^4-138430 \nu ^3+236922 \nu ^2-126451
  \nu +20022}{18240 \left(4 \nu ^3-10 \nu ^2+6 \nu -1\right)}x
- 0.26842133517043704\, x^2
\end{align}
\end{widetext}
The ``eulerlog'' functions $\text{eulerlog}_{m}(x)$ are defined as 
\begin{equation}
\label{eq:eulerlog}
\text{eulerlog}_m(x)  = \gamma_E + \log 2+\dfrac{1}{2}\log x + \log m
\ ,
\end{equation}
where $\gamma_{E}$ is Euler's constant, $\gamma_{E}=0.577215\dots$
and $\log(x)$ the natural logarithm function.

Let us now give the explicit expression of the residual phase corrections
$\delta_\lm$ that are implemented in the code. For $\delta_{2m}$, $\delta_{33}$ 
and $\delta_{31}$ we list here explicitly both their Taylor-expanded forms
(labeled with a ``Taylor'' superscript) and  their Pad\'e resummed ones. 
The $\delta_\lm$ for higher multipoles can be given only
in Taylor-expanded form and thus the label ``Taylor'' is omitted.
The terms in boldface are the highest-order known PN terms for $\nu=0$. 
They are omitted when $\nu\neq 0$, and in particular in
the computation of the Pad\'e approximants, but they are kept in 
the computation of the $\nu=0$ EOB waveform.
The Taylor-expanded $\delta_\lm$ read
 \begin{widetext}
 \begin{align}
\delta_{22}^{\rm Taylor} & = \dfrac{7}{3}y^{3/2} - 24\nu y^{5/2} +
\dfrac{428}{105}\pi y^3 \left(\dfrac{30995}{1134}\nu +
\dfrac{962}{135}\nu^2\right) y^{7/2}  \boldsymbol{+
  \left(\dfrac{1712}{315}\pi^2-\dfrac{2203}{81} \right) y^{9/2}}, \\ 
\delta_{21}^{\rm Taylor} & = \dfrac{2}{3}y^{3/2} - \dfrac{493}{4}\nu
y^{5/2}  + \dfrac{107}{105}\pi y^3
\boldsymbol{+\left(\dfrac{214}{315}\pi^2 -\dfrac{272}{81} \right)
  y^{9/2}},\\ 
\delta_{33}^{\rm Taylor} & = \dfrac{13}{10}y^{3/2} -
\dfrac{80897}{2430}\nu y^{5/2}   +   \dfrac{39}{7}\pi y^3 
\boldsymbol{+ \left(\dfrac{78}{7}\pi^2 -\dfrac{227827}{3000}\right)
  y^{9/2}},\\ 
\delta_{32} &=\dfrac{10+33\nu}{15(1-3\nu)}y^{3/2}+  \dfrac{52}{21}\pi
y^3  \boldsymbol{+\left(\dfrac{208}{63}\pi^2 - \dfrac{9112}{405}\right)y^{9/2}} ,\\
\delta_{31}^{\rm Taylor}
&=\dfrac{13}{30}y^{3/2}-\dfrac{17\nu}{10}y^{5/2}+\dfrac{13}{21}\pi
y^3+\boldsymbol{\left(\dfrac{26}{63}\pi^2 - \dfrac{227827}{81000} +
  \right)y^{9/2}},\\ 
\delta_{44} &=\dfrac{112+219\nu}{120(1-3\nu)}y^{3/2}+ \dfrac{25136}{3465}\pi y^3 \boldsymbol{+\left( \dfrac{201088}{10395}\pi^2 - \dfrac{55144}{375} \right)y^{9/2}},\\
 \delta_{43} &= \dfrac{486+4961\nu}{810(1-2\nu)}y^{3/2}+ \dfrac{1571}{385}\pi y^3,\\
\delta_{42} &=\dfrac{7(1+6\nu)}{15(1-3\nu)}y^{3/2}+  \dfrac{6284}{3465}\pi y^3 \boldsymbol{+ \left( \dfrac{25136}{10395}\pi^2 - \dfrac{6893}{375}\right)y^{9/2}},\\
 \delta_{41} &=\dfrac{2+507\nu}{10(1-2\nu)}y^{3/2} + \dfrac{1571}{3465} \pi y^3,\\
\delta_{55} &=\dfrac{96875+857528\nu}{131250(1-2\nu)}y^{3/2} \ .
\end{align}
\end{widetext}
Among these, we used $\delta_{32}$, $\delta_{4m}$ and $\delta_{55}$ in their Taylor-expanded form
indicated above. By contrast, for $\delta_{22}$, $\delta_{21}$, $\delta_{33}$ and $\delta_{31}$
we used (denoting $v_y\equiv\sqrt{y}$) the following Pad\'e resummed expressions 
(see Section~\ref{sec:resum_deltalm} for further details) 
\begin{widetext}
  \begin{align}
 \delta_{22} &= \dfrac{7}{3}v_y^3\,
 \dfrac{808920\nu\pi v_y + 137388\pi^2 v_y^2 + 35\nu^2(136080 + (154975
   - 1359276\nu) v_y^2)}
 {808920\nu\pi v_y + 137388\pi^2 v_y^2 + 35\nu^2(136080 + (154975 +
   40404\nu) v_y^2)},\\
 \delta_{21} &= \dfrac{2}{3}v_y^3\,  \dfrac{69020\nu +
   5992\pi v_y}{5992\pi v_y + 2456\nu (28+493\nu v_y^2)},\\
 \delta_{33} &= \dfrac{13}{10}v_y^3 \,
 \dfrac{1+94770\pi v_y/(566279\nu)}{1 +
   94770\pi v_y/(566279\nu) + 80897\nu v_y^2/3159},\\
\delta_{31}&=\dfrac{13}{30}v_y^3\,\dfrac{4641\nu + 1690\pi
  v_y}{4641\nu + 1690\pi v_y+18207 v_y^2\,\nu^2}.
\end{align}
\end{widetext}

\bibliography{refs20130415.bib}{}

\end{document}